\documentclass[11pt]{article}
\usepackage[british,UKenglish,USenglish,english,american]{babel}
\usepackage[utf8]{inputenc}
\usepackage{latexsym}
\usepackage{graphicx}
\usepackage{amsfonts}
\usepackage{amssymb}
\usepackage{amsmath,bm}
\usepackage{mathrsfs}
\usepackage{dsfont}
\usepackage{amsthm}
\usepackage{mdframed}
\usepackage{lipsum}
\usepackage{bbm}
\usepackage{color}
\usepackage{capt-of}
\usepackage{booktabs}
\usepackage{xcolor}
\usepackage{hyperref}
\usepackage[labelfont=bf]{caption}

\newcommand{\plus}{\raisebox{.4\height}{\scalebox{.75}{$+$}}}
\newcommand{\minus}{\raisebox{.4\height}{\scalebox{.75}{$-$}}}

\usepackage{subcaption}
\newtheorem{Prop}{Proposition}
\newtheorem{Lem}{Lemma}
\newtheorem{Cor}{Corollary}

\numberwithin{equation}{section}

\makeatletter
\newcommand*{\rom}[1]{\expandafter\@slowromancap\romannumeral #1@}
\makeatother
\title{Geometric Step Options with Jumps: Parity Relations, PIDEs, and Semi-Analytical Pricing}
\author{\sc \Large Walter Farkas \footnote{walter.farkas@bf.uzh.ch} \hspace{1.3em} Ludovic Mathys\footnote{ludovic.mathys@bf.uzh.ch} \vspace{0.5em} \\
       {\it Department of Banking and Finance, University of Zurich, Switzerland.} \vspace{0.2em} \\
			{\it Department of Mathematics, ETH Zurich, Switzerland.} \vspace{0.2em} \\
			{\it Swiss Finance Institute, Switzerland.} }
\date{}
\providecommand{\keywords}[1]{\textbf{Keywords:} #1}
\providecommand{\mscclass}[2]{\textbf{MSC (2010) Classification:} #1}
\providecommand{\jelclass}[3]{\textbf{JEL Classification:} #1}
\providecommand{\acknow}[4]{\textbf{Acknowledgements:} #1}

\begin{document}

\maketitle

\thispagestyle{empty}
% The abstract

\begin{abstract}
\noindent The present article studies geometric step options in exponential Lévy markets. Our contribution is manifold and extends several aspects of the geometric step option pricing literature. First, we provide symmetry and parity relations and derive various characterizations for both European-type and American-type geometric double barrier step options. In particular, we are able to obtain a jump-diffusion disentanglement for the early exercise premium of American-type geometric double barrier step contracts and its maturity-randomized equivalent as well as to characterize the diffusion and jump contributions to these early exercise premiums separately by means of partial integro-differential equations and ordinary integro-differential equations. As an application of our characterizations, we derive semi-analytical pricing results for (regular) European-type and American-type geometric down-and-out step call options under hyper-exponential jump-diffusion models. Lastly, we use the latter results to discuss the early exercise structure of geometric step options once jumps are added and to subsequently provide an analysis of the impact of jumps on the price and hedging parameters of (European-type and American-type) geometric step contracts.
\end{abstract}
$\;$ \vspace{2em} \\
\noindent \keywords{Geometric Step Options, American-Type Options, Lévy Markets, Jump-Diffusion Disentanglement, Maturity-Randomization.} \vspace{0.5em} \\
% Write down at least 3 Keywords
\noindent \mscclass{91-08, 91B25, 91B70, 91G20, 91G60, 91G80.}{} \vspace{0.5em}\\
\noindent \jelclass{C32, C61, C63, G13.} \vspace{-0.5em} \\
\newpage 
\section{Introduction}
\noindent Since their introduction in the seminal article of Linetsky (cf.~\cite{li99}) and their generalization in the subsequent work of Davydov and Linetsky (cf.~\cite{dl02}) geometric step options have constantly gained attention in both the financial industry and the academic literature (cf.~\cite{ccw10}, \cite{cmw13}, \cite{xy13}, \cite{lz16}, \cite{wzb17}, \cite{dlm19}). As a whole class of financial contracts written on an underlying asset, these options have the particularity to cumulatively and proportionally loose or gain value when the underlying asset price stays below or above a predetermined threshold and consequently offer a continuum of alternatives between standard options and (standard) barrier options. Especially when compared with the latter options, geometric step contracts bring clear advantages: Due to their immediate cancellation (or activation) when the barrier level is breached, (standard) barrier options are extremely sensitive to any (temporary) change in the underlying asset price near the barrier so that (delta-)hedging is not reasonably feasible in this region. Additionally, the immediate knock-out (or knock-in) feature inherent to (standard) barrier options may incentivize influential market participants to manipulate the underlying asset price close to the barrier, hence triggering cancellation (or activation) of these options. Switching from an immediate to a cumulative and proportional knock-out (or knock-in) feature instead substantially helps addressing these concerns. Indeed, in contrast to (standard) barrier options, the delta of geometric step contracts does not explode and is even continuous at the barrier. This already allows for typical delta-hedges across the barrier level. Furthermore, since it is more difficult to control underlying asset prices over an extended period of time, geometric step options are more robust to temporary market manipulations and therefore better protect their holders against adverse actions of market participants in the underlying asset. \vspace{1em} \\
\noindent The present article studies (European-type and American-type) geometric step contracts under exponential Lévy dynamics. Our paper's contribution is manifold and extends several aspects of the geometric step option pricing literature: Firstly, we establish symmetry and parity relations for geometric double barrier step contracts under exponential Lévy models. Since standard options are naturally embedded in the whole class of geometric double barrier step options, these results generalize in particular the ones obtained in~\cite{fm06}, \cite{fm14}. Secondly, we derive various characterizations for European-type and American-type geometric double barrier step contracts as well as for their respective maturity-randomized quantities. Most notably, we are able to derive a jump-diffusion disentanglement for the early exercise premium of American-type geometric double barrier step options and its maturity-randomized equivalent as well as to characterize the diffusion and jump contributions to these early exercise premiums separately by means of partial integro-differential equations (PIDEs) and ordinary integro-differential equations (OIDEs). Our results translate the formalism introduced in \cite{fmv19} to the setting of geometric double barrier step contracts and generalize at the same time the ideas introduced in \cite{cy13}, \cite{lv17} and \cite{cv18} to Lévy-driven markets. Next, as an application of these characterizations, we derive semi-analytical pricing results for (regular) European-type and American-type geometric down-and-out step call options under hyper-exponential jump-diffusion processes.\footnote{It is worth recalling that hyper-exponential jump-diffusion processes are particularly suitable for financial modeling since they are able to provide arbitrarily close approximations to Lévy processes having a completely monotone jump density. The latter processes form an important class of Lévy models and include popular market dynamics such as Variance Gamma (VG) processes (cf.~\cite{ms90}, \cite{mcc98}), the CGMY model (cf.~\cite{cgmy02}), and Normal Inverse Gaussian (NIG) processes (cf.~\cite{bn95}).}~Although semi-analytical pricing results for European-type geometric step options were already obtained by other authors under similar asset dynamics (cf.~\cite{ccw10}, \cite{lz16}, \cite{wzb17}), we note that these results employed double Laplace transform techniques while our method only relies on a one-dimensional \mbox{Laplace(-Carson)} transform. Additionally, the current geometric step option pricing literature seems to either study the Black~\&~Scholes framework (cf.~\cite{bs73}) or only European-type geometric step options under more advanced models. To the best of our knowledge, we are therefore the first to provide characterizations as well as (tractable) pricing results for American-type geometric step options. Lastly, we discuss the early exercise structure of geometric step options once jumps are added and subsequently provide an analysis of the impact of jumps on the price and hedging parameters of (European-type and American-type) geometric step contracts. As of now, no clear investigation of this sensitivity to jumps has been provided in the geometric step option pricing literature, which is mainly due to the scarcity of publications dealing with (American-type) geometric step options with jumps. \vspace{1em} \\
\noindent The remaining of this paper is structured as follows: In Section~\ref{SEC2}, we introduce (European-type and American-type) geometric step options under exponential Lévy markets and discuss symmetry and parity relations as well as PIDE and OIDE characterizations of these options. Section~\ref{SEC3} deals with geometric step contracts under hyper-exponential jump-diffusion models. Here, semi-analytical pricing results for both European-type and American-type contracts are derived by combining the derivations of Section~\ref{SEC2} with certain properties of the hyper-exponential distribution. These theoretical results are subsequently exemplified in Section~\ref{MR_NUMRES}, where structural and numerical properties of (regular) geometric down-and-out step call options with jumps are illustrated and a comparison to the respective results in the standard Black \& Scholes framework is provided. The paper concludes with Section~\ref{SecConCLUSION}. All proofs and complementary results are presented in the appendices (Appendix A and B).

\section{Geometric Step Options and Exponential Lévy Markets}
\label{SEC2}
\subsection{General Framework}
\label{GenLev}
We start with a filtered probability space $(\Omega, \mathcal{F}, \mathbf{F}, \mathbb{Q})$ -- a chosen risk-neutral probability space\footnote{It is well-known that exponential Lévy markets are incomplete as defined by Harrison and Pliska (cf.~\cite{hp81}). Specifying or discussing a particular choice of risk-neutral measure is not the sake of this article. Instead, we assume that a pricing measure under which our model has the required dynamics was previously fixed.} --, whose filtration $\mathbf{F} = \left( \mathcal{F}_{t} \right)_{t \geq 0}$ satisfies the usual conditions and consider two assets, a deterministic savings account $(B_{t}(r))_{t \geq 0}$ satisfying
\begin{equation}
B_{t}(r) = e^{r t}, \hspace{1.5em} r \geq 0, \, t \geq 0,
\label{market1}
\end{equation}
and a risky asset $(S_{t})_{t \geq 0}$, whose price dynamics, under $\mathbb{Q}$, are described by the following (ordinary) exponential Lévy model
\begin{equation}
S_{t} =  S_{0}e^{X_{t}}, \hspace{1.5em} S_{0}>0, \, t \geq 0.
\label{market2}
\end{equation}
\noindent \noindent Here, the process $(X_{t})_{t \geq 0}$ is an $\mathbf{F}$-Lévy process associated with a triplet $(b_{X}, \sigma_{X}^{2}, \Pi_{X})$, i.e.~a càdlàg (right-continuous with left limits) process having independent and stationary increments and Lévy-exponent $\Psi_{X}(\cdot)$ defined, for $\theta \in \mathbb{R}$, by
\begin{equation}
\Psi_{X}( \theta ) := - \log \left( \mathbb{E}^{\mathbb{Q}} \left[ e^{i \theta X_{1}} \right] \right)  =  -ib_{X} \theta + \frac{1}{2} \sigma_{X}^{2}\theta^{2} + \int \limits_{ \mathbb{R}} \big(1 - e^{i \theta y} + i \theta y \mathds{1}_{\{ | y | \leq 1\}} \big) \Pi_{X}( dy),
\label{CHARexp}
\end{equation}
\noindent where $\mathbb{E}^{\mathbb{Q}}[\cdot]$ refers to expectation with respect to the measure $\mathbb{Q}$. Numerous models in the financial literature fall into this framework. Important examples include hyper-exponential jump-diffusion (HEJD) models (cf.~\cite{ko02}, \cite{ca09}), Variance Gamma (VG) processes (cf.~\cite{ms90}, \cite{mcc98}), the CGMY model (cf.~\cite{cgmy02}) as well as Generalized Hyperbolic (GH) processes such as the popular Normal Inverse Gaussian (NIG) model (cf.~\cite{bn95}).\vspace{1em} \\
\noindent Applying standard results (cf.~\cite{sa}, \cite{Ap}), allows us to decompose $(X_{t})_{t \geq 0}$ in terms of its diffusion and jump parts as
\begin{equation}
X_{t} = b_{X}t + \sigma_{X} W_{t} + \int \limits_{\mathbb{R}} y \; \bar{N}_{X}(t,dy), \hspace{1.5em} t \geq 0,
\label{DecompX}
\end{equation}
\noindent where $(W_{t})_{t \geq 0}$ denotes an $\mathbf{F}$-Brownian motion, and $N_{X}$ refers to an independent Poisson random measure on $[0,\infty) \times \mathbb{R} \setminus \{0\}$ that has intensity measure given by $\Pi_{X}$. Here, we use for $t \geq 0$ and any Borel set $A \in \mathcal{B}(\mathbb{R}\setminus\{0 \})$ the following notation:
\begin{align*}
N_{X}(t,A) &:=N_{X}((0,t]\times A), \\
\tilde{N}_{X}(dt,dy) & :=  N_{X}(dt,dy) - \Pi_{X}(dy)dt ,\\
\bar{N}_{X}(dt,dy)&:= \left \{ \begin{array}{cc}
\tilde{N}_{X}(dt,dy), & \mbox{if} \; |y|\leq 1, \\
N_{X}(dt,dy), & \mbox{if} \; |y| > 1. \end{array}
\right. 
\end{align*}
\noindent Additionally, the Laplace exponent of the Lévy process $(X_{t})_{t \geq 0}$ can be defined for any $ \theta \in \mathbb{R}$ satisfying $ \mathbb{E}^{\mathbb{Q}} \left[ e^{\theta X_{1}} \right] < \infty$ and is then recovered from $\Psi_{X}(\cdot)$ via the following identity:
\begin{align}
\Phi_{X}( \theta )  :=  -\Psi_{X}(-i\theta) & =  b_{X} \theta + \frac{1}{2} \sigma_{X}^{2} \theta^{2} - \int \limits_{ \mathbb{R}} \big(1 - e^{ \theta y} +  \theta y \mathds{1}_{\{ | y | \leq 1\}} \big) \Pi_{X}( dy).
\end{align}
\noindent In the sequel, we always assume that $\Phi_{X}(\cdot)$ is at least for $\theta =1$ well-defined or, equivalently, that the price process $(S_{t})_{t \geq 0}$ is integrable. Additionally, we assume that the asset $(S_{t})_{t \geq 0}$ pays a proportional dividend with constant rate $\delta \geq 0$. In terms of the asset dynamics, this implies that the discounted cum-dividend price process $(e^{-(r-\delta)t} S_{t})_{t \geq 0}$ is a martingale under $\mathbb{Q}$, which then requires that
\begin{equation}
\Phi_{X}(1)= r - \delta.
\label{MartCond}
\end{equation}
\noindent In particular, rewriting (\ref{MartCond}) allows us to recover the following expression for $b_{X}$:
\begin{equation}
b_{X} = r- \delta - \frac{1}{2}\sigma_{X}^{2} + \int \limits_{\mathbb{R}} \big( 1- e^{y} + y \mathds{1}_{\{ | y | \leq 1\}} \big) \Pi_{X}(dy).
\label{bXequa}
\end{equation}
\noindent  Such dynamics are typically found when studying foreign exchange markets. In this case, holdings in the foreign currency can earn the foreign risk-free interest rate, which therefore corresponds, for each investment in the foreign currency, to a dividend payment of a certain amount $\delta \geq 0$ (cf.~\cite{jc02}, \cite{gk83}). \vspace{1em} \\
\noindent Finally, it should be noted that $(S_{t})_{t \geq 0}$ has a Markovian structure. Following standard theory of Markov processes, we therefore recall that its infinitesimal generator is a partial integro-differential operator given, for sufficiently smooth $V: [0,\infty) \times \mathbb{R} \rightarrow \mathbb{R}$, by
\begin{align}
\mathcal{A}_{S} V(\mathcal{T},x) & :=  \lim \limits_{t \downarrow 0}  \, \frac{\mathbb{E}^{\mathbb{Q}}_{x} \big[ V(\mathcal{T},S_{t})\big] - V(\mathcal{T},x)}{t} \nonumber \\
& =  \frac{1}{2} \sigma^{2}_{X} x^{2} \partial_{x}^{2} V(\mathcal{T},x) + \Phi_{X}(1) x \partial_{x} V(\mathcal{T},x) \hspace{15em} \nonumber \\
& \hspace{5.5em} + \int \limits_{\mathbb{R}} \big[ V(\mathcal{T},xe^{y}) - V(\mathcal{T},x) - x(e^{y}-1)\partial_{x} V(\mathcal{T},x) \big] \Pi_{X}(dy), 
\label{AINFA}
\end{align}
\noindent where $\mathbb{E}_{x}^{\mathbb{Q}}[ \cdot]$ denotes expectation under $\mathbb{Q}_{x}$, the pricing measure having initial distribution $S_{0}=x$. We will extensively make use of these notations in the upcoming sections.

\subsection{Characterizing Geometric Step Options}
\label{SECCHARA}
\noindent As mentioned in the introduction, geometric step options are financial contracts that are written on an underlying asset and that cumulatively and proportionally loose or gain value when the underlying's price stays above or below a certain, predetermined threshold. As such, these contracts are closely linked to the time the asset's price spends above or below a barrier level, so-called occupation times. To fix the notation, we define, for a time $t \geq 0$, the occupation time of asset $(S_{t})_{t \geq 0}$ below (\minus) and above (\plus) a constant barrier level $\ell > 0$ over the time interval $[0,t]$ via
\begin{equation}
\Gamma_{t,\ell}^{-} := \int \limits_{0}^{t} \mathds{1}_{(0,\ell)}(S_{r}) dr, \hspace{1.5em} \mbox{and} \hspace{1.9em} \Gamma_{t,\ell}^{+} := \int \limits_{0}^{t} \mathds{1}_{(\ell,\infty)}(S_{r}) dr.
\label{OT1}
\end{equation}
\noindent In addition, we set, for $\gamma \geq 0$,
\begin{equation}
\Gamma_{t,\ell}^{\pm}(\gamma) := \gamma + \Gamma_{t,\ell}^{\pm}
\label{OT2}
\end{equation}
\noindent and allow this way each of the occupation times $\Gamma_{t,\ell}^{-}$ and $\Gamma_{t,\ell}^{+}$ to start at a given initial value $\gamma \geq 0$. This generalization proves useful when valuing geometric step options over their entire lifetime. In this case, $\gamma$ refers to the occupation time the process $(S_{t})_{t \geq 0}$ has spent in the respective region from the establishment of the contract until the valuation date under consideration. \vspace{1em} \\
\noindent As for many other types of options, geometric step options can be found in various styles. Depending on the exercise specification, there exist European-type and American-type geometric step call and put options. Additionally, one can distinguish between ``knock-in'', ``knock-out'' as well as ``up'' and ``down'' features. Therefore, it is possible to construct a total of 32 different geometric step contracts, all of which can be studied in the unifying framework of geometric double barrier step options. A geometric double barrier step option with initial values $S_{0}=x \geq 0$ and $\Gamma_{0,L}^{-}(\gamma_{L}) = \gamma_{L} \geq 0$, $\Gamma_{0,H}^{+}(\gamma_{H}) = \gamma_{H} \geq 0$, strike price $K \geq 0$, barrier levels $0 \leq L \leq H < \infty$, and knock-out/knock-in rates $\rho_{L}, \rho_{H} \in \mathbb{R}$ pays off
\begin{equation}
e^{\rho_{L} \Gamma_{t,L}^{-}(\gamma_{L}) \, + \, \rho_{H} \Gamma_{t,H}^{+}(\gamma_{H})} \left(S_{t}-K \right)^{+} \hspace{0.7em} \mbox{(for a call)} \hspace{1.7em} \mbox{or} \hspace{2em} e^{\rho_{L} \Gamma_{t,L}^{-}(\gamma_{L}) \, + \, \rho_{H} \Gamma_{t,H}^{+}(\gamma_{H})} \left(K- S_{t}\right)^{+} \hspace{0.7em} \mbox{(for a put)}
\end{equation}  
\noindent at the exercise time $t \geq 0$. Here, any of the barrier levels, $\ell \in \{L,H \}$, is said to be of knock-out type whenever $\rho_{\ell} \leq 0$, while the case of $\rho_{\ell} >0$ is referred to as a knock-in feature. \vspace{1em} \\
\noindent Using standard valuation principles, probabilistic representations for the value of any type of geometric double barrier step options are readily obtained. For instance, the value of a European-type geometric double barrier knock-out step call defined on the exponential Lévy market (\ref{market1}), (\ref{market2}), (\ref{MartCond}) and having maturity $\mathcal{T} \geq 0$, initial values $S_{0}=x \geq 0$ and $\Gamma_{0,L}^{-}(\gamma_{L}) = \gamma_{L} \geq 0$, $\Gamma_{0,H}^{+}(\gamma_{H}) = \gamma_{H} \geq 0$, strike price $K \geq 0$, barrier levels $0 \leq L \leq H < \infty$, and knock-out rates $\rho_{L},\rho_H \leq 0$ is obtained as
\begin{equation}
\hspace{0.5em} \mathcal{DSC}_{E}\big( \mathcal{T},x, \gamma_{L}, \gamma_{H}; r,\delta, K, L, H,\rho_{L}, \rho_{H}, \Psi_{X}(\cdot)\big) : = \mathbb{E}_{x}^{\mathbb{Q}} \left[ B_{\mathcal{T}}(r)^{-1} \, e^{\rho_{L} \Gamma_{\mathcal{T},L}^{-}(\gamma_{L}) \, + \, \rho_{H} \Gamma_{\mathcal{T},H}^{+}(\gamma_{H})} \,\left(S_{\mathcal{T}} - K \right)^{+} \right],
\label{EuroDef}
\end{equation}
\noindent where we use the Lévy-exponent $\Psi_{X}(\cdot)$ to refer to the dynamics of the Lévy process (\ref{DecompX}) and therefore to further characterize the dynamics of the underlying price process $(S_{t})_{t \geq 0}$ specified in (\ref{market2}). Similarly, the value of a corresponding American-type geometric double barrier knock-out step call can be shown to have the representation 
\begin{equation}
\mathcal{DSC}_{A}\big( \mathcal{T},x, \gamma_{L}, \gamma_{H}; r,\delta, K, L, H,\rho_{L}, \rho_{H}, \Psi_{X}(\cdot)\big) : = \sup \limits_{ \tau \in \mathfrak{T}_{[0,\mathcal{T}]} } \mathbb{E}_{x}^{\mathbb{Q}} \left[ B_{\tau}(r)^{-1} \, e^{\rho_{L} \Gamma_{\tau,L}^{-}(\gamma_{L}) \, + \, \rho_{H} \Gamma_{\tau,H}^{+}(\gamma_{H})} \,\left(S_{\tau} - K \right)^{+} \right],
\label{AmerDef}
\end{equation}
where $\mathfrak{T}_{[0,\mathcal{T}]}$ denotes the set of stopping times that take values in the interval $[0,\mathcal{T}]$. Here, we note that both values (\ref{EuroDef}) and (\ref{AmerDef}) may be understood, for a given pair of times $(t,T)$ satisfying $0 \leq t \leq T < \infty$, as the time-$t$ value of the respective geometric step contract having maturity $T$, i.e.~we usually have in mind that $\mathcal{T} = T-t$ denotes the remaining time to maturity. \vspace{1em} \\
\noindent At this point, it is important to emphasize that other types of step options exist. Already in his seminal work, Linetsky introduced the class of arithmetic step options as other alternative to barrier options. Compared to standard call and put options, both geometric and arithmetic step options are characterized by an additional adjustment factor. However, while the adjustment factor of geometric step options is given as exponential function of (possibly one of) the occupation times defined in (\ref{OT2}), arithmetic step contracts are characterized by truncated linear adjustments. This implies in particular that, under comparable knock-out rates, arithmetic step contracts will knock-out faster than their geometric counterparts (cf.~\cite{li99}, \cite{dl02}). Clearly, our goal is not to discuss results for all existing types of step options. We will therefore mainly focus on geometric double barrier knock-out step calls and leverage on the fact that certain symmetry and parity relations hold between different geometric step contracts. Establishing these relations is the content of the next section. 

\subsection{Symmetry and Parity Relations}
\label{SECSYPA}
\noindent To allow for a simultaneous treatment of both European-type and American-type geometric step contracts, we start by introducing, for $T>0$, any stopping time $\tau \in \mathfrak{T}_{[0,T]}$, initial values $S_{0}=x \geq 0$ and \mbox{$\Gamma_{0,L}^{-}(\gamma_{L}) = \gamma_{L} \geq 0$,} $\Gamma_{0,H}^{+}(\gamma_{H}) = \gamma_{H} \geq 0$, strike price $K \geq 0$, barrier levels $0 \leq L \leq H < \infty$, and knock-out/knock-in rates $\rho_{L}, \rho_{H} \in \mathbb{R}$, the following quantities:
\begin{align}
\mathcal{DSC}\big( \tau ,x, \gamma_{L}, \gamma_{H}; r,\delta, K, L, H,\rho_{L}, \rho_{H}, \Psi_{X}(\cdot)\big) : = \mathbb{E}_{x}^{\mathbb{Q}} \left[ B_{\tau}(r)^{-1} \, e^{\rho_{L} \Gamma_{\tau,L}^{-}(\gamma_{L}) \, + \, \rho_{H} \Gamma_{\tau,H}^{+}(\gamma_{H})} \,\left(S_{\tau} - K \right)^{+} \right], \label{not1}\\
\mathcal{DSP}\big( \tau ,x, \gamma_{L}, \gamma_{H}; r,\delta, K, L, H,\rho_{L}, \rho_{H}, \Psi_{X}(\cdot)\big) : = \mathbb{E}_{x}^{\mathbb{Q}} \left[ B_{\tau}(r)^{-1} \, e^{\rho_{L} \Gamma_{\tau,L}^{-}(\gamma_{L}) \, + \, \rho_{H} \Gamma_{\tau,H}^{+}(\gamma_{H})} \,\left(K - S_{\tau}\right)^{+} \right]. \label{not2}
\end{align}
\noindent Using this notation, the next put-call-duality result can be derived. A proof is provided in Appendix~A.
\begin{Lem}[Duality of Geometric Step Contracts]
\label{lem1}
Consider an exponential Lévy market, as introduced in (\ref{market1}), (\ref{market2}) and (\ref{MartCond}), with driving process $(X_{t})_{t \geq 0}$ having Lévy exponent given as in (\ref{CHARexp}).
\noindent Then, under the notation (\ref{not1}) and (\ref{not2}), we have for any $T >0$ and stopping time $\tau \in \mathfrak{T}_{[0,T]}$ that
\begin{equation}
\mathcal{DSC}\big( \tau ,x, \gamma_{L}, \gamma_{H};  \,r,\delta, K, L, H,\rho_{L}, \rho_{H}, \Psi_{X}(\cdot) \big) = \mathcal{DSP}\Big(\tau,K,\gamma_{H}, \gamma_{L} ; \delta,r,x,\frac{xK}{H},\frac{xK}{L}, \rho_{H}, \rho_{L},\Psi_{Y}(\cdot)\Big),
\label{EqLem1}
\end{equation}
\noindent where $\Psi_{Y}(\cdot)$ represents the Lévy exponent of another Lévy process $(Y_{t})_{t \geq 0}$ driving an exponential Lévy market with
\begin{equation}
\Psi_{Y}(\theta) = \Psi_{X}(-(\theta + i)) + \Phi_{X}(1).
\label{Ysatisf}
\end{equation}
\noindent In particular, we obtain that the Lévy exponent $\Psi_{Y}(\cdot)$ is given by 
\begin{equation}
\Psi_{Y}(\theta) =  -ib_{Y} \theta + \frac{1}{2} \sigma_{Y}^{2}\theta^{2} + \int \limits_{ \mathbb{R}} \big(1 - e^{i \theta y} + i \theta y \mathds{1}_{\{ | y | \leq 1\}} \big) \Pi_{Y}( dy),
\label{Yequation}
\end{equation}
\noindent where $\left( b_{Y}, \sigma_{Y}^2, \Pi_{Y} \right)$ are obtained as
\begin{align}
b_{Y} & = \delta - r - \frac{1}{2}\sigma_{Y}^{2} + \int \limits_{\mathbb{R}} \big( 1 - e^{y} + y \mathds{1}_{\{ | y | \leq 1\}} \big) \Pi_{Y}(dy), \\
\sigma_{Y}^{2} & = \sigma_{X}^{2}, \\
\Pi_{Y}(dy) & = e^{-y} \, \Pi_{X}(-dy).
\label{inteMeas}
\end{align}
\end{Lem}
$\mbox{ }$ \vspace{0.9em} \\
\noindent \underline{\bf Remark 1.} \vspace{0.3em} \\
\noindent Our results in Lemma~\ref{lem1} are similar to Lemma 1 in \cite{fm06}. However, while these authors consider standard options, our results hold within the whole class of geometric double barrier step contracts. In particular, since geometric double barrier step options reduce to standard options for $\rho_{L} = \rho_{H} = 0$, Lemma \ref{lem1} offers a generalization of the derivations obtained in \cite{fm06}. Additionally, our proof reveals that similar results could be derived for other occupation time derivatives. Due to the focus of our article, we nevertheless refrain from discussing further duality results here. \\
$\mbox{}$ \hspace{44.8em} \scalebox{0.75}{$\blacklozenge$} \vspace{1em} \\
\noindent Combining Lemma \ref{lem1} with few simple transformations allows us to derive duality and symmetry relations for European-type and American-type geometric step options. The results are summarized in the next corollary, whose proof is given in Appendix A. 
\begin{Cor}[Duality and Symmetry of Geometric Step Contracts]
\label{coro1}
Consider an exponential Lévy market, as introduced in (\ref{market1}), (\ref{market2}) and (\ref{MartCond}), with driving process $(X_{t})_{t \geq 0}$ having Lévy exponent given as in (\ref{CHARexp}). Then, the following duality and symmetry results hold
\begin{align}
\mathcal{DSC}_{\bullet}\big( \mathcal{T} ,x, \gamma_{L}, \gamma_{H};  \,r,\delta, K, L, H,\rho_{L}, \rho_{H}, \Psi_{X}(\cdot) \big) = \mathcal{DSP}_{\bullet} \Big(\mathcal{T},K,\gamma_{H}, \gamma_{L} ; \delta,r,x,\frac{xK}{H},\frac{xK}{L}, \rho_{H}, \rho_{L},\Psi_{Y}(\cdot)\Big), \hspace{0.5em} \label{Toprove1a}\\
\mathcal{DSC}_{\bullet}\big( \mathcal{T} ,x, \gamma_{L}, \gamma_{H}; r,\delta, K, L, H,\rho_{L}, \rho_{H}, \Psi_{X}(\cdot) \big) = xK \cdot \mathcal{DSP}_{\bullet} \Big( \mathcal{T} ,\frac{1}{x}, \gamma_{H}, \gamma_{L}; \delta,r, \frac{1}{K}, \frac{1}{H}, \frac{1}{L},\rho_{H}, \rho_{L}, \Psi_{Y}(\cdot) \Big), \label{Toprove1b}
\end{align}
\noindent where the Lévy exponents $\Psi_{Y}(\cdot)$ is defined as in Lemma \ref{lem1} and $\bullet$ refers to the exercise specification of the options, i.e.~$\bullet \in \{E,A \}$.
\end{Cor}
%%%%%%%%%%%%%%%%%%%%%%%%%%%%%%%%%%%%%%%%%%%%%%%%%%%%%%%%%%%%%%%%%%%%%%%%%%%%%%%%%%%%%%%%%%%%%%
%%%%%%%%%%%%%%%%% ORIGINALLY DIVIDED INTO AMERICAN & EUROPEAN. NOW REGROUPED %%%%%%%%%%%%%%%%%
%%%%%%%%%%%%%%%%%%%%%%%%%%%%%%%%%%%%%%%%%%%%%%%%%%%%%%%%%%%%%%%%%%%%%%%%%%%%%%%%%%%%%%%%%%%%%%
% \begin{Cor}[Duality and Symmetry of American Step Contracts]
% \label{coro2}
% \begin{align}
% \mathcal{SC}_{A}^{\pm}(\mathcal{T},x,\gamma ; r,\delta,K,L,\rho,\Psi_{X}) = \mathcal{SP}_{A}^{\mp}\Big(\mathcal{T},K,\gamma ; \delta,r,x,\frac{xK}{L},\rho,\Psi_{Y}\Big) \label{Toprove2a}\\
% xK \cdot \mathcal{SC}_{A}^{\pm}(\mathcal{T},x,\gamma ; r,\delta,K,L,\rho,\Psi_{X}) = \mathcal{SP}_{A}^{\mp}\Big(\mathcal{T},K,\gamma ; \delta,r,x,\frac{xK}{L},\rho,\Psi_{Y}\Big) \label{Toprove2b}
% \end{align}
% \end{Cor}
%%%%%%%%%%%%%%%%%%%%%%%%%%%%%%%%%%%%%%%%%%%%%%%%%%%%%%%%%%%%%%%%%%%%%%%%%%%%%%%%%%%%%%%%%%%%%%
%%%%%%%%%%%%%%%%%%%%%%%%%%%%%%%%%  END ORIGINAL POST  %%%%%%%%%%%%%%%%%%%%%%%%%%%%%%%%%%%%%%%%%
%%%%%%%%%%%%%%%%%%%%%%%%%%%%%%%%%%%%%%%%%%%%%%%%%%%%%%%%%%%%%%%%%%%%%%%%%%%%%%%%%%%%%%%%%%%%%%
\subsection{Geometric Step Options and PIDEs}
\label{GEOPIDE}
\noindent We next turn to the pricing of geometric double barrier step contracts. As already mentioned in Section \ref{SECCHARA}, we focus from now on on geometric double barrier knock-out step call options, i.e.~we take $\rho_{L},\rho_{H} \leq 0$ and leverage on the relations obtained in Section~\ref{SECSYPA}. We emphasize however that the approach followed in the upcoming sections is general enough to produce similar results for other types of geometric step contracts and that only few, slight adaptions are needed. \vspace{1em} \\
\noindent In order to price both European-type as well as American-type double barrier step (call) options, it is sufficient to focus on corresponding step contracts that are initiated at the valuation date under consideration. This clearly follows since for $\bullet \in \{E,A \}$ and any $ \mathcal{T}, x, \gamma_{L}, \gamma_{H}, r,\delta, K, L, H,\rho_{L}, \rho_{H}$, and $\Psi_{X}(\cdot)$, we have that
\begin{align}
\mathcal{DSC}_{\bullet}\big( \mathcal{T},x, \gamma_{L}, \gamma_{H}; r,\delta, K, L, H,\rho_{L}, \rho_{H}, \Psi_{X}(\cdot) \big) =  e^{\rho_{L} \gamma_{L} \, + \, \rho_{H} \gamma_{H}} \cdot \mathcal{DSC}_{\bullet}\big( \mathcal{T} ,x, 0, 0;  \,r,\delta, K, L, H,\rho_{L}, \rho_{H}, \Psi_{X}(\cdot) \big).
\label{SimpliBUP}
\end{align}
\noindent Therefore, we assume from now on that an exponential Lévy market, described in terms of its characteristic exponent $\Psi_{X}(\cdot)$, has been pre-specified and concentrate, for $\bullet \in \{E,A \}$, on geometric step contracts of the form
\begin{equation}
\mathcal{DSC}^{\star}_{\bullet} (\mathcal{T},x; K, \bm{\ell},\bm{\rho}_{\bm{\ell}}) := \mathcal{DSC}_{\bullet}( \mathcal{T} ,x, 0, 0;  \,r,\delta, K, L, H,\rho_{L}, \rho_{H}, \Psi_{X}(\cdot)),
\label{SimpliBUP2}
\end{equation}
\noindent with $\bm{\ell}:= (L,H)$ and $\bm{\rho}_{\bm{\ell}} := (\rho_{L},\rho_{H})$.
\subsubsection{European-Type Contracts}
\noindent We first treat European-type contracts and characterize them by means of partial integro-differential equations (PIDEs). This is the content of the next proposition, whose proof is presented in Appendix A.
\begin{Prop}
\label{prop1}
\noindent For any fixed $T >0$, strike $K \geq 0$, barrier levels $0 \leq L \leq H < \infty$, and knock-out rates $ \rho_{L}, \rho_{H} \leq 0$, the value of the European-type geometric double barrier step call, $\mathcal{DSC}_{E}^{\star}(\cdot)$, is continuous on $[0,T] \times [0,\infty)$ and solves the partial integro-differential equation
\begin{equation}
- \partial_{\mathcal{T}} \mathcal{DSC}^{\star}_{E} (\mathcal{T},x; K, \bm{\ell},\bm{\rho}_{\bm{\ell}}) + \mathcal{A}_{S} \mathcal{DSC}^{\star}_{E} (\mathcal{T},x; K, \bm{\ell},\bm{\rho}_{\bm{\ell}}) -\bigg( r - \bm{\rho}_{\bm{\ell}} \cdot \bigg( \begin{array}{c}
 \mathds{1}_{(0,L)}(x) \\
 \mathds{1}_{(H,\infty)}(x) 
\end{array} \bigg) \bigg) \mathcal{DSC}^{\star}_{E} (\mathcal{T},x; K, \bm{\ell},\bm{\rho}_{\bm{\ell}}) = 0 , \label{GSCEuPIDE1}
\end{equation}
\noindent on $(0,T] \times [0,\infty)$ with initial condition
\begin{equation}
\mathcal{DSC}_{E}^{\star}(0,x; K, \bm{\ell},\bm{\rho}_{\bm{\ell}})  =  (x-K )^{+}, \hspace{1.5em}  x \in [0,\infty). \label{GSCEuPIDE2}
\end{equation}
\end{Prop}
\subsubsection{American-Type Contracts}
\label{GEOPIDEAmer}
\noindent We now discuss American-type contracts. First, as in the proof of Proposition \ref{prop1}, we note that American-type double barrier step call options can be re-expressed in the form
\begin{equation}
\mathcal{DSC}^{\star}_{A}(\mathcal{T},x; K, \bm{\ell}, \bm{\rho}_{\bm{\ell}} ) =  \sup \limits_{ \tau \in \mathfrak{T}_{[0,\mathcal{T}]} } \mathbb{E}_{x}^{\mathbb{Q}} \left[ \left(\bar{S}_{\tau} - K \right)^{+} \right],
\label{PROOOOB}
\end{equation}
\noindent where $(\bar{S}_{t})_{t \geq 0}$ refers to the (strong) Markov process obtained by ``killing''\footnote{The reader is referred, for further details, to the proof of Proposition \ref{prop1}.} the sample path of $(S_{t})_{t \geq 0}$ at the proportional rate $\lambda(x) := r - \bm{\rho}_{\bm{\ell}} \cdot \bigg( \begin{array}{c}
 \mathds{1}_{(0,L)}(x) \\
 \mathds{1}_{(H,\infty)}(x) 
\end{array} \bigg)$ and whose cemetery state is given, without loss of generality, by $\partial \equiv 0$. Therefore, using the fact that the payoff function $x \mapsto (x-K)^{+}$ is continuous as well as standard optimal stopping arguments (cf.~Corollary 2.9.~and Remark~2.10. in \cite{pe06}), we obtain that the continuation and stopping regions read for a (fixed) valuation horizon $[0,T]$, respectively
\begin{align}
\mathcal{D}_{c} & = \left \{ (\mathcal{T},x) \in [0,T] \times [0,\infty): \, \mathcal{DSC}^{\star}_{A}(\mathcal{T},x; K, \bm{\ell}, \bm{\rho}_{\bm{\ell}} ) > (x-K)^{+} \right \} , \label{Cregion} \\
\mathcal{D}_{s} & = \left \{ (\mathcal{T},x) \in [0,T] \times [0,\infty): \, \mathcal{DSC}^{\star}_{A}(\mathcal{T},x; K, \bm{\ell}, \bm{\rho}_{\bm{\ell}} ) = (x-K)^{+} \right \}, \label{Sregion}
\end{align}
\noindent and that, for any $\mathcal{T} \in [0,T]$, the first-entry time
\begin{equation}
\tau_{\mathcal{D}_{s}} := \inf \big \{ 0 \leq t \leq \mathcal{T}: \, (\mathcal{T}-t,\bar{S}_{t}) \in \mathcal{D}_{s} \big \}
\label{OTime}
\end{equation}
\noindent is optimal in (\ref{PROOOOB}). This subsequently allows us to make use of standard strong Markovian arguments to derive a characterization of the American-type contract, $\mathcal{DSC}^{\star}_{A}(\cdot)$, in terms of a Cauchy-type problem. This is the content of the next proposition, whose proof is provided in Appendix A.
\begin{Prop}
\label{prop2}
\noindent For any fixed $T>0$, strike $K \geq 0$, barrier levels $0 \leq L \leq H < \infty$, and knock-out rates $ \rho_{L}, \rho_{H} \leq 0$, the value of the American-type geometric double barrier step call, $\mathcal{DSC}_{A}^{\star}(\cdot)$, is continuous on $[0,T] \times [0,\infty)$ and satisfies the following Cauchy-type problem:
\begin{equation}
- \partial_{\mathcal{T}} \mathcal{DSC}^{\star}_{A} (\mathcal{T},x; K, \bm{\ell},\bm{\rho}_{\bm{\ell}}) + \mathcal{A}_{S} \mathcal{DSC}^{\star}_{A} (\mathcal{T},x; K, \bm{\ell},\bm{\rho}_{\bm{\ell}}) -\bigg( r - \bm{\rho}_{\bm{\ell}} \cdot \bigg( \begin{array}{c}
 \mathds{1}_{(0,L)}(x) \\
 \mathds{1}_{(H,\infty)}(x) 
\end{array} \bigg) \bigg) \mathcal{DSC}^{\star}_{A} (\mathcal{T},x; K, \bm{\ell},\bm{\rho}_{\bm{\ell}}) = 0,  \label{GSCAmePIDE1}
\end{equation}
\noindent for $ (\mathcal{T},x) \in \mathcal{D}_{c}$ with boundary condition
\begin{equation}
\mathcal{DSC}_{A}^{\star}(\mathcal{T},x; K, \bm{\ell},\bm{\rho}_{\bm{\ell}})  =  (x-K )^{+}, \hspace{1.5em}  \mbox{for} \; \,  (\mathcal{T},x) \in \mathcal{D}_{s}. \label{GSCAmePIDE2}
\end{equation}
\end{Prop}
\noindent Proposition \ref{prop1} and Proposition \ref{prop2} are of great practical importance since they both provide a characterization of the respective geometric step contracts in terms of a PIDE problem and therefore already allow for a simple treatment of the options $\mathcal{DSC}_{E}^{\star}(\cdot)$ and $\mathcal{DSC}_{A}^{\star}(\cdot)$ by means of standard numerical techniques. However, these results do not offer any additional insights on the early exercise structure of these options. Instead, an early exercise decomposition into diffusion and jump contributions can be specified and PIDE characterizations thereof can be derived by analyzing the early exercise premium, $\mathcal{E}_{\mathcal{DSC}}^{\star}(\cdot)$, that is defined, for any $\mathcal{T},x, K, \bm{\ell}$, and $\bm{\rho}_{\bm{\ell}}$, by
\begin{equation}
\mathcal{E}_{\mathcal{DSC}}^{\star}(\mathcal{T},x; K, \bm{\ell},\bm{\rho}_{\bm{\ell}}) := \mathcal{DSC}_{A}^{\star}(\mathcal{T},x; K, \bm{\ell},\bm{\rho}_{\bm{\ell}}) - \mathcal{DSC}_{E}^{\star}(\mathcal{T},x; K, \bm{\ell},\bm{\rho}_{\bm{\ell}}).
\label{DefNormalEEP}
\end{equation}
\noindent Deriving these characterizations is the content of the following discussion, where we restrict ourselves to jump distributions that are absolutely continuous with respect to the Lebesgue measure, i.e.~we only consider Lévy processes whose intensity measure takes the form 
\begin{equation}
\Pi_{X}(dy) = \pi_{X}(y)  dy
\label{JumpAssum}
\end{equation}
\noindent for a certain jump density $\pi_{X}(\cdot)$. This is to ensure that the upcoming decomposition stays meaningful. However, we emphasize that this assumption could be relaxed and additionally note that it does not constitute a real restriction since (almost) all Lévy processes studied in the financial literature satisfy this property. \vspace{1em} \\
\noindent We start our discussion by noting that the stopping region $\mathcal{D}_{s}$ is a closed and left-connected\footnote{We define left-connectedness in terms of the time to maturity and require the following property: $$\forall 0 \leq \mathcal{T}_{1} \leq \mathcal{T}_{2} \leq T, \; x \in [0,\infty) : \,\big( (\mathcal{T}_{2},x) \in \mathcal{D}_{s} \Rightarrow (\mathcal{T}_{1},x) \in \mathcal{D}_{s} \big).$$}~set in $[0,T] \times [0,\infty)$ that additionally has the following decomposition
\begin{equation}
\mathcal{D}_{s} = \mathcal{D}_{s}^{L} \cup \mathcal{D}_{s}^{H},
\label{SetsDec}
\end{equation}
 where $\mathcal{D}_{s}^{L}$ and $\mathcal{D}_{s}^{H}$ are themselves closed and left-connected sets in $[0,T] \times [0,\infty)$, with $\mathcal{D}_{s}^{L}$ and $\mathcal{D}_{s}^{H}\setminus \{L\}$ being disjoint. This can be seen from the following arguments: First, the closedness of $\mathcal{D}_{s}$ directly follows from the continuity of the function $(\mathcal{T},x) \mapsto \mathcal{DSC}^{\star}_{A}(\mathcal{T},x;K,\bm{\ell}, \bm{\rho}_{\bm{\ell}})$ on $[0,T] \times [0,\infty)$ for any $K,\bm{\ell}$, and $\bm{\rho}_{\bm{\ell}} $ (cf.~\cite{pe06}), while the fact that $\mathcal{T} \mapsto \mathcal{DSC}_{A}^{\star}(\mathcal{T},x; K, \bm{\ell},\bm{\rho}_{\bm{\ell}})$ is, for any $x, K, \bm{\ell}$, and $\bm{\rho}_{\bm{\ell}}$, non-decreasing implies, for $0\leq \mathcal{T}_{1} \leq \mathcal{T}_{2} \leq T$, that we have $(\mathcal{T}_{1},x) \in \mathcal{D}_{s}$ whenever $(\mathcal{T}_{2},x) \in \mathcal{D}_{s}$. This already gives what is often referred to as left-connectedness. Therefore, we only have to prove the disjointness of the sets $\mathcal{D}_{s}^{L}$, $\mathcal{D}_{s}^{H} \setminus \{L\}$ in the decomposition~(\ref{SetsDec}). To see this property, we note that, for any $\mathcal{T},x, K$, and $\bm{\ell}$, the following inequality holds
\begin{equation}
\mathcal{DSC}_{A}^{\star}(\mathcal{T},x; K, \bm{\ell},\bm{\tilde{\rho}}_{\bm{\ell}}) \leq \mathcal{DSC}_{A}^{\star}(\mathcal{T},x; K, \bm{\ell},\bm{\rho}_{\bm{\ell}}), \hspace{1.5em} \mbox{whenever} \; \, \bm{\tilde{\rho}}_{\bm{\ell}} \leq \bm{\rho}_{\bm{\ell}},
\end{equation} 
\noindent where $ (\tilde{\rho}_{L}, \tilde{\rho}_{H}) = \bm{\tilde{\rho}}_{\bm{\ell}} \leq \bm{\rho}_{\bm{\ell}} = (\rho_{L},\rho_{H})$ refers to the componentwise inequalities $\tilde{\rho}_{L} \leq \rho_{L} $ and $\tilde{\rho}_{H} \leq \rho_{H}$. Since standard options are recovered from geometric double barrier step options by replacing $\bm{\rho}_{\bm{\ell}}$ with $\bm{\rho^{S}}_{\bm{\ell}} := (0,0)$ in (\ref{SimpliBUP2}) and (standard) double barrier knock-out options can be understood as ``limit'' of geometric double barrier step contracts, e.g.~via the sequence $\big(\bm{\rho^{B}}_{n,\bm{\ell}}\big)_{n \in \mathbb{N}} := \big((-n, -n)\big)_{n \in \mathbb{N}}$, we obtain, in particular, that
\begin{equation}
\mathcal{DBC}_{A}(\mathcal{T},x; K, \bm{\ell}) \leq  \mathcal{DSC}_{A}^{\star}(\mathcal{T},x; K, \bm{\ell},\bm{\rho}_{\bm{\ell}}) \leq \mathcal{C}_{A}(\mathcal{T},x; K).
\end{equation}
\noindent Here, $\mathcal{C}_{A}(\cdot)$ and $\mathcal{DBC}_{A}(\cdot)$ refer to the (standard) American-type call and the (standard) American-type double barrier knock-out call, obtained by
\begin{align}
\mathcal{C}_{A}(\mathcal{T},x; K) & := \mathcal{DSC}_{A}^{\star}(\mathcal{T},x; K, \bm{\ell},\bm{\rho^{S}}_{\bm{\ell}}), \\
\mathcal{DBC}_{A}(\mathcal{T},x; K, \bm{\ell}) &:= \sup \limits_{\tau \in \mathfrak{T}_{[0,\mathcal{T}]}} \lim \limits_{n \uparrow \infty} \mathcal{DSC}^{\star}(\tau,x; K, \bm{\ell},\bm{\rho^{B}}_{n,\bm{\ell}}),
\end{align}
\noindent where $\mathcal{DSC}^{\star}(\tau,x; K, \bm{\ell}, \bm{\rho}_{\bm{\ell}}) = \mathcal{DSC}\big( \tau ,x, 0, 0;  \,r,\delta, K, L, H,\rho_{L}, \rho_{H}, \Psi_{X}(\cdot)\big)$ denotes the contract version of (\ref{not1}) that is initiated at the valuation date under consideration, i.e.~in the sense of the notation introduced in~(\ref{SimpliBUP2}). Hence, this gives that $\mathcal{D}_{S,s} \subseteq \mathcal{D}_{s} \subseteq \mathcal{D}_{B,s}$, with $\mathcal{D}_{S,s}$ and $\mathcal{D}_{B,s}$ denoting the stopping region of the corresponding (standard) American-type call and (standard) American-type double barrier knock-out call, respectively, i.e.
\begin{align}
\mathcal{D}_{S,s}  & = \left \{ (\mathcal{T},x) \in [0,T] \times [0,\infty): \, \mathcal{C}_{A}(\mathcal{T},x; K) = (x-K)^{+} \right \}, \\
\mathcal{D}_{B,s}  = & \big\{ (\mathcal{T},x) \in [0,T] \times [0,\infty): \, \mathcal{DBC}_{A}(\mathcal{T},x; K, \bm{\ell}) = (x-K)^{+} \big \}.
\end{align}
\noindent In particular, $\mathcal{D}_{S,s} \subseteq \mathcal{D}_{s}$ directly implies, for $\delta > 0$, the non-emptyness of the stopping region $\mathcal{D}_{s}$ (cf.~\cite{Ma18}), whereas combining well-known results for (standard) American-type double barrier options with the relation $\mathcal{D}_{s} \subseteq \mathcal{D}_{B,s}$ gives that early exercise of the geometric double barrier knock-out step call can only occur, for a fixed $\mathcal{T} \in [0,T]$, in subregions of the intervals $I_{1} := (K,L]$, whenever $L > K$, and $I_{2} := \big[\mathfrak{b}^{B}(\mathcal{T}),\infty\big)$, where $\mathfrak{b}^{B}(\mathcal{T}) \geq \max(K,L)$ denotes the early exercise up-boundary of the corresponding (standard) American-type double barrier knock-out call. This provides (\ref{SetsDec}). \vspace{1em} \\
\noindent Next, combining the closedness of $\mathcal{D}_{s}$ with its left-connectedness and decomposition (\ref{SetsDec}) leads to the following observations:\footnote{We refer the reader for similar ideas to \cite{fmv19}; see also \cite{lv17} and \cite{cv18}.}~First, any entry of the stopping region that is triggered by the diffusion part of the process $(S_{t})_{t \geq 0}$\footnote{Or, equivalently, by the diffusion part of the underlying Lévy process $(X_{t})_{t \geq 0}$.}~will happen by crossing the boundary $\partial \mathcal{D}_{s}$ of the set $\mathcal{D}_{s}$, where
$$ \partial \mathcal{D}_{s} := \Big\{ (\mathcal{T},x) \in \mathcal{D}_{s}: \, \forall \epsilon > 0: B_{\epsilon}\big((\mathcal{T},x)\big) \cap \mathcal{D}_{s} \neq \emptyset \; \land \; B_{\epsilon}\big(\big((\mathcal{T},x)\big) \cap \Big( \big( [0,T] \times [0,\infty) \big) \setminus \mathcal{D}_{s}\Big) \neq \emptyset \Big\} ,$$
\noindent and $B_{\epsilon}\big((\mathcal{T},x)\big)$ denotes the open ball around the (mid-)point $(\mathcal{T},x)$ and with radius $\epsilon > 0$. \noindent On the other hand, first-passage entries in the stopping region that are triggered by jumps will always occur at an interior point of the set $\mathcal{D}_{s}$, i.e.~within $\mathcal{D}_{s}^{\circ} := \mathcal{D}_{s} \setminus \partial \mathcal{D}_{s}$, whenever the $\mathcal{T}$-section \mbox{$\mathcal{D}_{s,\mathcal{T}} := \{ x \in [0,\infty): \, (\mathcal{T},x) \in \mathcal{D}_{s} \}$} contains, for all $\mathcal{T} \in [0,T]$, only finitely many $x$ with $(\mathcal{T},x) \in \partial \mathcal{D}_{s}$, i.e.~whenever we have for all $\mathcal{T} \in [0,T]$ that \mbox{$\# \left( \partial \mathcal{D}_{s} \cap \big(\{\mathcal{T} \} \times \mathcal{D}_{s,\mathcal{T}} \big) \right) < \infty$}. This is a direct consequence of Assumption (\ref{JumpAssum}), as this assumption implies that, conditional on a jump occuring at time $t$, events of the form $\{ S_{t} = \varphi + S_{t-}\}$ have for any fixed $\varphi \in \mathbb{R}$ zero probability. Additionally, in cases where $\# \left( \partial \mathcal{D}_{s} \cap \big( \{\mathcal{T}_{0} \} \times \mathcal{D}_{s,\mathcal{T}_{0}} \big)\right) = \infty$ holds for some $\mathcal{T}_{0} \in [0,T]$, the stopping region has the particularity to suddenly increase in size at this particular point in time $\mathcal{T}_{0}$ and any entry in $\partial \mathcal{D}_{s} \cap \big( \{ \mathcal{T}_{0} \} \times \mathcal{D}_{s,\mathcal{T}_{0}} \big)$ is very much due to the drastic change in the shape of the stopping region at this point. In particular, since Lévy processes are quasi left-continuous, i.e.~left-continuous over predictable stopping times, these stopping scenarios can only be due to the diffusion part of the process $(S_{t})_{t \geq 0}$. Consequently, these observations justify the usage of the sets $\partial \mathcal{D}_{s}$ and $\mathcal{D}_{s}^{\circ}$ to decompose the stopping region $\mathcal{D}_{s}$ into sub-regions where stopping is purely triggered by diffusion and by jumps, respectively. This subsequently results in a decomposition of the early exercise premium, $\mathcal{E}_{\mathcal{DSC}}^{\star}(\cdot)$, of the following form:
\begin{equation}
\mathcal{E}_{\mathcal{DSC}}^{\star}(\mathcal{T},x; K, \bm{\ell}, \bm{\rho}_{\bm{\ell}} ) = \mathcal{E}_{\mathcal{DSC}}^{0,\star}(\mathcal{T},x; K, \bm{\ell}, \bm{\rho}_{\bm{\ell}} ) + \mathcal{E}_{\mathcal{DSC}}^{\mathcal{J},\star}(\mathcal{T},x; K, \bm{\ell}, \bm{\rho}_{\bm{\ell}} ).
\end{equation}
\noindent Here, the premiums $\mathcal{E}_{\mathcal{DSC}}^{0,\star}(\cdot)$ and $\mathcal{E}_{\mathcal{DSC}}^{\mathcal{J},\star}(\cdot)$ refer to the early exercise contributions of the diffusion and jump parts, respectively, and are defined in the following way
\begin{align}
\mathcal{E}_{\mathcal{DSC}}^{0,\star}(\mathcal{T},x; K, \bm{\ell}, \bm{\rho}_{\bm{\ell}} ) & := \mathcal{DSC}^{0,\star}_{A}(\mathcal{T},x; K, \bm{\ell}, \bm{\rho}_{\bm{\ell}} ) - \mathcal{DSC}^{0,\star}_{E}(\mathcal{T},x; K, \bm{\ell}, \bm{\rho}_{\bm{\ell}} ), \\
\mathcal{E}_{\mathcal{DSC}}^{\mathcal{J},\star}(\mathcal{T},x; K, \bm{\ell}, \bm{\rho}_{\bm{\ell}} ) & := \mathcal{DSC}^{\mathcal{J},\star}_{A}(\mathcal{T},x; K, \bm{\ell}, \bm{\rho}_{\bm{\ell}} ) - \mathcal{DSC}^{\mathcal{J},\star}_{E}(\mathcal{T},x; K, \bm{\ell}, \bm{\rho}_{\bm{\ell}} ),
\end{align}
\noindent where the European-type functions $\mathcal{DSC}^{0,\star}_{E}(\cdot)$ and $\mathcal{DSC}^{\mathcal{J},\star}_{E}(\cdot)$ are given by
\begin{align}
\mathcal{DSC}^{0,\star}_{E}(\mathcal{T},x; K, \bm{\ell}, \bm{\rho}_{\bm{\ell}} ) & =  \mathbb{E}_{x}^{\mathbb{Q}} \left[ \left(\bar{S}_{\mathcal{T}} - K \right)^{+} \mathds{1}_{\partial \mathcal{D}_{s}}\big((\mathcal{T}-\tau_{\mathcal{D}_{s}}, \bar{S}_{\tau_{\mathcal{D}_{s}}}) \big) \right], \\
\mathcal{DSC}^{\mathcal{J},\star}_{E}(\mathcal{T},x; K, \bm{\ell}, \bm{\rho}_{\bm{\ell}} ) & =  \mathbb{E}_{x}^{\mathbb{Q}} \left[ \left(\bar{S}_{\mathcal{T}} - K \right)^{+} \mathds{1}_{ \mathcal{D}_{s}^{\circ}} \big((\mathcal{T}-\tau_{\mathcal{D}_{s}}, \bar{S}_{\tau_{\mathcal{D}_{s}}}) \big) \right],
\end{align}
\noindent and the American-type contributions $\mathcal{DSC}^{0,\star}_{A}(\cdot)$ and $\mathcal{DSC}^{\mathcal{J},\star}_{A}(\cdot)$ are defined, accordingly, as
\begin{align}
\mathcal{DSC}^{0,\star}_{A}(\mathcal{T},x; K, \bm{\ell}, \bm{\rho}_{\bm{\ell}} ) & =  \mathbb{E}_{x}^{\mathbb{Q}} \left[ \left(\bar{S}_{\tau_{\mathcal{D}_{s}}} - K \right)^{+} \mathds{1}_{\partial \mathcal{D}_{s}} \big( (\mathcal{T}-\tau_{\mathcal{D}_{s}}, \bar{S}_{\tau_{\mathcal{D}_{s}}}) \big) \right],  \\
\mathcal{DSC}^{\mathcal{J},\star}_{A}(\mathcal{T},x; K, \bm{\ell}, \bm{\rho}_{\bm{\ell}} ) & =  \mathbb{E}_{x}^{\mathbb{Q}} \left[ \left(\bar{S}_{\tau_{\mathcal{D}_{s}}} - K \right)^{+} \mathds{1}_{\mathcal{D}_{s}^{\circ}} \big( (\mathcal{T}-\tau_{\mathcal{D}_{s}}, \bar{S}_{\tau_{\mathcal{D}_{s}}}) \big) \right].
\end{align}
\noindent Combining these definitions with strong Markovian arguments finally allows us to derive PIDE characterizations of the early exercise contributions $\mathcal{E}_{\mathcal{DSC}}^{0,\star}(\cdot)$ and $\mathcal{E}_{\mathcal{DSC}}^{\mathcal{J},\star}(\cdot)$. This is the content of the next proposition, whose proof is presented in Appendix~A.
\begin{Prop}
\label{prop3}
\noindent For any fixed $T>0$, strike $K \geq 0$, barrier levels $0 \leq L \leq H < \infty$, and knock-out rates $ \rho_{L}, \rho_{H} \leq 0$, the value of the diffusion contribution to the early exercise premium of the geometric double barrier step call, $\mathcal{E}_{\mathcal{DSC}}^{0,\star}(\cdot)$, satisfies the following Cauchy-type problem:
\begin{equation}
- \partial_{\mathcal{T}} \mathcal{E}_{\mathcal{DSC}}^{0,\star} (\mathcal{T},x; K, \bm{\ell},\bm{\rho}_{\bm{\ell}}) + \mathcal{A}_{S} \mathcal{E}_{\mathcal{DSC}}^{0,\star} (\mathcal{T},x; K, \bm{\ell},\bm{\rho}_{\bm{\ell}}) -\bigg( r - \bm{\rho}_{\bm{\ell}} \cdot \bigg( \begin{array}{c}
 \mathds{1}_{(0,L)}(x) \\
 \mathds{1}_{(H,\infty)}(x) 
\end{array} \bigg) \bigg) \mathcal{E}_{\mathcal{DSC}}^{0,\star} (\mathcal{T},x; K, \bm{\ell},\bm{\rho}_{\bm{\ell}}) = 0 , \label{GSCAmeEEPPIDE1}
\end{equation}
\noindent for $ (\mathcal{T},x) \in \mathcal{D}_{c}$ with boundary conditions
\begin{align}
\mathcal{E}_{\mathcal{DSC}}^{0,\star}(\mathcal{T},x; K, \bm{\ell},\bm{\rho}_{\bm{\ell}})   =  (x-K )^{+} - & \mathcal{DSC}^{\star}_{E}(\mathcal{T},x; K, \bm{\ell}, \bm{\rho}_{\bm{\ell}} ), \hspace{1.5em}  \mbox{for} \; \,  (\mathcal{T},x) \in \partial \mathcal{D}_{s}, \label{GSCAmeEEPPIDE1-1} \\
\mathcal{E}_{\mathcal{DSC}}^{0,\star}(\mathcal{T},x; K, \bm{\ell},\bm{\rho}_{\bm{\ell}}) & = 0, \hspace{1.5em}  \mbox{for} \; \,  (\mathcal{T},x) \in \mathcal{D}_{s}^{\circ}. \label{GSCAmeEEPPIDE1-2}
\end{align}
\noindent Similarly, the value of the jump contribution to the early exercise premium of the geometric double barrier step call, $\mathcal{E}_{\mathcal{DSC}}^{\mathcal{J},\star}(\cdot)$, solves the following Cauchy-type problem:
\begin{equation}
- \partial_{\mathcal{T}} \mathcal{E}_{\mathcal{DSC}}^{\mathcal{J},\star} (\mathcal{T},x; K, \bm{\ell},\bm{\rho}_{\bm{\ell}}) + \mathcal{A}_{S} \mathcal{E}_{\mathcal{DSC}}^{\mathcal{J},\star} (\mathcal{T},x; K, \bm{\ell},\bm{\rho}_{\bm{\ell}}) -\bigg( r - \bm{\rho}_{\bm{\ell}} \cdot \bigg( \begin{array}{c}
 \mathds{1}_{(0,L)}(x) \\
 \mathds{1}_{(H,\infty)}(x) 
\end{array} \bigg) \bigg) \mathcal{E}_{\mathcal{DSC}}^{\mathcal{J},\star} (\mathcal{T},x; K, \bm{\ell},\bm{\rho}_{\bm{\ell}}) = 0 , \label{GSCAmeEEPPIDE2}
\end{equation}
\noindent for $ (\mathcal{T},x) \in \mathcal{D}_{c}$ with boundary conditions
\begin{align}
\mathcal{E}_{\mathcal{DSC}}^{\mathcal{J},\star}(\mathcal{T},x; K, \bm{\ell},\bm{\rho}_{\bm{\ell}})  & =  0, \hspace{1.5em}  \mbox{for} \; \,  (\mathcal{T},x) \in \partial \mathcal{D}_{s}, \label{GSCAmeEEPPIDE2-1} \\
\mathcal{E}_{\mathcal{DSC}}^{\mathcal{J},\star}(\mathcal{T},x; K, \bm{\ell},\bm{\rho}_{\bm{\ell}}) = (x-K )^{+} - & \mathcal{DSC}^{\star}_{E}(\mathcal{T},x; K, \bm{\ell}, \bm{\rho}_{\bm{\ell}} ), \hspace{1.5em}  \mbox{for} \; \,  (\mathcal{T},x) \in \mathcal{D}_{s}^{\circ}. \label{GSCAmeEEPPIDE2-2}
\end{align}
\end{Prop}
$\mbox{ }$ \vspace{-0.7em} \\
\noindent \underline{\bf Remark 2.} \vspace{0.3em} \\
\noindent Although Proposition \ref{prop3} provides a meaningful characterization of diffusion and jump contributions to the early exercise premium of geometric step options, one may have the impression that these results are lacking applicability. In particular, it seems difficult to make use of these characterizations in practice since the sets~$\mathcal{D}_{s}$, $\partial \mathcal{D}_{s}$, and $\mathcal{D}_{s}^{\circ}$ are usually not known in advance. However, we will see that Proposition~\ref{prop3} and the upcoming results of Section~\ref{MROIDE} will play a crucial role in Section \ref{SEC3}, where they will allow for a derivation of semi-analytical diffusion and jump contributions to the early exercise premium of geometric down-and-out step call options under hyper-exponential jump-diffusion markets. \\
$\mbox{}$ \hspace{44.8em} \scalebox{0.75}{$\blacklozenge$} \\
\subsection{Maturity-Randomization and OIDEs}
\label{MROIDE}
\noindent We next deal with maturity-randomized geometric step contracts. To this end, we consider for a function $g: \mathbb{R}^{+} \rightarrow \mathbb{R}$ satisfying
\begin{equation}
\int \limits_{0}^{\infty} e^{-\vartheta t} |g(t)| \, dt < \infty, \hspace{1.5em} \forall \vartheta >0,
\label{GAVGAVEquLCTransform0}
\end{equation}
\noindent the Laplace-Carson transform $\widehat{g}(\cdot)$ defined via  
\begin{align}
\widehat{g}(\vartheta) & := \int \limits_{0}^{\infty} \vartheta e^{-\vartheta t } \, g(t) \,dt
\label{EquLCTransform1}
\end{align}
\noindent and note that this transform has several desirable properties.\footnote{We refer the interested reader to \cite{kw03}, \cite{ki10}, \cite{lv17}, and \cite{fmv19} for a discussion of some of these properties.}~In particular, applying the Laplace-Carson transform in the context of mathematical finance allows to randomize the maturity of~(certain) financial contracts, i.e.~to switch from objects with deterministic maturity to corresponding objects with stochastic maturity. This last property offers various approaches to the valuation of financial positions and has therefore led to a wide adoption of the Laplace-Carson transform in the option pricing literature, with \cite{ca98} being one of the seminal articles in this context. \vspace{1em} \\
\noindent Once an (analytical or numerical) expression for the Laplace-Carson transform has been obtained, inversion is carried out numerically through an inversion algorithm. One possible choice is the Gaver-Stehfest algorithm that has the advantage to allow for an inversion of the transform on the real line and that has been successfully used by several authors in the option pricing literature (cf.~\cite{kw03}, \cite{ki10}, \cite{wz10}, \cite{hm13}, \cite{lv17}, \cite{cv18}, \cite{lv19}). We will also rely on this algorithm, i.e.~we set
\begin{equation}
g_{N}(t) := \sum \limits_{k=1}^{2N} \zeta_{k,N} \, \mathcal{LC}\big(g\big)\left( \frac{k \log(2)}{t}\right), \hspace{1.5em} N \in \mathbb{N}, \; t > 0,
\end{equation}
\noindent where the coefficients are given by
\begin{equation}
\label{zetaEQUA}
\zeta_{k,N} := \frac{(-1)^{N+k}}{k} \sum \limits_{j = \lfloor (k+1)/2 \rfloor }^{\min \{k,N \}} \frac{j^{N+1}}{N!} \binom{N}{j} \binom{2j}{j} \binom{j}{k-j}, \hspace{1.5em} N \in \mathbb{N}, \; 1 \leq k \leq 2N,
\end{equation}
\noindent with $\lfloor a \rfloor := \sup \{z \in \mathbb{Z}: \, z \leq a \}$, and will recover the original function $g(\cdot)$ by means of the following relation
\begin{equation}
\lim \limits_{N \rightarrow \infty} g_{N}(t) = g(t).
\label{CONVer}
\end{equation}
\noindent More technical details around the Gaver-Stehfest inversion as well as formal proofs of the convergence result~(\ref{CONVer}) for ``sufficiently well-behaved functions'' are provided in \cite{va04}, \cite{aw06}, \cite{ku13}, and references therein. 
\subsubsection{European-Type Contracts}
\label{EUROtypeCo}
\noindent To start, we focus on maturity-randomized versions of the European-type geometric step option $\mathcal{DSC}_{E}^{\star}(\cdot)$, i.e.~we consider geometric step contracts of the form
\begin{equation}
\widehat{\mathcal{DSC}_{E}^{\star}}(\vartheta,x;K,\bm{\ell}, \bm{\rho}_{\bm{\ell}}) := \mathbb{E}_{x}^{\mathbb{Q}} \left[ \left(\bar{S}_{\mathcal{T}_{\vartheta}} - K \right)^{+} \right],
\label{MREuro1}
\end{equation}
\noindent where $(\bar{S}_{t})_{t \geq 0}$ refers, once again, to the (strong) Markov process obtained by ``killing'' the sample path of $(S_{t})_{t \geq 0}$ at the proportional rate $\lambda(x) := r - \bm{\rho}_{\bm{\ell}} \cdot \bigg( \begin{array}{c}
 \mathds{1}_{(0,L)}(x) \\
 \mathds{1}_{(H,\infty)}(x) 
\end{array} \bigg)$ and whose cemetery state is given by $\partial \equiv 0$, and $\mathcal{T}_{\vartheta}$ denotes an exponentially distributed random time of intensity $\vartheta > 0$ that is independent of $(S_{t})_{t \geq 0}$. It is not hard to see that (\ref{MREuro1}) re-writes as
\begin{equation}
\widehat{\mathcal{DSC}_{E}^{\star}}(\vartheta,x;K,\bm{\ell}, \bm{\rho}_{\bm{\ell}}) = \mathbb{E}_{x}^{\mathbb{Q}} \left[ \, \mathbb{E}_{x}^{\mathbb{Q}} \left[ \left(\bar{S}_{\mathcal{T}_{\vartheta}} - K \right)^{+} \big| \mathcal{T}_{\vartheta} \right] \, \right] = \int \limits_{0}^{\infty} \vartheta e^{-\vartheta t } \, \mathcal{DSC}_{E}^{\star}(t,x;K,\bm{\ell}, \bm{\rho}_{\bm{\ell}}) \,dt,
\label{LCTMREuro1}
\end{equation}
\noindent and therefore that the maturity-randomized versions (\ref{MREuro1}) correspond, for any fixed $x, K, \bm{\ell},$ and $\bm{\rho}_{\bm{\ell}}$, to a strict application of the Laplace-Carson transform to the function $\mathcal{T} \mapsto \mathcal{DSC}_{E}^{\star}(\mathcal{T},x;K,\bm{\ell}, \bm{\rho}_{\bm{\ell}})$. Additionally, we note that this transform is well-defined. Indeed, this was already shown in a slightly different context for standard (European- and American-type) options in \cite{Ma18} and directly follows from these results, for $\bm{\rho}_{\bm{\ell}} \leq 0$ and $\bullet \in \{E,A\}$, by means of the inequality 
\begin{equation}
\mathcal{DSC}_{\bullet}^{\star}(\mathcal{T},x;K,\bm{\ell}, \bm{\rho}_{\bm{\ell}}) \leq \mathcal{DSC}_{\bullet}^{\star}(\mathcal{T},x;K,\bm{\ell}, (0,0)) =: \mathcal{C}_{\bullet}(\mathcal{T},x;K).
\end{equation}
\noindent Consequently, combining these properties with arguments similarly used in the proof of Proposition~\ref{prop1} allows to obtain an OIDE characterization of the maturity-randomized European-type contracts~(\ref{MREuro1}). This is the content of the next proposition, whose proof is provided in Appendix A.
\begin{Prop}
\label{prop4}
\noindent For any intensity $\vartheta >0$, strike $K \geq 0$, barrier levels $0 \leq L \leq H < \infty$, and knock-out rates $ \rho_{L}, \rho_{H} \leq 0$, the value of the maturity-randomized European-type geometric double barrier step call, $\widehat{\mathcal{DSC}_{E}^{\star}}(\cdot)$, is continuous on $[0,\infty)$ and solves the ordinary integro-differential equation
\begin{equation}
\vartheta (x-K)^{+} + \mathcal{A}_{S} \widehat{\mathcal{DSC}^{\star}_{E}} (\vartheta,x; K, \bm{\ell},\bm{\rho}_{\bm{\ell}}) -\bigg( (r + \vartheta) - \bm{\rho}_{\bm{\ell}} \cdot \bigg( \begin{array}{c}
 \mathds{1}_{(0,L)}(x) \\
 \mathds{1}_{(H,\infty)}(x) 
\end{array} \bigg) \bigg) \widehat{\mathcal{DSC}^{\star}_{E}} (\vartheta,x; K, \bm{\ell},\bm{\rho}_{\bm{\ell}}) = 0 , \label{MRGSCEuOIDE1}
\end{equation}
\noindent on $(0,\infty)$ with initial condition
\begin{equation}
\widehat{\mathcal{DSC}_{E}^{\star}}(\vartheta,0; K, \bm{\ell},\bm{\rho}_{\bm{\ell}})  =  0. \label{MRGSCEuOIDE2}
\end{equation}
\end{Prop}
\subsubsection{American-Type Contracts}
Lastly, we discuss maturity-randomized versions of the American-type geometric step option $\mathcal{DSC}_{A}^{\star}(\cdot)$, i.e.~we consider the following geometric step contracts
\begin{equation}
\widehat{\mathcal{DSC}_{A}^{\star}}(\vartheta,x;K,\bm{\ell}, \bm{\rho}_{\bm{\ell}}) := \sup \limits_{\tau \in \mathfrak{T}_{[0,\infty)}} \mathbb{E}_{x}^{\mathbb{Q}} \left[ \left(\bar{S}_{\mathcal{T}_{\vartheta} \wedge \tau} - K \right)^{+} \right],
\label{MRAmer1}
\end{equation}
\noindent where we use the notation introduced in Section~\ref{MROIDE}.\ref{EUROtypeCo} Due to their complex early exercise structure, these maturity-randomized contracts do not anymore coincide with a strict application of the Laplace-Carson transform to their deterministic counterparts $\mathcal{T} \mapsto \mathcal{DSC}_{A}^{\star}(\mathcal{T},x;K,\bm{\ell}, \bm{\rho}_{\bm{\ell}})$. Instead, conditioning on the (independent) exponential random time $\mathcal{T}_{\vartheta}$ only leads to the following expression
\begin{equation}
\widehat{\mathcal{DSC}_{A}^{\star}}(\vartheta,x;K,\bm{\ell}, \bm{\rho}_{\bm{\ell}}) = \sup \limits_{\tau \in \mathfrak{T}_{[0,\infty)}} \mathbb{E}_{x}^{\mathbb{Q}} \left[ \, \mathbb{E}_{x}^{\mathbb{Q}} \left[ \left(\bar{S}_{\mathcal{T}_{\vartheta} \wedge \tau} - K \right)^{+}  \big | \mathcal{T}_{\vartheta} \right] \, \right] = \sup \limits_{\tau \in \mathfrak{T}_{[0,\infty)}} \int \limits_{0}^{\infty} \vartheta e^{-\vartheta t } \, \mathcal{DSC}^{\star}(t \wedge \tau,x;K,\bm{\ell}, \bm{\rho}_{\bm{\ell}}) \,dt, 
\label{RandTimeInt}
\end{equation}
\noindent where $\mathcal{DSC}^{\star}(\tau,x; K, \bm{\ell}, \bm{\rho}_{\bm{\ell}}) = \mathcal{DSC}\big( \tau ,x, 0, 0;  \,r,\delta, K, L, H,\rho_{L}, \rho_{H}, \Psi_{X}(\cdot) \big) $ denotes, as earlier, for any $T >0$ and stopping time $\tau \in \mathfrak{T}_{[0,T]}$, the contract version of (\ref{not1}) that is initiated at the valuation date under consideration, i.e.~in the sense of the notation introduced in~(\ref{SimpliBUP2}). Nevertheless, the same arguments as in Section~\ref{MROIDE}.\ref{EUROtypeCo} (cf.~\cite{Ma18}) directly show that the right-hand side in (\ref{RandTimeInt}) is well-defined for $\bm{\rho}_{\bm{\ell}} \leq 0$ and any $\vartheta >0$. Furthermore, OIDE characterizations of the maturity-randomized American-type contract $\widehat{\mathcal{DSC}_{A}^{\star}}(\cdot)$ as well as of the respective early exercise premiums can be derived using strong Markovian arguments. This is the content of the following discussion. \vspace{1em} \\
To start, we recall that the (independent) exponential random time $\mathcal{T}_{\vartheta}$ can be interpreted as the (first) jump time of a corresponding (independent) Poisson process $(N_{t})_{t \geq 0}$ with intensity $\vartheta > 0$ and that this can be used to re-express the optimal stopping problem in a slightly different form. In particular, we can consider, for any $\vartheta > 0 $ and initial value $z = (n,x) \in \mathbb{N}_{0} \times [0,\infty)$, the process $(Z_{t})_{t \geq 0}$ defined on the state domain $\mathcal{D} := \mathbb{N}_{0} \times [0,\infty)$ via $Z_{t} := (n + N_{t} , \bar{S}_{t})$, $\bar{S}_{0} = x$, as well as its stopped version, $(Z_{t}^{\mathcal{S}_{J}})_{t \geq 0}$, defined, for $t \geq 0$, via
\begin{equation}
Z_{t}^{\mathcal{S}_{J}} := Z_{t \wedge \tau_{\mathcal{S}_{J}}}, \hspace{1.5em}  \mbox{with} \hspace{1.7em} \tau_{\mathcal{S}_{J}} := \inf \{t \geq 0 : \, Z_{t} \in \mathcal{S}_{J} \}, \hspace{1.5em}  \mbox{and} \hspace{1.7em} \mathcal{S}_{J} := \mathbb{N} \times [0,\infty).
\label{STOPPEDprocDEF}
\end{equation} 
\noindent Clearly, the process $(Z_{t}^{\mathcal{S}_{J}})_{t \geq 0}$ behaves exactly like the process $(Z_{t})_{t \geq 0}$ for all times $t < \tau_{\mathcal{S}_{J}}$, which implies that most of the properties of $(Z_{t})_{t \geq 0}$ naturally extend to $(Z_{t}^{\mathcal{S}_{J}})_{t \geq 0}$.\footnote{In particular, the process $(Z_{t}^{\mathcal{S}_{J}})_{t \geq 0}$ is again strongly Markovian on the state domain~$\mathcal{D}$.}~Additionally, $\widehat{\mathcal{DSC}_{A}^{\star}}(\cdot)$ can be re-expressed, for $\vartheta, K,\bm{\ell}$ and $\bm{\rho}_{\bm{\ell}}$, in the form
\begin{equation}
\widehat{\mathcal{DSC}^{\star}_{A}}(\vartheta,x; K, \bm{\ell}, \bm{\rho}_{\bm{\ell}} ) = \widehat{V_{A}}\big((0,x)\big),
\label{MRAmerIMeq}
\end{equation}
\noindent where the value function $\widehat{V_{A}}(\cdot)$ has the following representation under the measure $\mathbb{Q}_{z}^{Z}$ having initial distribution $Z_{0} = z$:
\begin{align}
\widehat{V_{A}}(z) : = \sup \limits_{\tau \in \mathfrak{T}_{[0,\infty)}} \mathbb{E}^{\mathbb{Q}^{Z}}_{z} \big[ G(Z_{\tau}^{\mathcal{S}_{J}}) \big], \hspace{1.5em}  G(z)  := (x - K)^{+}.
\label{NEWproBL}
\end{align}
\noindent Therefore, using the fact that the payoff function $x \mapsto (x-K)^{+}$ is continuous as well as standard optimal stopping arguments (cf.~Corollary 2.9.~and Remark~2.10. in \cite{pe06}), we can infer that the continuation and stopping regions to (the more general) Problem (\ref{NEWproBL}) read, respectively
\begin{align}
\widehat{\mathcal{D}_{c}^{Gen.}} = \left \{ z \in \mathcal{D}: \, \widehat{V_{A}}(z) > G(z) \right \} , \hspace{1.5em} \mbox{and} \hspace{1.7em}  \widehat{\mathcal{D}_{s}^{Gen.}} = \left \{ z \in \mathcal{D}: \, \widehat{V_{A}}(z) = G(z) \right \}, \label{NEWregion}
\end{align}
\noindent and that the first-entry time
\begin{equation}
\tau_{\widehat{\mathcal{D}_{s}^{Gen.}}} := \inf \Big \{ t \geq 0: \, Z_{t}^{\mathcal{S}_{J}} \in \widehat{\mathcal{D}_{s}^{Gen.}} \Big \}
\label{NeWWOTime}
\end{equation}
\noindent is optimal in (\ref{NEWproBL}).\footnote{Note that the finiteness of this stopping time directly follows from the finiteness of the first moment of any exponential distribution and the fact that $\mathcal{S}_{J} \subseteq \widehat{\mathcal{D}_{s}^{Gen.}}$.}~This then allows us to make use of standard strong Markovian arguments to derive a characterization of the American-type contract $\widehat{\mathcal{DSC}^{\star}_{A}}(\cdot)$ in terms of a Cauchy-type problem and leads via Relation (\ref{MRAmerIMeq}) and the following continuation and stopping regions
\begin{align}
\widehat{\mathcal{D}}_{\vartheta,c} & = \left \{ x \in [0,\infty): \, \widehat{\mathcal{DSC}^{\star}_{A}}(\vartheta,x; K, \bm{\ell}, \bm{\rho}_{\bm{\ell}} ) > (x-K)^{+} \right \} , \label{MRCregion} \\
\widehat{\mathcal{D}}_{\vartheta,s} & = \left \{ x \in [0,\infty): \, \widehat{\mathcal{DSC}^{\star}_{A}}(\vartheta,x; K, \bm{\ell}, \bm{\rho}_{\bm{\ell}} ) = (x-K)^{+} \right \}, \label{MRSregion}
\end{align}
to the next proposition. A proof is presented in Appendix A.
\begin{Prop}
\label{prop5}
\noindent For any intensity $\vartheta >0$, strike $K \geq 0$, barrier levels $0 \leq L \leq H < \infty$, and knock-out rates $\rho_{L}, \rho_{H} \leq 0$, the value of the maturity-randomized American-type geometric double barrier step \mbox{call, $\widehat{\mathcal{DSC}_{A}^{\star}}(\cdot)$,} is continuous on $[0,\infty)$ and satisfies the following Cauchy-type problem:
\begin{equation}
\vartheta (x-K)^{+} + \mathcal{A}_{S} \widehat{\mathcal{DSC}^{\star}_{A}} (\vartheta,x; K, \bm{\ell},\bm{\rho}_{\bm{\ell}}) -\bigg( (r + \vartheta) - \bm{\rho}_{\bm{\ell}} \cdot \bigg( \begin{array}{c}
 \mathds{1}_{(0,L)}(x) \\
 \mathds{1}_{(H,\infty)}(x) 
\end{array} \bigg) \bigg) \widehat{\mathcal{DSC}^{\star}_{A}} (\vartheta,x; K, \bm{\ell},\bm{\rho}_{\bm{\ell}}) = 0 , \label{MRGSCAmerOIDE1}
\end{equation}
\noindent for $x \in \widehat{\mathcal{D}}_{\vartheta,c}$ with boundary condition
\begin{equation}
\widehat{\mathcal{DSC}_{A}^{\star}}(\vartheta,x; K, \bm{\ell},\bm{\rho}_{\bm{\ell}})  =  (x-K)^{+}, \hspace{1.5em}  \mbox{for} \; \, x \in \widehat{\mathcal{D}}_{\vartheta,s}. \label{MRGSCAmerOIDE2}
\end{equation}
\end{Prop}
\noindent To finalize our discussion, we aim to characterize diffusion and jump contributions to the maturity-randomized early exercise premium of geometric double barrier step contracts, that is defined for $\vartheta, x, K, \bm{\ell},$ and $\bm{\rho}_{\bm{\ell}}$ via
\begin{equation}
\widehat{\mathcal{E}_{\mathcal{DSC}}^{\star}}(\vartheta,x; K, \bm{\ell},\bm{\rho}_{\bm{\ell}}) := \widehat{\mathcal{DSC}_{A}^{\star}}(\vartheta,x; K, \bm{\ell},\bm{\rho}_{\bm{\ell}}) - \widehat{\mathcal{DSC}_{E}^{\star}}(\vartheta,x; K, \bm{\ell},\bm{\rho}_{\bm{\ell}}).
\label{EEP_MR1bup}
\end{equation}
\noindent For simplicity of the exposition, we directly rely on the continuation and stopping regions introduced in (\ref{MRCregion}), (\ref{MRSregion}) and note that the maturity-randomized American-type option $\widehat{\mathcal{DSC}_{A}^{\star}}(\cdot)$ can be equivalently written as
\begin{equation}
\widehat{\mathcal{DSC}_{A}^{\star}}(\vartheta,x; K, \bm{\ell},\bm{\rho}_{\bm{\ell}}) = \mathbb{E}_{x}^{\mathbb{Q}} \left[ \left(\bar{S}_{\mathcal{T}_{\vartheta} \wedge \tau_{\widehat{\mathcal{D}}_{\vartheta,s}}} - K \right)^{+} \right],
\end{equation}
\noindent since the first-entry time $\tau_{\widehat{\mathcal{D}}_{\vartheta,s}} := \inf \left \{ t \geq 0: \, \bar{S}_{t} \in \widehat{\mathcal{D}}_{\vartheta,s} \right \}$ clearly inherits the optimality of its counterpart~(\ref{NeWWOTime}) in the more general problem~(\ref{NEWproBL}). Then, following the line of the arguments provided in Section~\ref{GEOPIDE}.\ref{GEOPIDEAmer}, we can make use of the sets $\partial \widehat{\mathcal{D}}_{\vartheta,s}$ and $\widehat{\mathcal{D}}_{\vartheta,s}^{\circ}$ to decompose the stopping region into sub-regions where (early) stopping is purely due to diffusion and jumps, respectively, and subsequently derive a decomposition of the maturity-randomized early exercise premium, $\widehat{\mathcal{E}_{\mathcal{DSC}}^{\star}}(\cdot)$, of the form
\begin{equation}
\widehat{\mathcal{E}_{\mathcal{DSC}}^{\star}}(\vartheta,x; K, \bm{\ell},\bm{\rho}_{\bm{\ell}}) = \widehat{\mathcal{E}_{\mathcal{DSC}}^{0,\star}}(\vartheta,x; K, \bm{\ell},\bm{\rho}_{\bm{\ell}}) + \widehat{\mathcal{E}_{\mathcal{DSC}}^{\mathcal{J},\star}}(\vartheta,x; K, \bm{\ell},\bm{\rho}_{\bm{\ell}}).
\end{equation}
\noindent Here, the premiums $\widehat{\mathcal{E}_{\mathcal{DSC}}^{0,\star}}(\cdot)$ and $\widehat{\mathcal{E}_{\mathcal{DSC}}^{\mathcal{J},\star}}(\cdot)$ refer to the maturity-randomized early exercise contributions of the diffusion and jump parts, respectively, and are defined via
\begin{align}
\widehat{\mathcal{E}_{\mathcal{DSC}}^{0,\star}}(\vartheta,x; K, \bm{\ell}, \bm{\rho}_{\bm{\ell}} ) & := \widehat{\mathcal{DSC}_{A}^{0,\star}}(\vartheta,x; K, \bm{\ell}, \bm{\rho}_{\bm{\ell}} ) - \widehat{\mathcal{DSC}_{E}^{0,\star}}(\vartheta,x; K, \bm{\ell}, \bm{\rho}_{\bm{\ell}} ), \\
\widehat{\mathcal{E}_{\mathcal{DSC}}^{\mathcal{J},\star}}(\vartheta,x; K, \bm{\ell}, \bm{\rho}_{\bm{\ell}} ) & := \widehat{\mathcal{DSC}_{A}^{\mathcal{J},\star}}(\vartheta,x; K, \bm{\ell}, \bm{\rho}_{\bm{\ell}} ) - \widehat{\mathcal{DSC}_{E}^{\mathcal{J},\star}}(\vartheta,x; K, \bm{\ell}, \bm{\rho}_{\bm{\ell}} ),
\end{align}
\noindent where the maturity-randomized European-type functions $\widehat{\mathcal{DSC}_{E}^{0,\star}}(\cdot)$ and $\widehat{\mathcal{DSC}_{E}^{\mathcal{J},\star}}(\cdot)$ are given by
\begin{align}
\widehat{\mathcal{DSC}_{E}^{0,\star}}(\vartheta,x; K, \bm{\ell}, \bm{\rho}_{\bm{\ell}} ) & =  \mathbb{E}_{x}^{\mathbb{Q}} \left[ \left(\bar{S}_{\mathcal{T}_{\vartheta} } - K \right)^{+} \mathds{1}_{\partial \widehat{\mathcal{D}}_{\vartheta,s}}\big(\bar{S}_{\mathcal{T}_{\vartheta} \wedge \tau_{\widehat{\mathcal{D}}_{\vartheta,s}}} \big) \right], \\
\widehat{\mathcal{DSC}_{E}^{\mathcal{J},\star}}(\vartheta,x; K, \bm{\ell}, \bm{\rho}_{\bm{\ell}} ) & =  \mathbb{E}_{x}^{\mathbb{Q}} \left[ \left(\bar{S}_{\mathcal{T}_{\vartheta}} - K \right)^{+} \mathds{1}_{\widehat{\mathcal{D}}_{\vartheta,s}^{\circ}}\big(\bar{S}_{\mathcal{T}_{\vartheta} \wedge \tau_{\widehat{\mathcal{D}}_{\vartheta,s}}} \big) \right],
\end{align}
\noindent and the maturity-randomized American-type contributions $\widehat{\mathcal{DSC}_{A}^{0,\star}}(\cdot)$ and $\widehat{\mathcal{DSC}_{A}^{\mathcal{J},\star}}(\cdot)$ are defined accordingly, as
\begin{align}
\widehat{\mathcal{DSC}_{A}^{0,\star}}(\mathcal{T},x; K, \bm{\ell}, \bm{\rho}_{\bm{\ell}} ) & = \mathbb{E}_{x}^{\mathbb{Q}} \left[ \Big(\bar{S}_{\mathcal{T}_{\vartheta} \wedge \tau_{\widehat{\mathcal{D}}_{\vartheta,s}} } - K \Big)^{+} \mathds{1}_{\partial \widehat{\mathcal{D}}_{\vartheta,s}}\big(\bar{S}_{\mathcal{T}_{\vartheta} \wedge \tau_{\widehat{\mathcal{D}}_{\vartheta,s}}} \big) \right],  \\
\widehat{\mathcal{DSC}_{A}^{\mathcal{J},\star}}(\mathcal{T},x; K, \bm{\ell}, \bm{\rho}_{\bm{\ell}} )  & = \mathbb{E}_{x}^{\mathbb{Q}} \left[ \Big(\bar{S}_{\mathcal{T}_{\vartheta} \wedge \tau_{\widehat{\mathcal{D}}_{\vartheta,s}} } - K \Big)^{+} \mathds{1}_{\widehat{\mathcal{D}}_{\vartheta,s}^{\circ}}\big(\bar{S}_{\mathcal{T}_{\vartheta} \wedge \tau_{\widehat{\mathcal{D}}_{\vartheta,s}}} \big) \right].
\end{align}
\noindent Combining these definitions with strong Markovian arguments similarly used in the proof of the previous propositions and the memorylessness of the exponential distribution finally allows us to derive OIDE characterizations of the early exercise contributions $\widehat{\mathcal{E}_{\mathcal{DSC}}^{0,\star}}(\cdot)$ and $\widehat{\mathcal{E}_{\mathcal{DSC}}^{\mathcal{J},\star}}(\cdot)$. This is the content of the next proposition, whose proof is provided in Appendix~A.
\begin{Prop}
\label{prop6}
\noindent For any intensity $\vartheta > 0$, strike $K \geq 0$, barrier levels $0 \leq L \leq H < \infty$, and knock-out rates $ \rho_{L}, \rho_{H} \leq 0$, the value of the diffusion contribution to the maturity-randomized early exercise premium of the geometric double barrier step call, $\widehat{\mathcal{E}_{\mathcal{DSC}}^{0,\star}}(\cdot)$, satisfies the following Cauchy-type problem:
\begin{equation}
\mathcal{A}_{S} \widehat{\mathcal{E}_{\mathcal{DSC}}^{0,\star}} (\vartheta,x; K, \bm{\ell},\bm{\rho}_{\bm{\ell}}) -\bigg( (r + \vartheta)- \bm{\rho}_{\bm{\ell}} \cdot \bigg( \begin{array}{c}
 \mathds{1}_{(0,L)}(x) \\
 \mathds{1}_{(H,\infty)}(x) 
\end{array} \bigg) \bigg) \widehat{\mathcal{E}_{\mathcal{DSC}}^{0,\star}} (\vartheta,x; K, \bm{\ell},\bm{\rho}_{\bm{\ell}}) = 0 , \label{GSCAmeEEPOIDE1}
\end{equation}
\noindent for $ x \in \widehat{\mathcal{D}}_{\vartheta,c}$ with boundary conditions
\begin{align}
\widehat{\mathcal{E}_{\mathcal{DSC}}^{0,\star}}(\vartheta,x; K, \bm{\ell},\bm{\rho}_{\bm{\ell}})   =  (x-K )^{+} - & \widehat{\mathcal{DSC}_{E}^{\star}}(\vartheta,x; K, \bm{\ell}, \bm{\rho}_{\bm{\ell}} ), \hspace{1.5em}  \mbox{for} \; \,  x \in \partial \widehat{\mathcal{D}}_{\vartheta,s}, \label{GSCAmeEEPOIDE1-1} \\
\widehat{\mathcal{E}_{\mathcal{DSC}}^{0,\star}}(\vartheta,x; K, \bm{\ell},\bm{\rho}_{\bm{\ell}}) & = 0, \hspace{1.5em}  \mbox{for} \; \,  x \in \widehat{\mathcal{D}}_{\vartheta,s}^{\circ}. \label{GSCAmeEEPOIDE1-2}
\end{align}
\noindent Similarly, the value of the jump contribution to the maturity-randomized early exercise premium of the geometric double barrier step call, $\widehat{\mathcal{E}_{\mathcal{DSC}}^{\mathcal{J},\star}}(\cdot)$, solves the following Cauchy-type problem:
\begin{equation}
\mathcal{A}_{S} \widehat{\mathcal{E}_{\mathcal{DSC}}^{\mathcal{J},\star}}(\vartheta,x; K, \bm{\ell},\bm{\rho}_{\bm{\ell}}) -\bigg( (r + \vartheta)- \bm{\rho}_{\bm{\ell}} \cdot \bigg( \begin{array}{c}
 \mathds{1}_{(0,L)}(x) \\
 \mathds{1}_{(H,\infty)}(x) 
\end{array} \bigg) \bigg) \widehat{\mathcal{E}_{\mathcal{DSC}}^{\mathcal{J},\star}}(\vartheta,x; K, \bm{\ell},\bm{\rho}_{\bm{\ell}}) = 0 , \label{GSCAmeEEPOIDE2}
\end{equation}
\noindent for $ x \in \widehat{\mathcal{D}}_{\vartheta,c}$ with boundary conditions
\begin{align}
\widehat{\mathcal{E}_{\mathcal{DSC}}^{\mathcal{J},\star}}(\vartheta,x; K, \bm{\ell},\bm{\rho}_{\bm{\ell}})  & =  0, \hspace{1.5em}  \mbox{for} \; \,  x \in \partial \widehat{\mathcal{D}}_{\vartheta,s}, \label{GSCAmeEEPOIDE2-1} \\
\widehat{\mathcal{E}_{\mathcal{DSC}}^{\mathcal{J},\star}}(\vartheta,x; K, \bm{\ell},\bm{\rho}_{\bm{\ell}}) = (x-K )^{+} - & \widehat{\mathcal{DSC}_{E}^{\star}}(\vartheta,x; K, \bm{\ell}, \bm{\rho}_{\bm{\ell}} ), \hspace{1.5em}  \mbox{for} \; \,  x \in \widehat{\mathcal{D}}_{\vartheta,c}^{\circ}. \label{GSCAmeEEPOIDE2-2}
\end{align}
\end{Prop}
$\mbox{ }$ \vspace{-0.7em} \\
\noindent \underline{\bf Remark 3.} \vspace{0.2em} \\
\noindent Although maturity-randomized American-type contracts and maturity-randomized early exercise premiums do not anymore coincide with a strict application of the Laplace-Carson transform to their deterministic counterparts, they exhibit a very similar structure. This becomes clear when comparing Equations (\ref{RandTimeInt}) and (\ref{EEP_MR1bup}) with Identity (\ref{LCTMREuro1}). Hence, once (analytical or numerical) results are obtained for these quantities, a very natural pricing algorithm consists in dealing with their results as if they would actually correspond to proper Laplace-Carson applications and therefore in inverting them via an algorithm such as the one proposed in the Gaver-Stehfest inversion. This has been already investigated by other authors in a similar context (cf.~\cite{wz10}, \cite{lv17}, \cite{cv18}) where this approach has proven to deliver a very good pricing accuracy. We will follow the idea of this literature and will provide in Section \ref{MR_NUMRES} numerical results for geometric down-and-out step call options under hyper-exponential jump-diffusion markets based on this ansatz. This also justifies our slight abuse of notation in the current section, where we intentionally used for both maturity-randomized American-type contracts as well as maturity-randomized early exercise premiums the same notation as for Laplace-Carson transforms. \\
$\mbox{}$ \hspace{44.8em} \scalebox{0.75}{$\blacklozenge$} \\

\section{Geometric Step Options and Hyper-Exponential Jump-Diffusion Markets}
\label{SEC3}
As an application of the theory developed in Section \ref{SEC2}, we derive semi-analytical pricing results for (regular) geometric down-and-out step call options under hyper-exponential jump-diffusion markets, i.e.~we fix in~(\ref{market1}), (\ref{market2}) hyper-exponential jump-diffusion dynamics $(X_{t})_{t \geq 0}$ and consider geometric step options of the form
\begin{equation}
\mathcal{DOSC}^{\star}_{\bullet}(\mathcal{T},x; K, L, \rho_{L} ) := \mathcal{DSC}^{\star}_{\bullet} \big(\mathcal{T},x; K, (L,L), (\rho_{L},0) \big),
\end{equation}
\noindent for $\bullet \in \{E, A \}$, time to maturity $\mathcal{T} \geq 0$, initial value $x \geq 0$, strike $K \geq 0$, lower barrier $0 \leq L \leq K < \infty$ and knock-out rate $\rho_{L} \leq 0$.
\subsection{Generalities on Hyper-Exponential Jump-Diffusion Markets}
\noindent We recall that a hyper-exponential jump-diffusion market is a Lévy market consisting of a deterministic savings account $(B_{t}(r))_{t \geq 0}$ (cf.~(\ref{market1})) and a risky asset $(S_{t})_{t \geq 0}$ (cf.~(\ref{market2})) whose driving process $(X_{t})_{t \geq 0}$ combines a Brownian diffusion with hyper-exponentially distributed jumps. In particular, the underlying dynamics $(X_{t})_{t \geq 0}$ have the usual jump-diffusion structure, i.e.~they can be characterized on a filtered probability space $(\Omega, \mathcal{F}, \mathbf{F}, \mathbb{P})$ via
\begin{equation}
X_{t} = \Big( r -\delta -\lambda \zeta - \frac{1}{2}\sigma_{X}^{2} \Big) t + \sigma_{X} W_{t} + \sum \limits_{i=1}^{N_{t}} J_{i}, \hspace{1.5em} t \geq 0,
\label{DYNhypexp}
\end{equation}
\noindent where $(W_{t})_{t \geq 0}$ denotes an $\mathbf{F}$-Brownian motion and $(N_{t})_{t \geq 0}$ is an $\mathbf{F}$-Poisson process having intensity parameter $\lambda >0$. The constants $\zeta := \mathbb{E}^{\mathbb{Q}} \left[ e^{J_{1}} - 1 \right]$ and $\sigma_{X} > 0$ express the average (percentage) jump size and the volatility of the diffusion part, respectively. Additionally, the jumps $(J_{i})_{i \in \mathbb{N}}$ are assumed to be independent of $(N_{t})_{t \geq 0}$ and to form a sequence of independent and identically distributed random variables following a hyper-exponential distribution, i.e.~their (common) density function $f_{J_{1}}(\cdot)$ is given by
\begin{equation}
f_{J_{1}}(y) = \sum \limits_{i=1}^{m} p_{i} \xi_{i}e^{-\xi_{i} y} \mathds{1}_{ \{ y \geq 0 \}} + \sum \limits_{j=1}^{n} q_{j} \eta_{j} e^{\eta_{j} y } \mathds{1}_{ \{ y < 0 \} },
\label{hypexpDens}
\end{equation}
\noindent where $p_{i} >0$ and $\xi_{i} > 1$ for $i \in \{1,\ldots, m\}$ and $q_{j}>0$ and $\eta_{j} >0$ for $j \in \{ 1, \ldots, n \}$. Here, the parameters $(p_{i})_{i \in \{1,\ldots,m \}}$ and $(q_{j})_{j \in \{1,\ldots,n \}}$ represent the proportion of jumps that are attributed to particular jump types and are therefore assumed to satisfy the condition $\sum \limits_{i=1}^{m} p_{i} + \sum \limits_{j=1}^{n} q_{j} = 1$. For notational simplicity, we require that the intensity parameters $(\xi_{i})_{i \in \{1, \ldots, m \}}$ and $(\eta_{j})_{j \in \{1, \ldots, n \}}$ are ordered in the sense that
\begin{equation}
\xi_{1} < \xi_{2} < \cdots < \xi_{m} \hspace{2.5em} \mbox{and} \hspace{2.5em} \eta_{1} < \eta_{2} < \cdots < \eta_{n}
\end{equation}
\noindent and note that this does not consist in a loss of generality. \vspace{1em} \\
\noindent As special class of Lévy markets, hyper-exponential jump-diffusion markets can be equivalently characterized in terms of their Lévy triplet $\left(b_{X},\sigma_{X}^{2},\Pi_{X} \right)$, where $b_{X}$ and $\Pi_{X}$ are then obtained as 
\begin{align}
b_{X}  := \Big(r -\delta -\lambda \zeta - \frac{1}{2}\sigma_{X}^{2}\Big) + \int \limits_{\{ |y|\leq 1 \}} y \, \Pi_{X}(dy) \hspace{2em} \mbox{and} \hspace{2.3em} \Pi_{X}(dy) := \lambda f_{J_{1}}(y) dy .
\label{INTENSmeasure}
\end{align}
\noindent Combining these results with Equation (\ref{CHARexp}), their Lévy exponent, $\Psi_{X}(\cdot)$, is then easily derived as
\begin{align}
\Psi_{X}(\theta) = -i \Big( r -\delta -\lambda \zeta - \frac{1}{2}\sigma_{X}^{2} \Big) \theta + \frac{1}{2} \sigma_{X}^2 \theta^2  - \lambda \left( \sum \limits_{i=1}^{m} \frac{p_{i} \xi_{i}}{\xi_{i} - i \theta} + \sum \limits_{j=1}^{n} \frac{q_{j} \eta_{j}}{\eta_{j} + i \theta} - 1\right).
\end{align}
\noindent Similarly, their Laplace exponent, $\Phi_{X}(\cdot)$, is well-defined for $\theta \in (-\eta_{1}, \xi_{1}) $ and equals
\begin{equation}
\Phi_{X}(\theta) = \Big( r -\delta -\lambda \zeta - \frac{1}{2}\sigma_{X}^{2} \Big) \theta + \frac{1}{2} \sigma_{X}^2 \theta^2  + \lambda \left( \sum \limits_{i=1}^{m} \frac{p_{i} \xi_{i}}{\xi_{i} - \theta} + \sum \limits_{j=1}^{n} \frac{q_{j} \eta_{j}}{\eta_{j} + \theta} - 1\right).
\label{LAP1}
\end{equation}
\noindent In what follows, we will consider the Laplace exponent as standalone function on the extended real domain $\Phi_{X}: \mathbb{R} \setminus \{\xi_{1}, \ldots, \xi_{m}, -\eta_{1}, \ldots, -\eta_{n} \} \rightarrow \mathbb{R}$. This quantity will play a central role in the upcoming derivations. In fact, many distributional properties of hyper-exponential jump-diffusion markets (and of their generalizations) are closely linked to the roots of the equation $\Phi_{X}(\theta) = \alpha$, for $\alpha \geq 0$. This was already used in various articles dealing with option pricing and risk management within the class of mixed-exponential jump-diffusion models (cf.~among others \cite{ca09}, \cite{cc09}, \cite{ck11}, \cite{ck12}). In this context, the following (important) lemma was derived in \cite{ca09} under hyper-exponential jump-diffusion models. The interested reader is referred for a proof to the latter article.
\begin{Lem}
\label{SimpleCAIlem}
Let $\sigma_{X} > 0$ and $\Phi_{X}(\cdot)$ be defined as in (\ref{LAP1}). Then, for any $\alpha >0$, the equation $\Phi_{X}(\theta) = \alpha$ has $(m+n+2)$ real roots $\beta_{1,\alpha}, \ldots, \beta_{m+1,\alpha}$ and $\gamma_{1,\alpha}, \ldots, \gamma_{n+1,\alpha}$ that satisfy
\begin{align}
-\infty < \gamma_{n+1,\alpha} < -\eta_{n} < \gamma_{n,\alpha} < -\eta_{n-1} < \cdots < \gamma_{2,\alpha} < -\eta_{1} < \gamma_{1,\alpha} < 0, \\
0 < \beta_{1,\alpha} < \xi_{1} < \beta_{2,\alpha} < \cdots < \xi_{m-1} < \beta_{m,\alpha} < \xi_{m} < \beta_{m+1,\alpha} < \infty. \hspace{0.5em} 
\end{align}
\end{Lem} 
$\mbox{ }$ \vspace{-0.7em} \\
\noindent \underline{\bf Remark 4.}
\begin{itemize} \setlength \itemsep{-0.1em}
\item[i)] At this point, one should note that the roots in Lemma~\ref{SimpleCAIlem} are only known in analytical form in very few cases. Nevertheless, this does not impact the importance and practicability of this result since all roots can be anyway recovered using standard numerical techniques. 
\item[ii)] Similar characterizations to the one presented in Lemma~\ref{SimpleCAIlem} can be derived under the assumption that $\sigma_{X} =0$ (cf.~\cite{fmv19}) and combining these characterizations with the upcoming derivations of Section~\ref{SECmaRaHYPER} subsequently allows to derive semi-analytical pricing results under hyper-exponential jump-diffusion markets with $\sigma_{X} = 0$. However, since the main techniques do not substantially differ from the ones presented in this article, we refrain from discussing this type of results and focus on the more important case where $\sigma_{X} > 0$.\footnote{A discussion of results for $\sigma_{X} = 0$ in a slightly different context is provided in \cite{fmv19}.}
\end{itemize}
$\mbox{}$ \hspace{44.8em} \scalebox{0.75}{$\blacklozenge$} \\
\subsection{Maturity-Randomization and OIDEs}
\label{SECmaRaHYPER}
\noindent We now go back to the OIDE characterizations of Proposition~\ref{prop4}, Proposition~\ref{prop5}, and Proposition~\ref{prop6}, and consider the respective problems (\ref{MRGSCEuOIDE1})-(\ref{MRGSCEuOIDE2}), (\ref{MRGSCAmerOIDE1})-(\ref{MRGSCAmerOIDE2}), and (\ref{GSCAmeEEPOIDE1})-(\ref{GSCAmeEEPOIDE2-2}) for (regular) geometric down-and-out step call options under hyper-exponential jump-diffusion markets with $\sigma_{X} > 0$. First, we note that the infinitesimal generator (\ref{AINFA}) simplifies in this case to
\begin{equation}
\mathcal{A}_{S} V(\mathcal{T},x)  = \frac{1}{2} \sigma^{2}_{X} x^{2} \partial_{x}^{2} V(\mathcal{T},x) +(r-\delta-\lambda \zeta) x \partial_{x} V(\mathcal{T},x) +  \lambda \int \limits_{\mathbb{R}}  \big( V(\mathcal{T},xe^{y}) - V(\mathcal{T},x) \big) f_{J_{1}}(y)dy.
\label{HyperINFIG}
\end{equation}
\noindent Together with the properties of the hyper-exponential density $f_{J_{1}}(\cdot)$, this allows us to uniquely solve the problems (\ref{MRGSCEuOIDE1})-(\ref{MRGSCEuOIDE2}), (\ref{MRGSCAmerOIDE1})-(\ref{MRGSCAmerOIDE2}), and (\ref{GSCAmeEEPOIDE1})-(\ref{GSCAmeEEPOIDE2-2}), and to derive closed-form expressions for the (regular) maturity-randomized geometric down-and-out step contracts $\widehat{\mathcal{DOSC}^{\star}_{E}}(\cdot)$, $\widehat{\mathcal{DOSC}^{\star}_{A}}(\cdot)$, and corresponding early exercise premiums $\widehat{\mathcal{E}_{\mathcal{DOSC}}^{\star}}(\cdot)$, $\widehat{\mathcal{E}_{\mathcal{DOSC}}^{0,\star}}(\cdot)$, and $\widehat{\mathcal{E}_{\mathcal{DOSC}}^{\mathcal{J},\star}}(\cdot)$. This is discussed next.\vspace{1em} \\
\noindent We start by dealing with the maturity-randomized European-type contract $\widehat{\mathcal{DOSC}^{\star}_{E}}(\cdot)$. Here, upon imposing a natural smooth-fit condition (cf.~among others \cite{ccw10}, \cite{xy13}, \cite{lz16}), the following characterization of the (regular) maturity-randomized European-type geometric down-and-out step call option $\widehat{\mathcal{DOSC}^{\star}_{E}}(\cdot)$ can be obtained. A proof is provided in Appendix B.  
\begin{Prop}
\label{PropEurHEJD}
\noindent Consider a hyper-exponential jump-diffusion market as described by (\ref{market1}), (\ref{market2}), and (\ref{DYNhypexp}), (\ref{hypexpDens}). Then, for any intensity parameter $\vartheta >0$, the (regular) maturity-randomized European-type geometric down-and-out step call, $\widehat{\mathcal{DOSC}^{\star}_{E}}(\cdot)$, has the following representation
\begin{equation}
\widehat{\mathcal{DOSC}^{\star}_{E}}(\vartheta,x; K, L, \rho_{L}) = \left \{ \begin{array}{lc}
\sum \limits_{s=1}^{m+1} A_{s}^{+} \Big(\frac{x}{L} \Big)^{\beta_{s,(r+\vartheta-\rho_{L})}}, &  0 \leq x < L,\\
\sum \limits_{s=1}^{m+1} B_{s}^{+} \Big(\frac{x}{L} \Big)^{\beta_{s,(r+\vartheta)}} + \sum \limits_{u=1}^{n+1} B_{u}^{-} \Big(\frac{x}{K} \Big)^{\gamma_{u,(r+\vartheta)}}, & L \leq x \leq K, \\
 \sum \limits_{u=1}^{n+1} C_{u}^{-} \Big(\frac{x}{K} \Big)^{\gamma_{u,(r+\vartheta)}} + \vartheta \left( \frac{x}{\delta + \vartheta } -\frac{K}{r + \vartheta} \right) , & K < x < \infty ,
\end{array} \right.
\end{equation}
\noindent where the vector of coefficients $\mathbf{v} := (A_{1}^{+}, \ldots, A_{m+1}^{+},B_{1}^{+}, \ldots, B_{m+1}^{+}, B_{1}^{-}, \ldots, B_{n+1}^{-}, C_{1}^{-}, \ldots, C_{n+1}^{-})^{\intercal}$ solves the system of equations given in (\ref{AppendixBSysEq}) of Appendix B.
\end{Prop}
\noindent We next derive (semi-)analytical results for the (regular) maturity-randomized American-type geometric down-and-out step call contract $\widehat{\mathcal{DOSC}^{\star}_{A}}(\cdot)$. Having already obtained a closed-form expression for the European-type option $\widehat{\mathcal{DOSC}^{\star}_{E}}(\cdot)$, we can now focus on the maturity-randomized early exercise pricing problem instead. Indeed, although a direct application of the techniques developed in the proof of Proposition~\ref{PropEurHEJD} to $\widehat{\mathcal{DOSC}^{\star}_{A}}(\cdot)$ is equally feasible, switching to the maturity-randomized early exercise pricing problem substantially reduces the complexity of the resulting equations. We therefore follow this approach and decompose the American-type contract $\widehat{\mathcal{DOSC}^{\star}_{A}}(\cdot)$ as sum of the European-type option $\widehat{\mathcal{DOSC}^{\star}_{E}}(\cdot)$ and the early exercise premium $\widehat{\mathcal{E}_{\mathcal{DOSC}}^{\star}}(\cdot)$. Additionally, since we have seen in Section~\ref{SEC2} that the stopping region of a (maturity-randomized) geometric knock-out option is a sub-domain of the stopping region for the respective (maturity-randomized) barrier-type knock-out option, we can follow the ansatz in \cite{xy13} (cf.~\cite{lv17}, \cite{cv18}) and conjecture that the early-exercise region is delimited by a free-boundary $\mathfrak{b}_{s} > K$, whose value has to be found. Combining these observations, we therefore arrive at the next proposition, whose proof is provided in Appendix B.
\begin{Prop}
\label{PropAmerHEJD}
\noindent Consider a hyper-exponential jump-diffusion market as described by (\ref{market1}), (\ref{market2}), and (\ref{DYNhypexp}), (\ref{hypexpDens}). Then, for any intensity parameter $\vartheta >0$, the (regular) maturity-randomized American-type geometric down-and-out step call option, $\widehat{\mathcal{DOSC}^{\star}_{A}}(\cdot)$, is given by
\begin{equation}
\widehat{\mathcal{DOSC}^{\star}_{A}}(\vartheta,x; K, L, \rho_{L}) = \widehat{\mathcal{DOSC}^{\star}_{E}}(\vartheta,x; K, L, \rho_{L}) + \widehat{\mathcal{E}_{\mathcal{DOSC}}^{\star}}(\vartheta,x; K, L, \rho_{L}) ,
\end{equation}
\noindent where the maturity-randomized early exercise premium to the (regular) geometric down-and-out step call, $\widehat{\mathcal{E}_{\mathcal{DOSC}}^{\star}}(\cdot)$, has the following representation:
\begin{equation}
\widehat{\mathcal{E}_{\mathcal{DOSC}}^{\star}}(\vartheta,x; K, L, \rho_{L})  = \left \{ \begin{array}{lc}
\sum \limits_{s=1}^{m+1} D_{s}^{+} \Big(\frac{x}{L} \Big)^{\beta_{s,(r+\vartheta-\rho_{L})}}, &  0 \leq x < L,\\
\sum \limits_{s=1}^{m+1} F_{s}^{+} \Big(\frac{x}{L} \Big)^{\beta_{s,(r+\vartheta)}} + \sum \limits_{u=1}^{n+1} F_{u}^{-} \Big(\frac{x}{\mathfrak{b}_{s}} \Big)^{\gamma_{u,(r+\vartheta)}}, & L \leq x < \mathfrak{b}_{s}, \\
 x-K-\widehat{\mathcal{DOSC}^{\star}_{E}}(\vartheta,x; K, L, \rho_{L}) , & \mathfrak{b}_{s} \leq x < \infty .
\end{array} \right.
\end{equation}
\noindent Here, the vector of coefficients $\mathbf{w} := (D_{1}^{+}, \ldots, D_{m+1}^{+},F_{1}^{+}, \ldots, F_{m+1}^{+}, F_{1}^{-}, \ldots, F_{n+1}^{-})^{\intercal}$ solves the system of equations given in (\ref{AppendixBAmerSysEq}) of Appendix B and the early exercise boundary $\mathfrak{b}_{s}$ is implicitly given by combining (\ref{AppendixBAmerSysEq}) with Equation (\ref{SPuseful}).
\end{Prop}
\noindent To complete our derivations, we lastly generalize the results obtained in \cite{lv17} to American-type geometric step contracts and provide a jump-diffusion disentanglement of the maturity-randomized early exercise premium to the (regular) geometric down-an-out step call. Here, combining our results in Proposition~\ref{prop6} with ideas similarly employed in~\cite{lv17}, \cite{cv18}, and \cite{fmv19}, allows us to derive (semi-)analytical expressions for $\widehat{\mathcal{E}_{\mathcal{DOSC}}^{0,\star}}(\cdot)$ and $\widehat{\mathcal{E}_{\mathcal{DOSC}}^{\mathcal{J},\star}}(\cdot)$, the maturity-randomized early exercise contribution of the diffusion and jump parts to the geometric down-and-out step call option. This leads to our final proposition, whose proof is provided in Appendix B. 
\begin{Prop}
\label{PropEepHEJD}
\noindent Consider a hyper-exponential jump-diffusion market as described by (\ref{market1}), (\ref{market2}), and (\ref{DYNhypexp}), (\ref{hypexpDens}). Then, for any intensity parameter $\vartheta >0$, the maturity-randomized early exercise premium to the (regular) geometric down-and-out step call, $\widehat{\mathcal{E}_{\mathcal{DOSC}}^{\star}}(\cdot)$, has the following decomposition
\begin{equation}
\widehat{\mathcal{E}_{\mathcal{DOSC}}^{\star}}(\vartheta,x; K, L, \rho_{L}) = \widehat{\mathcal{E}_{\mathcal{DOSC}}^{0,\star}}(\vartheta,x; K, L, \rho_{L}) + \widehat{\mathcal{E}_{\mathcal{DOSC}}^{\mathcal{J},\star}}(\vartheta,x; K, L, \rho_{L}).
\end{equation}
\noindent Here, the premiums $\widehat{\mathcal{E}_{\mathcal{DOSC}}^{0,\star}}(\cdot)$ and $\widehat{\mathcal{E}_{\mathcal{DOSC}}^{\mathcal{J},\star}}(\cdot)$ refer to the maturity-randomized early exercise contributions of the diffusion and jump parts, respectively, and are given by
\begin{equation}
\widehat{\mathcal{E}_{\mathcal{DOSC}}^{0,\star}}(\vartheta,x; K, L, \rho_{L})  = \left \{ \begin{array}{lc}
\sum \limits_{s=1}^{m+1} D_{s}^{0,+} \Big(\frac{x}{L} \Big)^{\beta_{s,(r+\vartheta-\rho_{L})}}, &  0 \leq x < L,\\
\sum \limits_{s=1}^{m+1} F_{s}^{0,+} \Big(\frac{x}{L} \Big)^{\beta_{s,(r+\vartheta)}} + \sum \limits_{u=1}^{n+1} F_{u}^{0-} \Big(\frac{x}{\mathfrak{b}_{s}} \Big)^{\gamma_{u,(r+\vartheta)}}, & L \leq x < \mathfrak{b}_{s}, \\
 x-K-\widehat{\mathcal{DOSC}^{\star}_{E}}(\vartheta,x; K, L, \rho_{L}) , & x = \mathfrak{b}_{s}, \\
0, & \mathfrak{b}_{s} < x < \infty,
\end{array} \right. 
\label{DiffEEP_MR1}
\end{equation}
\begin{equation}
\widehat{\mathcal{E}_{\mathcal{DOSC}}^{\mathcal{J},\star}}(\vartheta,x; K, L, \rho_{L})  = \left \{ \begin{array}{lc}
\sum \limits_{s=1}^{m+1} D_{s}^{\mathcal{J},+} \Big(\frac{x}{L} \Big)^{\beta_{s,(r+\vartheta-\rho_{L})}}, &  0 \leq x < L,\\
\sum \limits_{s=1}^{m+1} F_{s}^{\mathcal{J},+} \Big(\frac{x}{L} \Big)^{\beta_{s,(r+\vartheta)}} + \sum \limits_{u=1}^{n+1} F_{u}^{\mathcal{J},-} \Big(\frac{x}{\mathfrak{b}_{s}} \Big)^{\gamma_{u,(r+\vartheta)}}, & L \leq x < \mathfrak{b}_{s}, \\
0, & x = \mathfrak{b}_{s}, \\
 x-K-\widehat{\mathcal{DOSC}^{\star}_{E}}(\vartheta,x; K, L, \rho_{L}) , & \mathfrak{b}_{s} < x < \infty ,
\end{array} \right.
\label{JumpEEP_MR1}
\end{equation}
\noindent where the two vectors of coefficients $\mathbf{w_{0}} := (D_{1}^{0,+}, \ldots, D_{m+1}^{0,+},F_{1}^{0,+}, \ldots, F_{m+1}^{0,+}, F_{1}^{0,-}, \ldots, F_{n+1}^{0,-})^{\intercal}$ and $\mathbf{w_{J}} := (D_{1}^{\mathcal{J},+}, \ldots, D_{m+1}^{\mathcal{J},+},F_{1}^{\mathcal{J},+}, \ldots, F_{m+1}^{\mathcal{J},+}, F_{1}^{\mathcal{J},-}, \ldots, F_{n+1}^{\mathcal{J},-})^{\intercal}$ solve the systems of equations given in (\ref{AppendixBEepSysEq}).
\end{Prop}

\section{Numerical Results}
\label{MR_NUMRES}
\noindent To complement the theoretical results of Section \ref{SEC2} and Section \ref{SEC3}, we lastly illustrate structural and numerical properties of (regular) geometric down-and-out step call options under hyper-exponential jump-diffusion markets. For simplicity of the exposition as well as to allow for a better comparability of our results with the existing literature, we rely on Kou's double-exponential jump-diffusion model (cf.~\cite{ko02}) as class representative and combine a variety of parameters that were similarly used in the following related articles:~\cite{li99}, \cite{kw04}, \cite{cc09}, \cite{ccw10}, \cite{ck12}, \cite{lz16}, \cite{lv17}, \cite{cv18}, and \cite{dlm19}. All our numerical results are obtained using Matlab R2017b on an Intel CORE i7 processor.
\begin{center}
\captionof{table}{Theoretical (down-and-out) call values and diffusion contributions to the early exercise premium for $r=0.05$, $\delta= 0.07$, $S_{0}=100$, $K=100$, $L=95$, $\rho_{L}=-26.34$, $p=0.7$, $\xi=25$ and $\eta = 50$.}
\label{table1STEP}
\scalebox{0.764}{
\begin{tabular}{lrrrrrrrrrrrrr}  
\toprule
\multicolumn{11}{c}{\bf (Down-and-Out) Call Option Prices} \\
%\cmidrule{1-12}
\bottomrule
 \multicolumn{2}{c}{\it Parameters}   & \multicolumn{3}{c}{\it Standard Call Price} & \multicolumn{3}{c}{\it Step Call Price} & \multicolumn{3}{c}{\it Barrier Call Price} \\
\cmidrule(r){1-2} \cmidrule(r){3-5} \cmidrule(r){6-8} \cmidrule(l){9-11}
  &     $\lambda$    &  \it Euro  &  \it Amer  &   \it DC (\%)     &    \it Euro     &  \it Amer  &  \it DC (\%)  & \it Euro &  \it Amer   &  \it DC (\%)   \\
	\midrule
	\midrule
  & $1$ & $6.833$ & $7.040$  & $91.52 \%$ & $4.596$ & $4.789$  & $91.71 \% $ & $3.374$ & $3.551$ & $91.88 \%$ \\
 $S_{0} = 100$ & $0.1$ & $6.622$ & $6.822$  & $99.07 \%$ & $4.519$ & $4.706$  & $99.09 \%$ & $3.338$ & $3.514$ & $99.12 \%$ \\
 $\sigma_{X}=0.2$ & $0.01$ & $6.600$ & $6.800$  & $99.91 \% $ & $4.511$ & $4.698$  & $99.91 \%$ & $3.334$ & $3.510$  & $99.91 \%$ \\
 $\mathcal{T} = 1.0$ & $0.001$ & $6.598$ & $6.797$  & $99.99 \%$ & $4.510$ & $4.697$  & $99.99 \%$ & $3.333$ & $3.509$  & $99.99 \%$ \\
 & $0.0001$ & $6.598$ & $6.797$  & $100.00 \%$ & $4.510$ & $4.697$  & $100.00\% $ & $3.333$ & $3.509$ & $100.00 \% $ \\
\bottomrule
{\bf B\&S Values} & {\bf --} & {\bf 6.698} & {\bf 6.885}  & {\bf --} & {\bf 4.511} & {\bf 4.745} & {\bf --} & {\bf 3.332} & {\bf 3.529} & {\bf --} \\	
\mbox{\bf Rel. Error (\%)} & {\bf --} & {\bf 0.001\%} & {\bf -1.277\%}  & {\bf --} & {\bf 0.015\%} & {\bf -1.025\%} & {\bf --} & {\bf 0.025\%} & {\bf -0.568\%} & {\bf --} \\	
\bottomrule
\end{tabular}
}
\end{center}
$\mbox{}$ \vspace{-0.8em} \\
\subsection{Geometric Step Options and Limiting Contracts}
\noindent We start our illustrations by investigating the convergence of geometric knock-out step call options to their limiting contracts. As already pointed out in Section~\ref{SEC2}, standard and (standard) barrier-type options can be understood as extremities on a continuum of geometric double barrier knock-out step contracts, namely when the knock-out rates are chosen as $\bm{\rho}_{\bm{\ell}} = (0,0)$ and $\bm{\rho}_{\bm{\ell}}=(-\infty, -\infty)$, respectively. Furthermore, since hyper-exponential jump-diffusion markets reduce to the Black \& Scholes market (cf.~\cite{bs73}) when the jump intensity $\lambda $ is zero, our results should be consistent in the limit $\lambda \downarrow 0$ with those obtained e.g.~in \cite{li99} and \cite{dlm19}. We verify these results in Table~\ref{table1STEP}, where we compare the value of (regular) European-type and American-type geometric down-and-out step call options for $\rho_{L}=0$ (``Standard Call Price''), $\rho_{L} = -26.34$ (``Step Call Price''), and $\rho_{L} = -50'000'000$ (``Barrier Call Price'') with the respective Black~\&~Scholes values (``B\&S Values'').\footnote{We compute the value of the American-type contracts under the Black \& Scholes model using the algorithm in \cite{dlm19} as well as Ritchken's trinomial tree method with $5'000$ time steps.}~As in these papers (cf.~also \cite{lz16}), we take $\mathcal{T}=1.0$, $\sigma_{X} = 0.2$, $r=0.05$, $\delta=0.07$, $S_{0}=100$, $K=100$, $L=95$, and $\rho_{L} = -26.34$. Furthermore, we align the parameters of the double-exponential distribution to frequent choices in the literature and fix the probability of an up-jump with $p=0.7$ (cf.~\cite{lv17}) and positive and negative jump parameters with $\xi=25$ and $\eta=50$, respectively (cf.~\cite{kw04},\cite{ck12}, \cite{lv17}, \cite{cv18}). Finally, as in \cite{lz16} the convergence to the Black \& Scholes values is investigated via $\lambda \in \{1, 0.1, 0.01, 0.001, 0.0001 \}$. \vspace{1em} \\
%%%%%%%%%%%%%%%
%%%%%%%%%%%%%%%
%%%%%%%%%%%%%%%
\noindent As expected, the results in Table~\ref{table1STEP} show that standard options, geometric step options, and (standard) barrier-type options under the Black \& Scholes market can be recovered by means of their respective contracts under double-exponential jump-diffusion markets as $\lambda \downarrow 0$. Furthermore, our results confirm the convergence of geometric down-and-out step call options to barrier-type down-and-out call contracts as $\rho_{L} \downarrow -\infty$. This becomes evident when looking at the ``Barrier Call Price'' of Table~\ref{table1STEP} while recalling that the Black \& Scholes value is a true barrier-type value that was obtained using Ritchken's trinomial tree method and that the converging values correspond to those of geometric down-and-out step call options with $\rho_{L} = -50'000'000$. Finally, we note that our results are in line with the observations in \cite{lv17}, where the pricing accuracy of the Gaver-Stehfest inversion algorithm for European-type options was very high\footnote{In this article, the relative pricing errors of the Gaver-Stehfest inversion algorithm for European-type contracts never exceeded $\pm 0.22 \%$.}~and the relative pricing errors of the same inversion method applied to American-type options instead ranged from roughly $\pm 0.33 \%$ to $\pm 1.38 \%$. As explained in Remark~3, this is mainly due to the fact that maturity-randomized American-type contracts as well as maturity-randomized early exercise premiums do not anymore coincide with a strict application of the Laplace-Carson transform but are regardless treated as such.
\begin{center}
\captionof{table}{Theoretical (down-and-out) call values and structure of the early exercise premium for $r=0.05$, $\delta= 0.07$, $K=100$, $L=95$, $\rho_{L} = -26.34$, $p=0.5$, $\xi=50$ and $\eta = 25$.}
\label{tableSTEP:HEJD1}
\scalebox{0.764}{
\begin{tabular}{lrrrrrrrrrrrrr}  
\toprule
\multicolumn{14}{c}{\bf (Down-and-Out) Call Option Prices} \\
%\cmidrule{1-12}
\bottomrule
 \multicolumn{2}{c}{\it Parameters}   & \multicolumn{4}{c}{\it Standard Call Price} & \multicolumn{4}{c}{\it Step Call Price} & \multicolumn{4}{c}{\it Barrier Call Price} \\
\cmidrule(r){1-2} \cmidrule(r){3-6} \cmidrule(r){7-10} \cmidrule(l){11-14}
  &     $S_{0}$    &  \it Euro  &  \it EEP  &  \it EEP (\%)      &  \it DC (\%)     &    \it Euro     &  \it EEP   & \it EEP (\%) &  \it DC (\%)  & \it Euro &  \it EEP   &  \it EEP (\%) &  \it DC (\%)   \\
	\midrule
	\midrule
 & $90$ & $3.500$ & $0.062$ & $1.74 \%$ & $94.20 \%$ & $0.268$ & $0.009$ & $3.07 \% $ & $94.32 \%$ & $0$ & $0$ & {\bf --} & {\bf --}  \\
 (1) & $95$ & $5.241$ & $0.112$ & $2.09 \%$ & $94.27 \%$ & $1.757$ & $0.059$ & $3.23 \%$ & $94.33 \% $ & $0$ & $0$ & {\bf --} & {\bf --} \\
$\sigma_{X}=0.2$ & $100$ & $7.416$ & $0.190$ & $2.50 \%$ & $94.34 \%$ & $4.992$ & $0.178$ & $3.45 \%$ & $94.36 \%$ & $3.686$ & $0.165$ & $4.28 \%$ & $94.37 \%$ \\
 $\lambda = 5.0$ & $105$ & $10.011$ & $0.305$ & $2.96 \% $ & $94.40 \% $ & $8.309$ & $0.330$ & $3.82 \%$ & $94.39 \%$ & $7.305$ & $0.353$ & $4.61 \%$ & $94.40 \%$ \\
 $\mathcal{T} = 1.0$ & $110$ & $12.992$ & $0.469$ & $3.48 \% $ & $94.46 \%$ & $11.804$ & $0.535$ & $4.34 \%$ & $94.44 \%$ & $11.037$ & $0.597$ & $5.13 \%$ & $94.44 \%$ \\
 & $115$ & $16.314$ & $0.691$ & $4.07 \% $ & $94.52 \%$ & $15.492$ & $0.811$ & $4.98 \%$ & $94.50 \% $ & $14.914$ & $0.920$ & $5.81 \%$ & $94.54 \% $ \\
\midrule
\midrule
 & $90$ & $4.098$ & $0.065$ & $1.57 \%$ & $89.68 \%$ & $0.344$ & $0.010$ & $2.79 \% $ & $89.87 \%$ & $0$ & $0$ & {\bf --} & {\bf --}  \\
 (2) & $95$ & $5.933$ & $0.113$ & $1.87 \%$ & $89.80 \%$ & $2.012$ & $0.061$ & $2.93 \%$ & $89.89 \% $ & $0$ & $0$ & {\bf --} & {\bf --} \\
$\sigma_{X}=0.2$ & $100$ & $8.169$ & $0.186$ & $2.22 \%$ & $89.90 \%$ & $5.413$ & $0.175$ & $3.12 \%$ & $89.93 \%$ & $3.990$ & $0.161$ & $3.88 \%$ & $89.95 \%$ \\
 $\lambda = 10.0$ & $105$ & $10.791$ & $0.290$ & $2.62 \% $ & $90.00 \% $ & $8.791$ & $0.314$ & $3.44 \%$ & $89.99 \%$ & $7.683$ & $0.334$ & $4.17 \%$ & $89.99 \%$ \\
 $\mathcal{T} = 1.0$ & $110$ & $13.767$ & $0.435$ & $3.06 \% $ & $90.08 \%$ & $12.313$ & $0.497$ & $3.88 \%$ & $90.05 \%$ & $11.442$ & $0.552$ & $4.60 \%$ & $90.05 \%$ \\
 & $115$ & $17.056$ & $0.628$ & $3.55 \% $ & $90.17 \%$ & $16.004$ & $0.738$ & $4.41 \%$ & $90.12 \% $ & $15.325$ & $0.835$ & $5.16 \%$ & $90.13 \% $ \\
\bottomrule
\end{tabular}
}
\end{center}
$\mbox{}$ \vspace{-0.8em} \\
\subsection{Early Exercise Structure of Geometric Step Options with Jumps}
\noindent Having verified the convergence of geometric step options to their limiting contracts, we next investigate the early exercise structure of (regular) geometric down-and-out step call options. To this end, we start by computing absolute European-type values (``Euro''), absolute early exercise premiums (``EEP''), relative early exercise contributions\footnote{The relative early exercise contribution is expressed as percentage of the American-type geometric step option price.}~(``EEP\%''), and diffusion contributions to the early exercise premium (``DC\%'') for standard call options (``Standard Call Price''), (regular) geometric down-and-out step call options (``Step Call Price'') and (regular) pseudo barrier-type down-and-out call options (``Barrier Call Price'').\footnote{As earlier, we rely on results for geometric down-and-out step call contracts with $\rho_{L} = -50'000'000$ to derive pseudo barrier-type down-and-out call option values.}~Here, we combine again the parameter choices in \cite{li99} and \cite{dlm19} with frequent jump specifications in the literature. More specifically, we choose $\mathcal{T}=1.0$, $\sigma_{X} = 0.2$, $r=0.05$, $\delta = 0.07$, $S_{0} \in \{90,95,100,105,110,115 \}$, $K=100$, $L = 95$, $\rho_{L}=-26.24$ and fix the intensity measure $\Pi_{X}$ in (\ref{INTENSmeasure}) by taking $\lambda \in \{5,10 \}$ (cf.~\cite{lv17}, \cite{cv18}), $p=0.5$ (cf.~\cite{cc09}, \cite{ccw10}, \cite{lz16}, \cite{cv18}), and $(\xi,\eta) \in \{(50,25), (50,50), (25,50), (25,25) \}$ (cf.~\cite{kw04}, \cite{ck12}, \cite{lv17},\cite{cv18}). The results are presented in Tables~\ref{tableSTEP:HEJD1}-\ref{tableSTEP:HEJD4}.

\begin{figure}[!t]
\begin{subfigure}{.5\linewidth}
\centering
\includegraphics[scale=0.16]{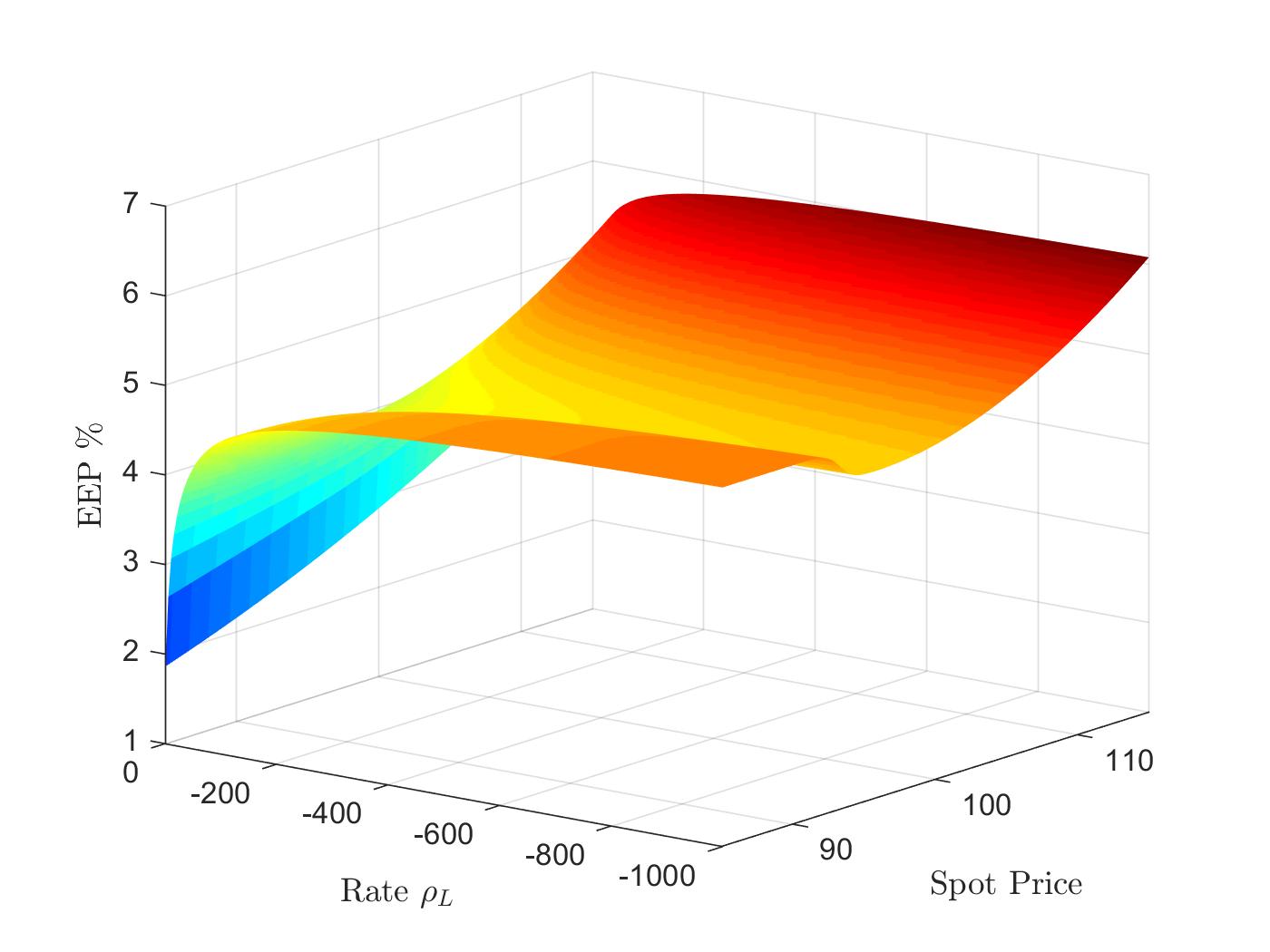}
\caption{Relative Early Exercise Contribution.}
\label{HEJD:FIG1:sub11}
\end{subfigure}%
\begin{subfigure}{.5\linewidth}
\centering
\includegraphics[scale=.16]{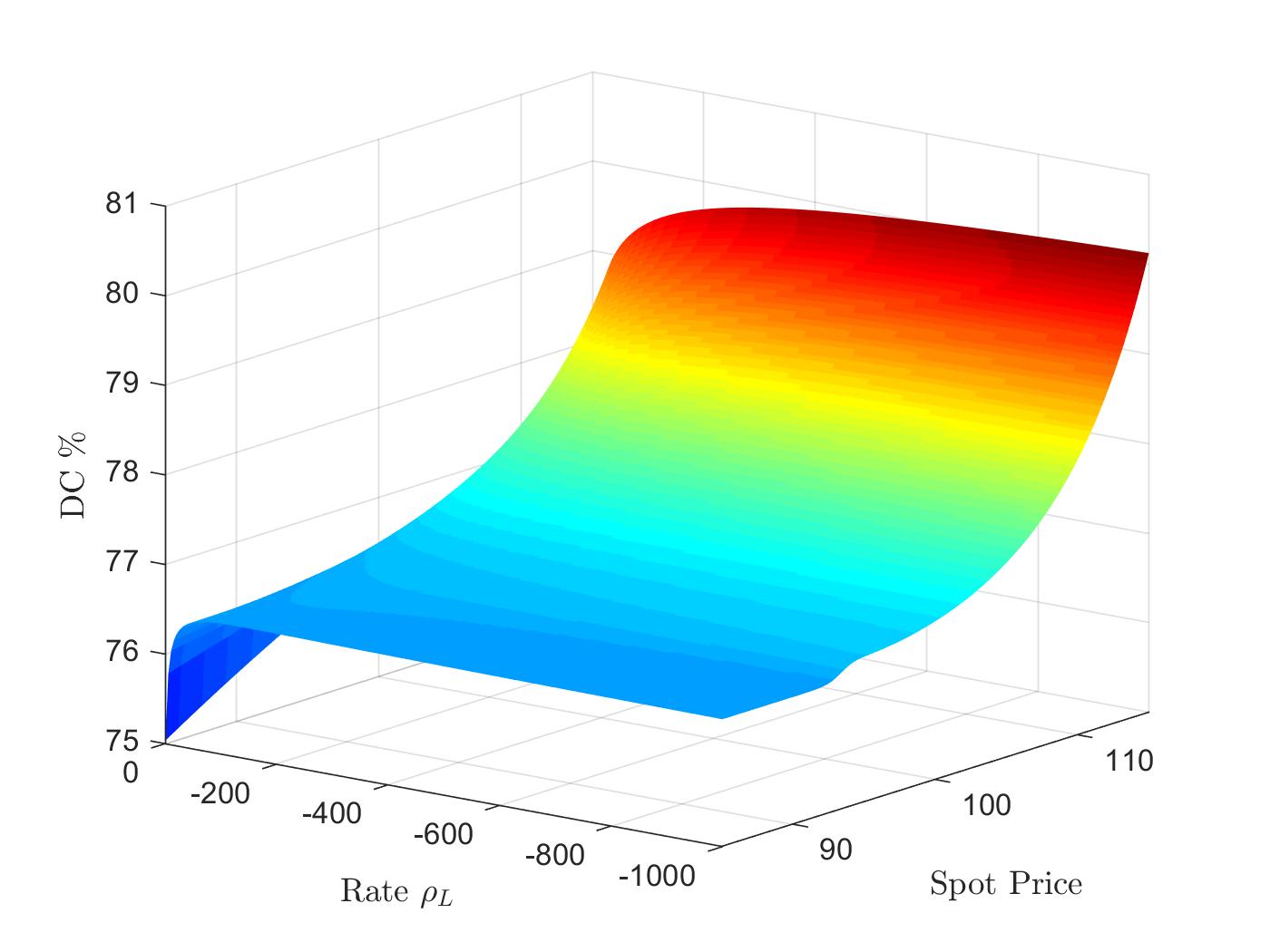}
\caption{Diffusion Contribution.}
\label{HEJD:FIG1:sub12}
\end{subfigure}\\[1ex]
\begin{subfigure}{\linewidth}
\centering
\includegraphics[scale=.16]{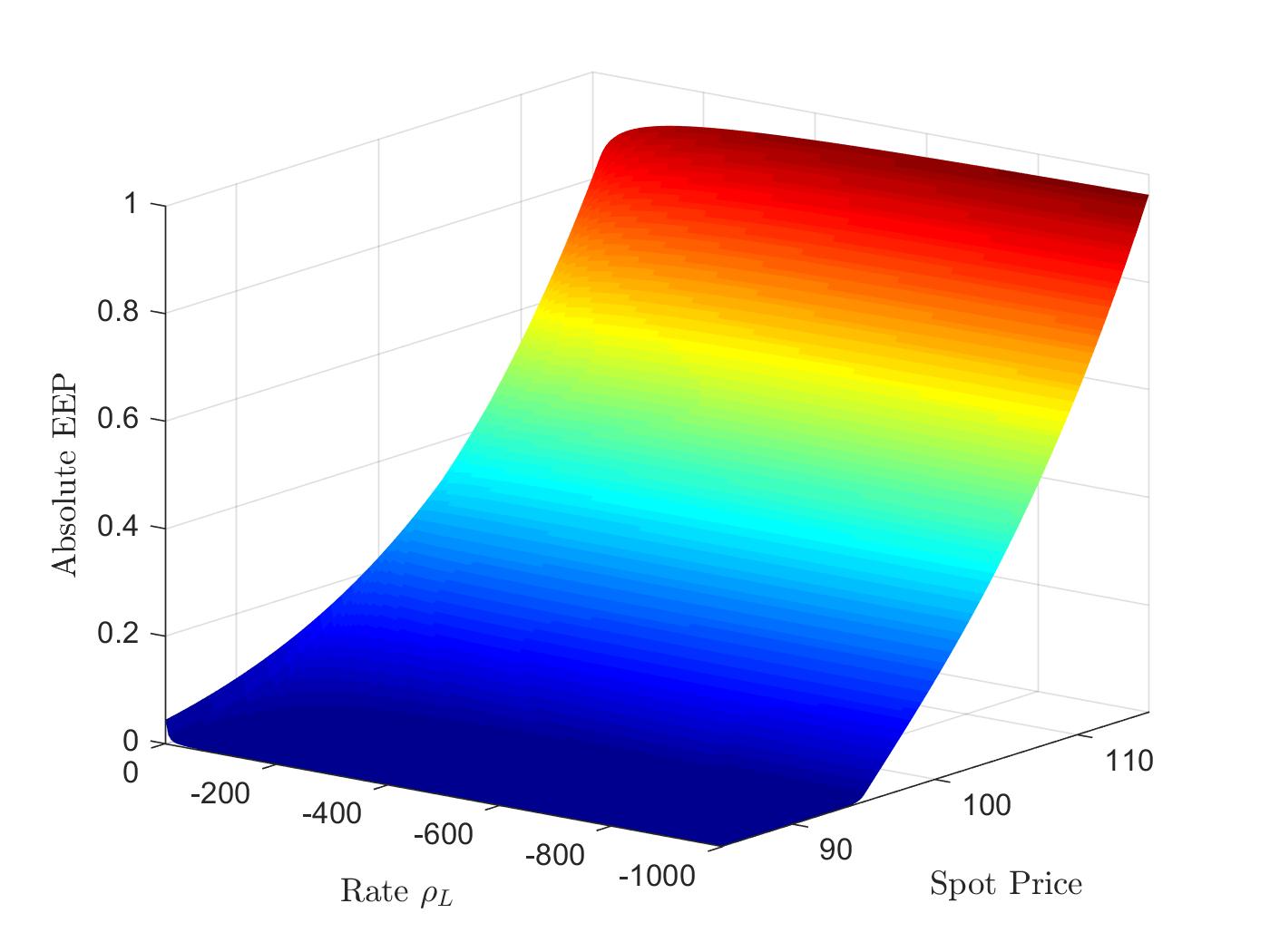}
\caption{Absolute Early Exercise Premium.}
\label{HEJD:FIG1:sub13}
\end{subfigure}
\caption{Relative early exercise contribution, diffusion contribution to the early exercise premium, and absolute early exercise premium of the geometric down-and-out step call as functions of the underlying price $S_{0} \in [85,115]$ and the knock-out rate $\rho_{L} \in [-1000,0]$, when the remaining parameters are chosen as: $\mathcal{T} = 1.0$, $\sigma_{X} =0.2$, $r=0.05$, $\delta =0.07$, $K=100$, $L =95$, $\lambda = 5$, $p=0.5$, $\xi = 25$, $\eta = 50$.}
\label{HEJD:FIG1}
\end{figure}

\begin{center}
\captionof{table}{Theoretical (down-and-out) call values and structure of the early exercise premium for $r=0.05$, $\delta= 0.07$, $K=100$, $L=95$, $\rho_{L}=-26.34$, $p=0.5$, $\xi=50$ and $\eta = 50$.}
\label{tableSTEP:HEJD2}
\scalebox{0.764}{
\begin{tabular}{lrrrrrrrrrrrrr}  
\toprule
\multicolumn{14}{c}{\bf (Down-and-Out) Call Option Prices} \\
%\cmidrule{1-12}
\bottomrule
 \multicolumn{2}{c}{\it Parameters}   & \multicolumn{4}{c}{\it Standard Call Price} & \multicolumn{4}{c}{\it Step Call Price} & \multicolumn{4}{c}{\it Barrier Call Price} \\
\cmidrule(r){1-2} \cmidrule(r){3-6} \cmidrule(r){7-10} \cmidrule(l){11-14}
  &     $S_{0}$    &  \it Euro  &  \it EEP  &  \it EEP (\%)      &  \it DC (\%)     &    \it Euro     &  \it EEP   & \it EEP (\%) &  \it DC (\%)  & \it Euro &  \it EEP   &  \it EEP (\%) &  \it DC (\%)   \\
	\midrule
	\midrule
 & $90$ & $3.163$ & $0.064$ & $1.98 \%$ & $93.97 \%$ & $0.232$ & $0.008$ & $3.46 \% $ & $94.10 \%$ & $0$ & $0$ & {\bf --} & {\bf --}  \\
 (1) & $95$ & $4.835$ & $0.117$ & $2.37 \%$ & $94.05 \%$ & $1.588$ & $0.060$ & $3.63 \%$ & $94.12 \% $ & $0$ & $0$ & {\bf --} & {\bf --} \\
$\sigma_{X}=0.2$ & $100$ & $6.958$ & $0.202$ & $2.82 \%$ & $94.12 \%$ & $4.679$ & $0.188$ & $3.87 \%$ & $94.14 \%$ & $3.432$ & $0.174$ & $4.82\%$ & $94.15 \%$ \\
 $\lambda = 5.0$ & $105$ & $9.523$ & $0.328$ & $3.33 \% $ & $94.19 \% $ & $7.949$ & $0.355$ & $4.28 \%$ & $94.18 \%$ & $6.983$ & $0.382$ & $5.18 \%$ & $94.18 \%$ \\
 $\mathcal{T} = 1.0$ & $110$ & $12.498$ & $0.509$ & $3.91 \% $ & $94.25 \%$ & $11.430$ & $0.583$ & $4.85 \%$ & $94.23 \%$ & $10.702$ & $0.654$ & $5.76 \%$ & $94.24 \%$ \\
 & $115$ & $15.835$ & $0.758$ & $4.57 \% $ & $94.33 \%$ & $15.122$ & $0.891$ & $5.57 \%$ & $94.31 \% $ & $14.586$ & $1.017$ & $6.52 \%$ & $94.43 \% $ \\
\midrule
\midrule
 & $90$ & $3.441$ & $0.068$ & $1.94 \%$ & $88.95 \%$ & $0.268$ & $0.009$ & $3.40 \% $ & $89.16 \%$ & $0$ & $0$ & {\bf --} & {\bf --}  \\
 (2) & $95$ & $5.155$ & $0.121$ & $2.30 \%$ & $89.08 \%$ & $1.685$ & $0.062$ & $3.55 \%$ & $89.19 \% $ & $0$ & $0$ & {\bf --} & {\bf --} \\
$\sigma_{X}=0.2$ & $100$ & $7.303$ & $0.204$ & $2.72 \%$ & $89.20 \%$ & $4.836$ & $0.190$ & $3.77 \%$ & $89.23 \%$ & $3.522$ & $0.174$ & $4.71 \%$ & $89.25 \%$ \\
 $\lambda = 10.0$ & $105$ & $9.875$ & $0.325$ & $3.19 \% $ & $89.30 \% $ & $8.138$ & $0.352$ & $4.14 \%$ & $89.29 \%$ & $7.107$ & $0.377$ & $5.04 \%$ & $89.29 \%$ \\
 $\mathcal{T} = 1.0$ & $110$ & $12.839$ & $0.497$ & $3.72 \% $ & $89.40 \%$ & $11.636$ & $0.569$ & $4.66 \%$ & $89.36 \%$ & $10.845$ & $0.638$ & $5.56 \%$ & $89.37 \%$ \\
 & $115$ & $16.152$ & $0.729$ & $4.32 \% $ & $89.53 \%$ & $15.330$ & $0.859$ & $5.31 \%$ & $89.46 \% $ & $14.737$ & $0.981$ & $6.24 \%$ & $89.57 \% $ \\
\bottomrule
\end{tabular}
}
\end{center}
$\mbox{}$ \vspace{-0.8em} \\
\begin{center}
\captionof{table}{Theoretical (down-and-out) call values and structure of the early exercise premium for $r=0.05$, $\delta= 0.07$, $K=100$, $L=95$, $\rho_{L}= -26.34$, $p=0.5$, $\xi=25$ and $\eta = 50$.}
\label{tableSTEP:HEJD3}
\scalebox{0.764}{
\begin{tabular}{lrrrrrrrrrrrrr}  
\toprule
\multicolumn{14}{c}{\bf (Down-and-Out) Call Option Prices} \\
%\cmidrule{1-12}
\bottomrule
 \multicolumn{2}{c}{\it Parameters}   & \multicolumn{4}{c}{\it Standard Call Price} & \multicolumn{4}{c}{\it Step Call Price} & \multicolumn{4}{c}{\it Barrier Call Price} \\
\cmidrule(r){1-2} \cmidrule(r){3-6} \cmidrule(r){7-10} \cmidrule(l){11-14}
  &     $S_{0}$    &  \it Euro  &  \it EEP  &  \it EEP (\%)      &  \it DC (\%)     &    \it Euro     &  \it EEP   & \it EEP (\%) &  \it DC (\%)  & \it Euro &  \it EEP   &  \it EEP (\%) &  \it DC (\%)   \\
	\midrule
	\midrule
 & $90$ & $3.645$ & $0.080$ & $2.15 \%$ & $75.53 \%$ & $0.294$ & $0.012$ & $3.75 \% $ & $76.36 \%$ & $0$ & $0$ & {\bf --} & {\bf --}  \\
 (1) & $95$ & $5.362$ & $0.137$ & $2.49 \%$ & $75.97 \%$ & $1.685$ & $0.067$ & $3.82 \%$ & $76.45 \% $ & $0$ & $0$ & {\bf --} & {\bf --} \\
$\sigma_{X}=0.2$ & $100$ & $7.501$ & $0.222$ & $2.88 \%$ & $76.37 \%$ & $4.854$ & $0.202$ & $3.99 \%$ & $76.61 \%$ & $3.506$ & $0.182$ & $4.94\%$ & $76.81 \%$ \\
 $\lambda = 5.0$ & $105$ & $10.054$ & $0.345$ & $3.31 \% $ & $76.71 \% $ & $8.177$ & $0.368$ & $4.31 \%$ & $76.94 \%$ & $7.110$ & $0.391$ & $5.21 \%$ & $77.25 \%$ \\
 $\mathcal{T} = 1.0$ & $110$ & $12.994$ & $0.514$ & $3.80 \% $ & $76.98 \%$ & $11.685$ & $0.585$ & $4.77 \%$ & $77.55 \%$ & $10.861$ & $0.652$ & $5.67 \%$ & $78.24 \%$ \\
 & $115$ & $16.279$ & $0.740$ & $4.35 \% $ & $77.14 \%$ & $15.381$ & $0.870$ & $5.35 \%$ & $78.79 \% $ & $14.759$ & $0.989$ & $6.28 \%$ & $80.55 \% $ \\
\midrule
\midrule
 & $90$ & $4.347$ & $0.096$ & $2.16 \%$ & $62.58 \%$ & $0.391$ & $0.015$ & $3.78 \% $ & $63.44 \%$ & $0$ & $0$ & {\bf --} & {\bf --}  \\
 (2) & $95$ & $6.141$ & $0.155$ & $2.45 \%$ & $63.00 \%$ & $1.865$ & $0.074$ & $3.82 \%$ & $63.47 \% $ & $0$ & $0$ & {\bf --} & {\bf --} \\
$\sigma_{X}=0.2$ & $100$ & $8.321$ & $0.238$ & $2.78 \%$ & $63.38 \%$ & $5.152$ & $0.212$ & $3.94 \%$ & $63.56 \%$ & $3.649$ & $0.188$ & $4.89\%$ & $63.68 \%$ \\
 $\lambda = 5.0$ & $105$ & $10.878$ & $0.354$ & $3.15 \% $ & $63.72 \% $ & $8.549$ & $0.374$ & $4.19 \%$ & $63.76 \%$ & $7.328$ & $0.393$ & $5.09 \%$ & $63.90 \%$ \\
 $\mathcal{T} = 1.0$ & $110$ & $13.788$ & $0.508$ & $3.55 \% $ & $64.01 \%$ & $12.099$ & $0.577$ & $4.55 \%$ & $64.10 \%$ & $11.126$ & $0.640$ & $5.44 \%$ & $64.46 \%$ \\
 & $115$ & $17.019$ & $0.709$ & $4.00 \% $ & $64.18 \%$ & $15.809$ & $0.834$ & $5.01 \%$ & $64.74 \% $ & $15.047$ & $0.946$ & $5.91 \%$ & $65.93 \% $ \\
\bottomrule
\end{tabular}
}
\end{center}
$\mbox{}$ \vspace{1em} \\
\noindent The results in Tables~\ref{tableSTEP:HEJD1}-\ref{tableSTEP:HEJD4} show that the early exercise premium comprises a substantial part of the price of American-type geometric step contracts even if the option is out of the money. Additionally, they suggest that the absolute early exercise premium is for any rate $\rho_{L}$ increasing in the underlying price $S_{0}$ and that the relative early exercise contribution tends to increase with more severe (i.e.~more negative) knock-out rates. This is intuitively clear, since increasing the magnitude of the knock-out rate widens the early exercise domain of the American-type geometric step option and therefore further incentivizes early stopping. This subsequently raises the importance of the early exercise premium in the American-type geometric step option value and consequently increases its relative contribution. Next, we note that the diffusion contribution to the early exercise premium is a non-decreasing function of the underlying price $S_{0}$ and that this similarly seems to hold for the relative early exercise contribution. However, this last suggestion is wrong as can be seen in Figure~\ref{HEJD:FIG1:sub11} where we have plotted the relative early exercise contribution of the geometric down-and-out step call as a function of the underlying price $S_{0} \in [85,115]$ and the knock-out rate $\rho_{L} \in [-1000,0]$ using the following standard parameters: $\mathcal{T} = 1.0$, $\sigma_{X} =0.2$, $r=0.05$, $\delta =0.07$, $K=100$, $L =95$, $\lambda = 5$, $p=0.5$, $\xi = 25$, $\eta = 50$. As it turns out, the general behavior of the relative early exercise contribution depends on the location of the spot price relative to the barrier level $L$. In particular, while the relative early exercise contribution is increasing in the underlying price $S_{0}$ above the barrier $L=95$, it may be decreasing below the barrier for severe (i.e.~large negative) knock-out rates $\rho_{L}$. Nevertheless, we note that the results in Figure~\ref{HEJD:FIG1} also confirm many of the properties already discussed. In particular, the monotonicity of the relative early exercise premium as function of the knock-out rate is clearly documented here. Additionally, Figure~\ref{HEJD:FIG1:sub12} provides further evidence for the monotonicity of the diffusion contribution to the early exercise premium as function of the underlying price $S_{0}$, while Figure~\ref{HEJD:FIG1:sub13} confirms the monotonicity of the absolute early exercise premium as function of the underlying price $S_{0}$.
\begin{center}
\captionof{table}{Theoretical (down-and-out) call values and structure of the early exercise premium for $r=0.05$, $\delta= 0.07$, $K=100$, $L=95$, $\rho_{L}=-26.34$, $p=0.5$, $\xi=25$ and $\eta = 25$.}
\label{tableSTEP:HEJD4}
\scalebox{0.764}{
\begin{tabular}{lrrrrrrrrrrrrr}  
\toprule
\multicolumn{14}{c}{\bf (Down-and-Out) Call Option Prices} \\
%\cmidrule{1-12}
\bottomrule
 \multicolumn{2}{c}{\it Parameters}   & \multicolumn{4}{c}{\it Standard Call Price} & \multicolumn{4}{c}{\it Step Call Price} & \multicolumn{4}{c}{\it Barrier Call Price} \\
\cmidrule(r){1-2} \cmidrule(r){3-6} \cmidrule(r){7-10} \cmidrule(l){11-14}
  &     $S_{0}$    &  \it Euro  &  \it EEP  &  \it EEP (\%)      &  \it DC (\%)     &    \it Euro     &  \it EEP   & \it EEP (\%) &  \it DC (\%)  & \it Euro &  \it EEP   &  \it EEP (\%) &  \it DC (\%)   \\
	\midrule
	\midrule
 & $90$ & $3.966$ & $0.077$ & $1.91 \%$ & $76.61 \%$ & $0.330$ & $0.012$ & $3.37 \% $ & $77.40 \%$ & $0$ & $0$ & {\bf --} & {\bf --}  \\
 (1) & $95$ & $5.745$ & $0.131$ & $2.23 \%$ & $77.04 \%$ & $1.845$ & $0.066$ & $3.44 \%$ & $77.48 \% $ & $0$ & $0$ & {\bf --} & {\bf --} \\
$\sigma_{X}=0.2$ & $100$ & $7.931$ & $0.210$ & $2.58 \%$ & $77.42 \%$ & $5.151$ & $0.192$ & $3.60 \%$ & $77.62 \%$ & $3.748$ & $0.174$ & $4.44\%$ & $77.79 \%$ \\
 $\lambda = 5.0$ & $105$ & $10.514$ & $0.323$ & $2.98 \% $ & $77.75 \% $ & $8.516$ & $0.346$ & $3.90 \%$ & $77.89 \%$ & $7.415$ & $0.366$ & $4.70 \%$ & $78.11 \%$ \\
 $\mathcal{T} = 1.0$ & $110$ & $13.463$ & $0.479$ & $3.43 \% $ & $78.01 \%$ & $12.037$ & $0.544$ & $4.32 \%$ & $78.35 \%$ & $11.178$ & $0.603$ & $5.12 \%$ & $78.81 \%$ \\
 & $115$ & $16.740$ & $0.685$ & $3.93 \% $ & $78.17 \%$ & $15.730$ & $0.803$ & $4.85 \%$ & $79.21 \% $ & $15.069$ & $0.907$ & $5.68 \%$ & $80.34 \% $ \\
\midrule
\midrule
 & $90$ & $4.950$ & $0.091$ & $1.81 \%$ & $64.97 \%$ & $0.468$ & $0.016$ & $3.20 \% $ & $65.77 \%$ & $0$ & $0$ & {\bf --} & {\bf --}  \\
 (2) & $95$ & $6.842$ & $0.144$ & $2.07 \%$ & $65.37 \%$ & $2.166$ & $0.073$ & $3.25 \%$ & $65.81 \% $ & $0$ & $0$ & {\bf --} & {\bf --} \\
$\sigma_{X}=0.2$ & $100$ & $9.098$ & $0.220$ & $2.36 \%$ & $65.74 \%$ & $5.678$ & $0.198$ & $3.37 \%$ & $65.91 \%$ & $4.077$ & $0.177$ & $4.16\%$ & $66.01 \%$ \\
 $\lambda = 5.0$ & $105$ & $11.704$ & $0.322$ & $2.68 \% $ & $66.08 \% $ & $9.138$ & $0.341$ & $3.60 \%$ & $66.09 \%$ & $7.852$ & $0.358$ & $4.35 \%$ & $66.17 \%$ \\
 $\mathcal{T} = 1.0$ & $110$ & $14.634$ & $0.458$ & $3.03 \% $ & $66.37 \%$ & $12.709$ & $0.518$ & $3.92 \%$ & $66.34 \%$ & $11.668$ & $0.571$ & $4.67 \%$ & $66.50 \%$ \\
 & $115$ & $17.857$ & $0.632$ & $3.42 \% $ & $66.60 \%$ & $16.419$ & $0.741$ & $4.32 \%$ & $66.73 \% $ & $15.583$ & $0.834$ & $5.08 \%$ & $67.20 \% $ \\
\bottomrule
\end{tabular}
}
\end{center}
$\mbox{}$ \vspace{-0.8em} \\

\subsection{The Impact of Jumps on Geometric Step Options}

\begin{figure}
\begin{subfigure}{.5\linewidth}
\centering
\includegraphics[scale=.16]{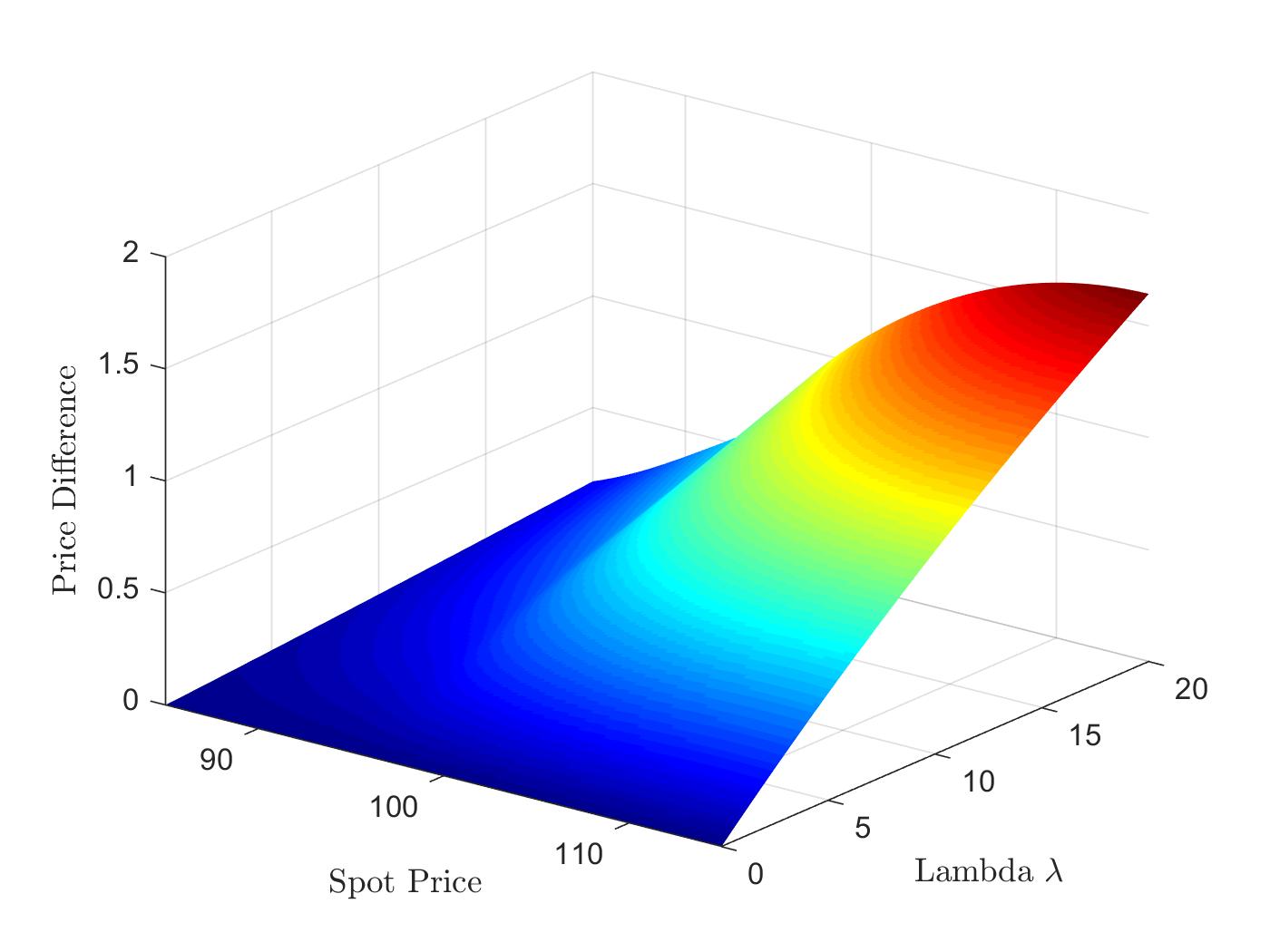}
\caption{European Price Difference}
\label{HEJD:FIG2:sub21}
\end{subfigure}%
\begin{subfigure}{.5\linewidth}
\centering
\includegraphics[scale=.16]{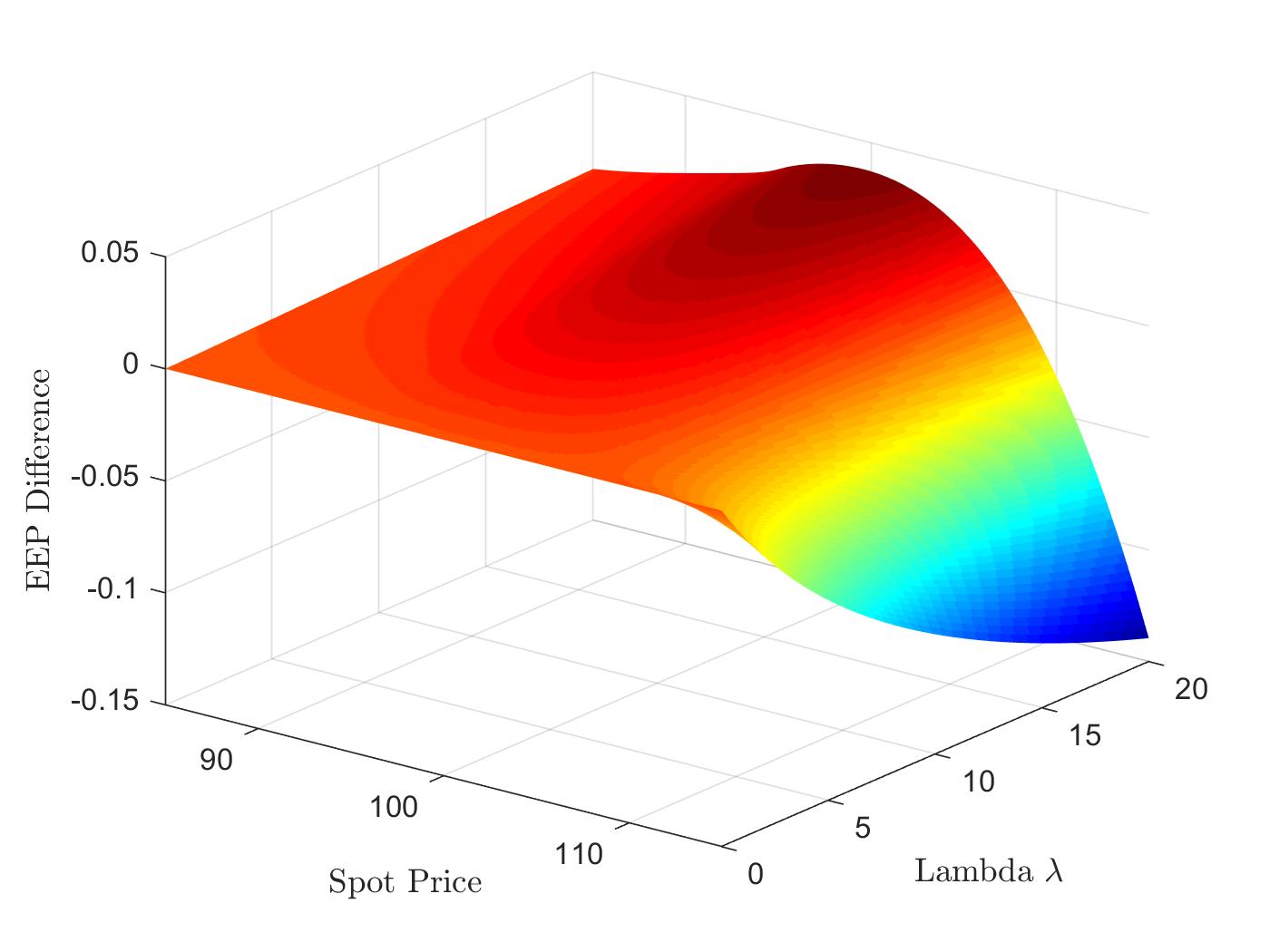}
\caption{EEP Price Difference}
\label{HEJD:FIG2:sub22}
\end{subfigure}\\[1ex]
\begin{subfigure}{.5\linewidth}
\centering
\includegraphics[scale=.16]{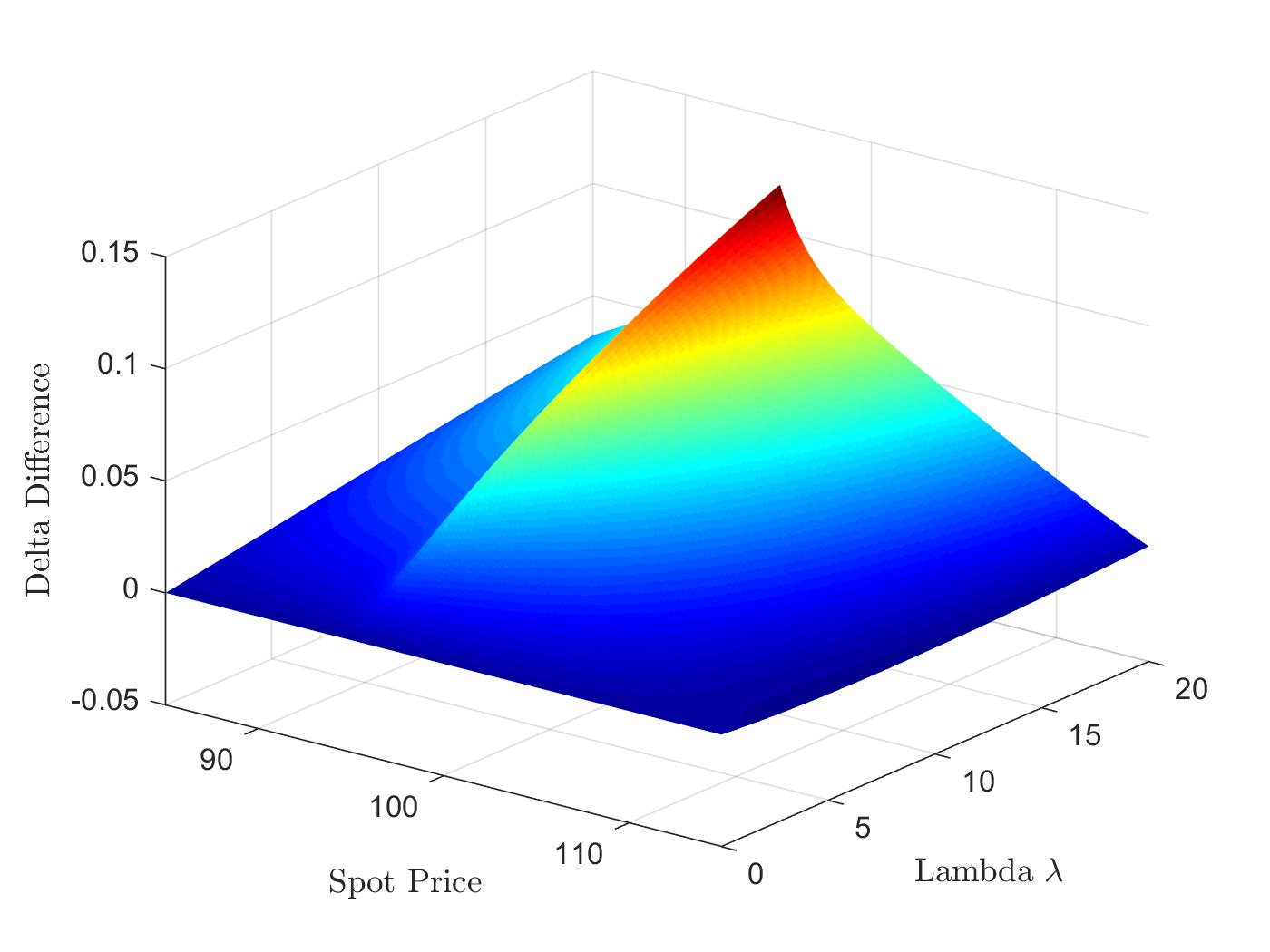}
\caption{European Delta Difference}
\label{HEJD:FIG2:sub23}
\end{subfigure}
\begin{subfigure}{.5\linewidth}
\centering
\includegraphics[scale=.16]{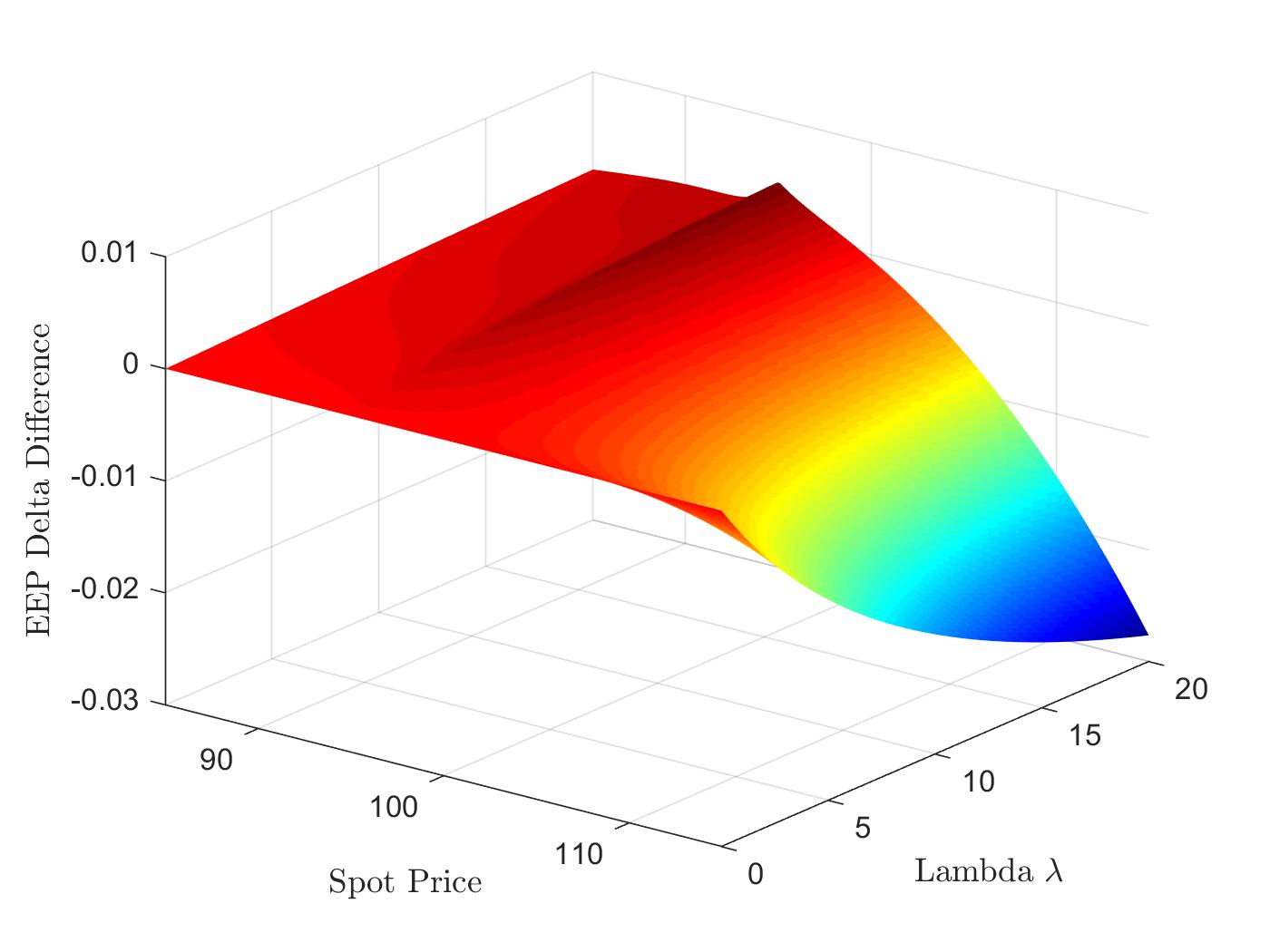}
\caption{EEP Delta Difference}
\label{HEJD:FIG2:sub24}
\end{subfigure}
\begin{subfigure}{.5\linewidth}
\centering
\includegraphics[scale=.16]{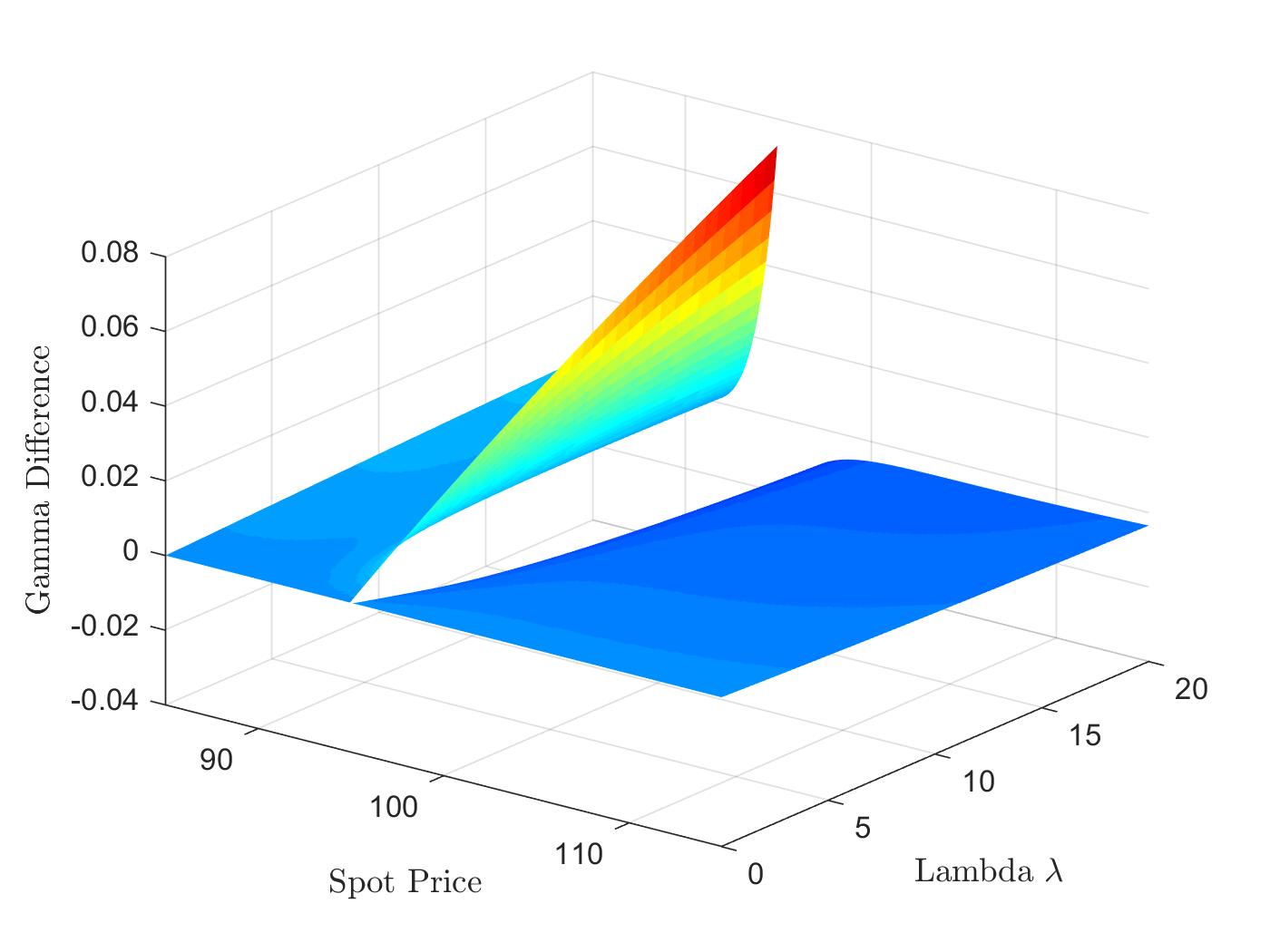}
\caption{European Gamma Difference}
\label{HEJD:FIG2:sub25}
\end{subfigure}
\begin{subfigure}{.5\linewidth}
\centering
\includegraphics[scale=.16]{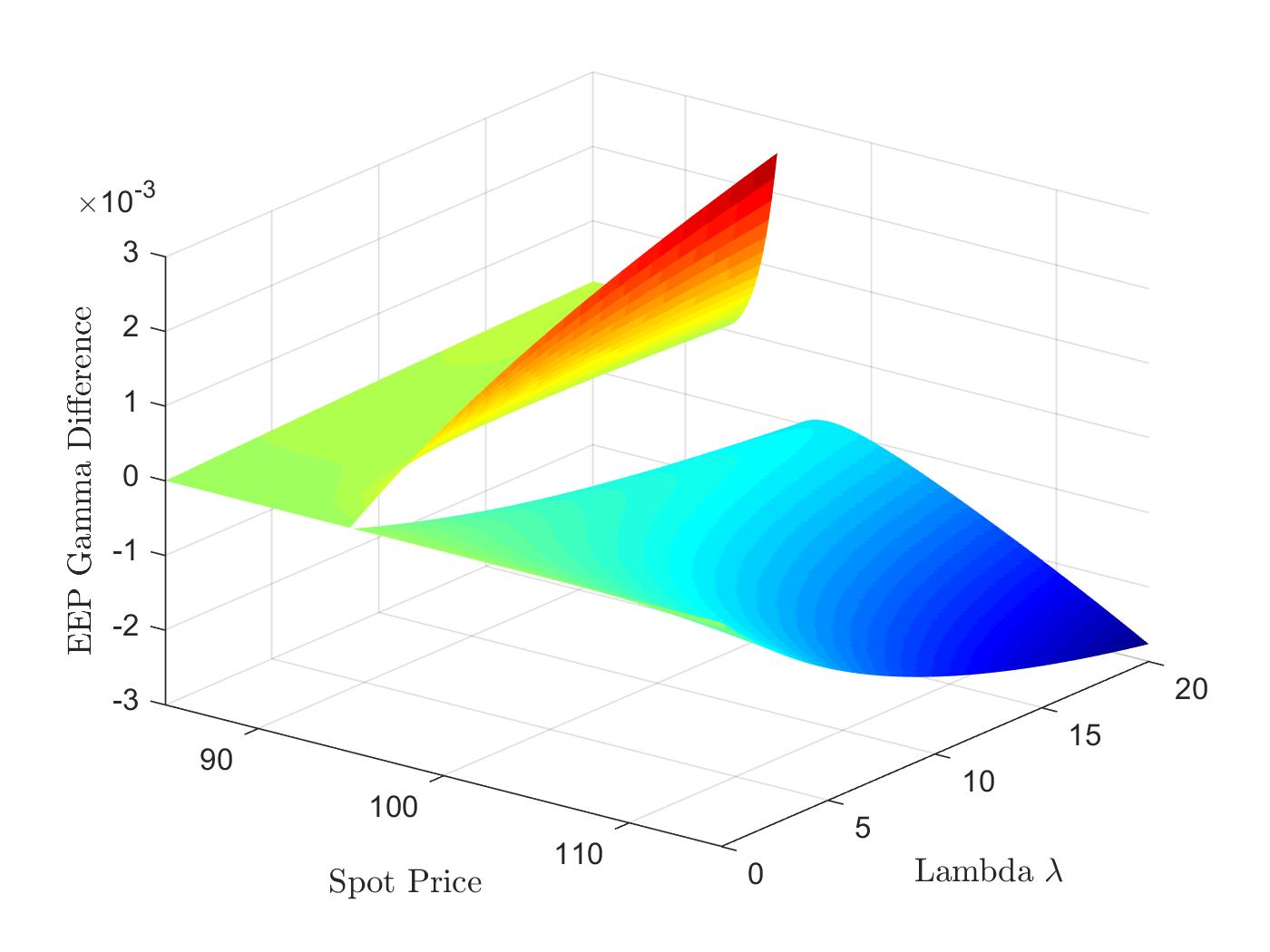}
\caption{EEP Gamma Difference}
\label{HEJD:FIG2:sub26}
\end{subfigure}
\caption{Difference in the prices, deltas, and gammas for the geometric down-and-out step calls with and without jumps as functions of the underlying price $S_{0} \in [85.115]$ and the intensity parameter $\lambda \in [0,20]$, when the remaining parameters are chosen as: $\mathcal{T} = 1.0$, $\sigma_{X} =0.2$, $r=0.05$, $\delta =0.07$, $K=100$, $L =95$, $\rho_{L} = -26.34$, $p=0.5$, $\xi = 25$, $\eta = 50$.}
\label{HEJD:FIG2}
\end{figure}
\begin{figure}
\begin{subfigure}{.5\linewidth}
\centering
\includegraphics[scale=.16]{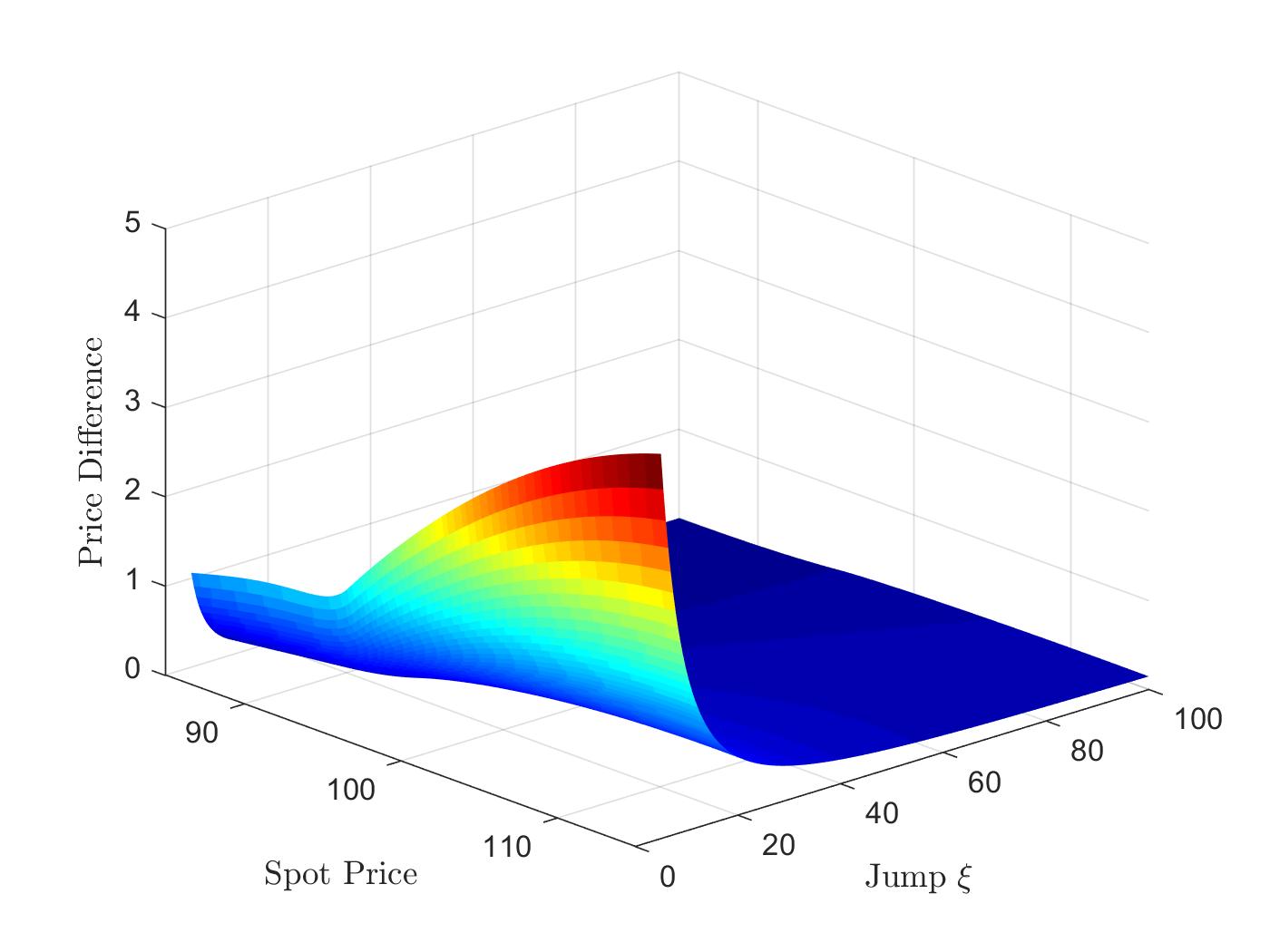}
\caption{European Price Difference}
\label{HEJD:FIG3:sub31}
\end{subfigure}%
\begin{subfigure}{.5\linewidth}
\centering
\includegraphics[scale=.16]{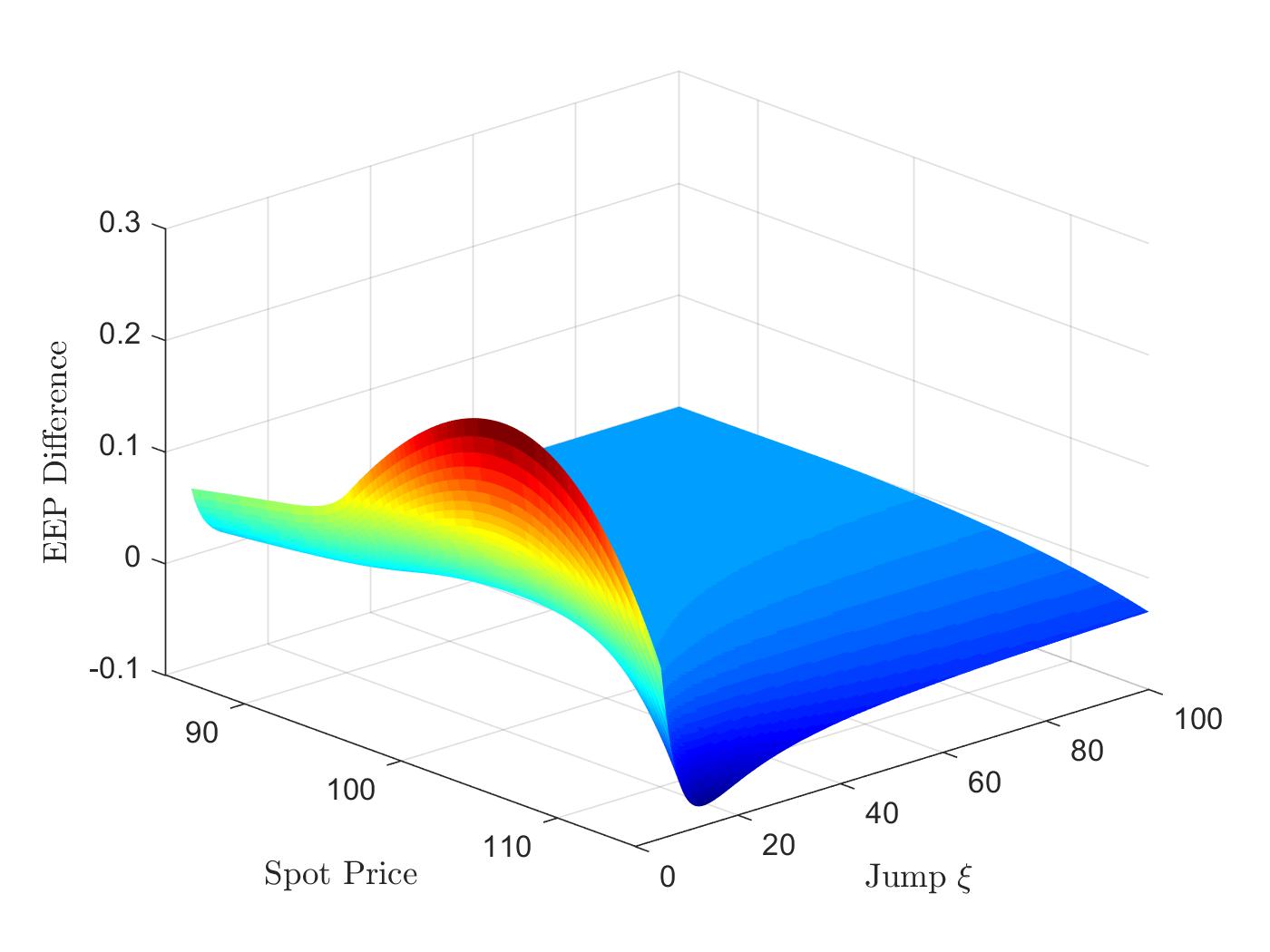}
\caption{EEP Price Difference}
\label{HEJD:FIG3:sub32}
\end{subfigure}\\[1ex]
\begin{subfigure}{.5\linewidth}
\centering
\includegraphics[scale=.16]{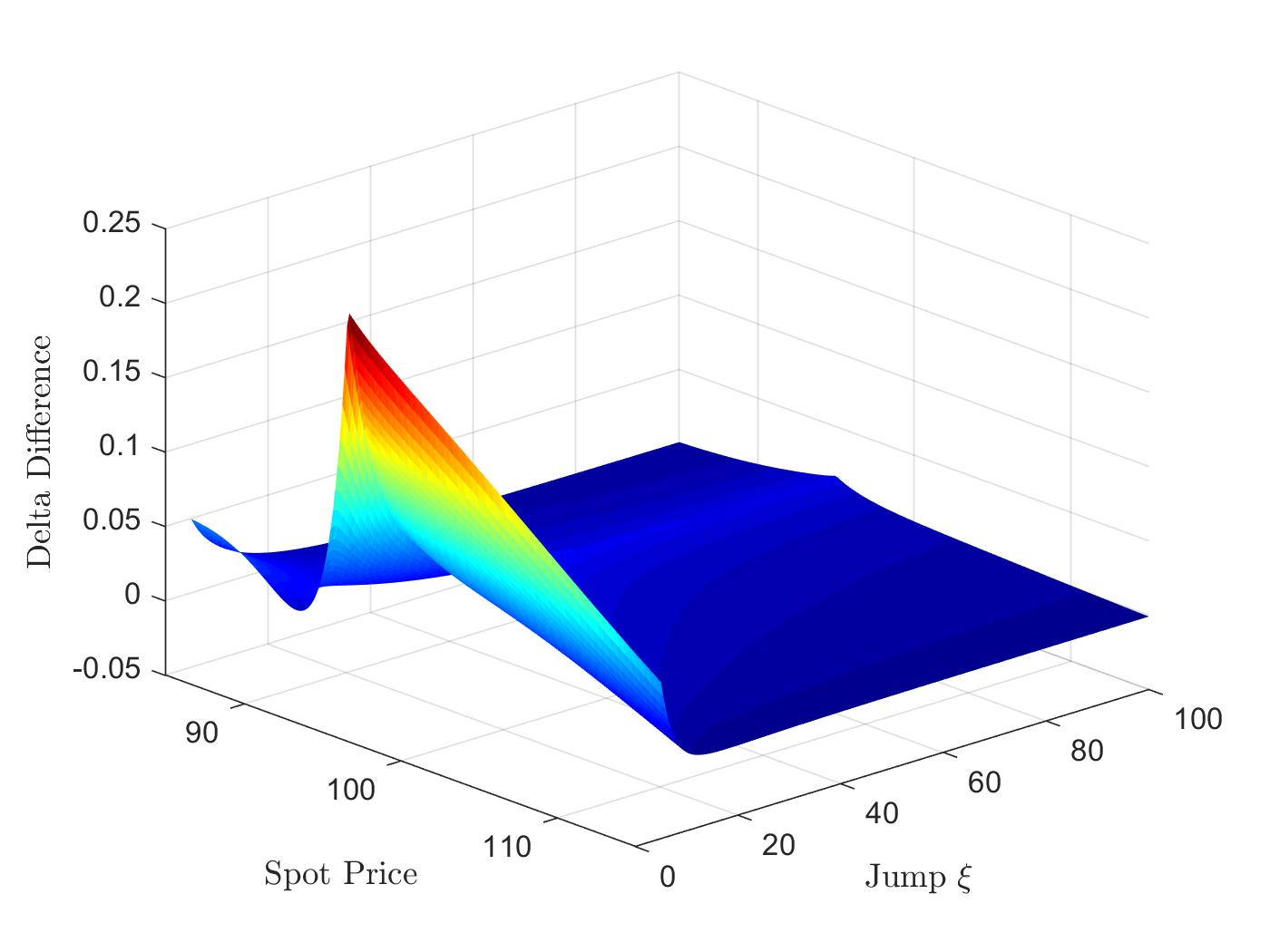}
\caption{European Delta Difference}
\label{HEJD:FIG3:sub33}
\end{subfigure}
\begin{subfigure}{.5\linewidth}
\centering
\includegraphics[scale=.16]{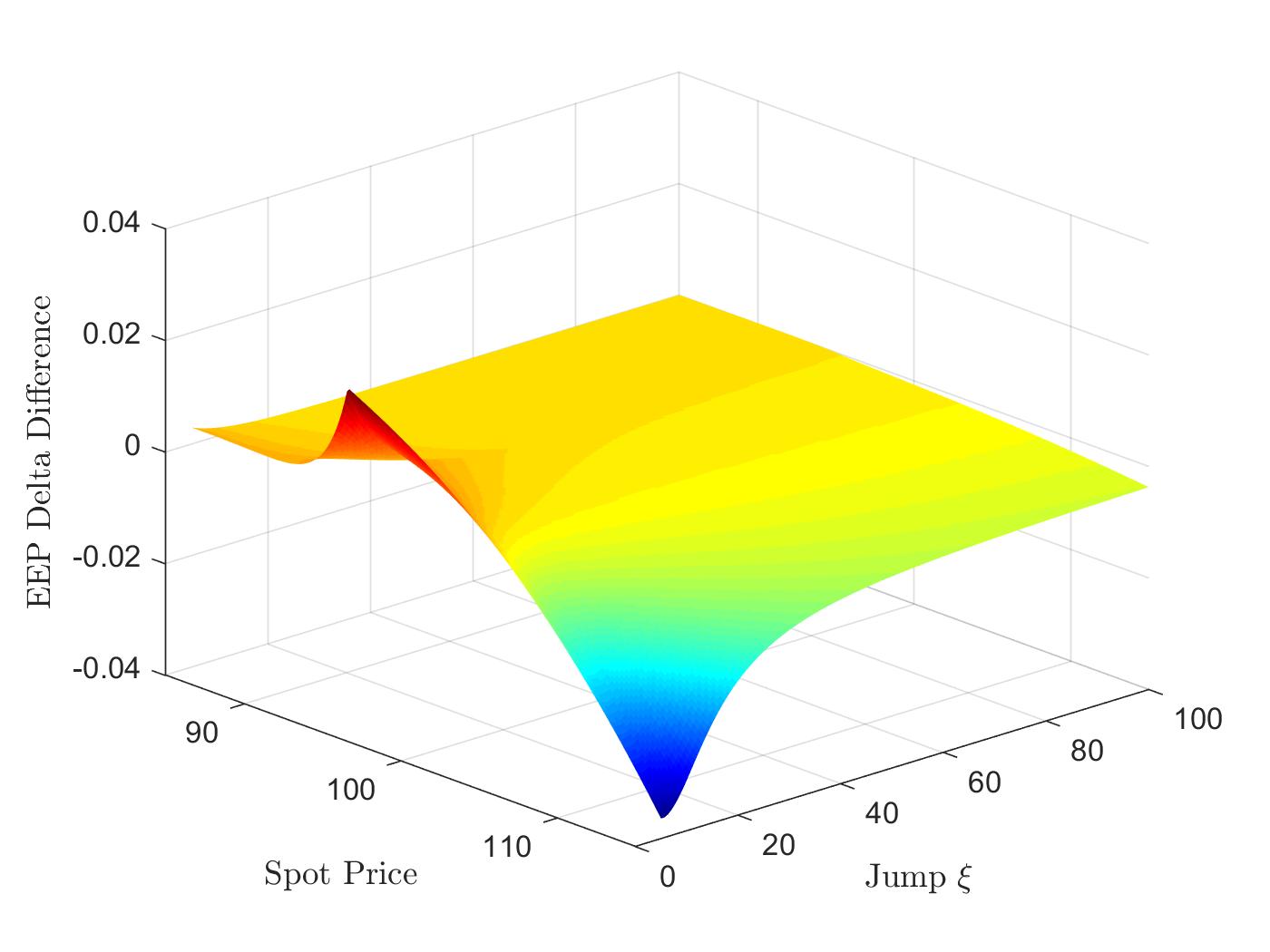}
\caption{EEP Delta Difference}
\label{HEJD:FIG3:sub34}
\end{subfigure}
\begin{subfigure}{.5\linewidth}
\centering
\includegraphics[scale=.16]{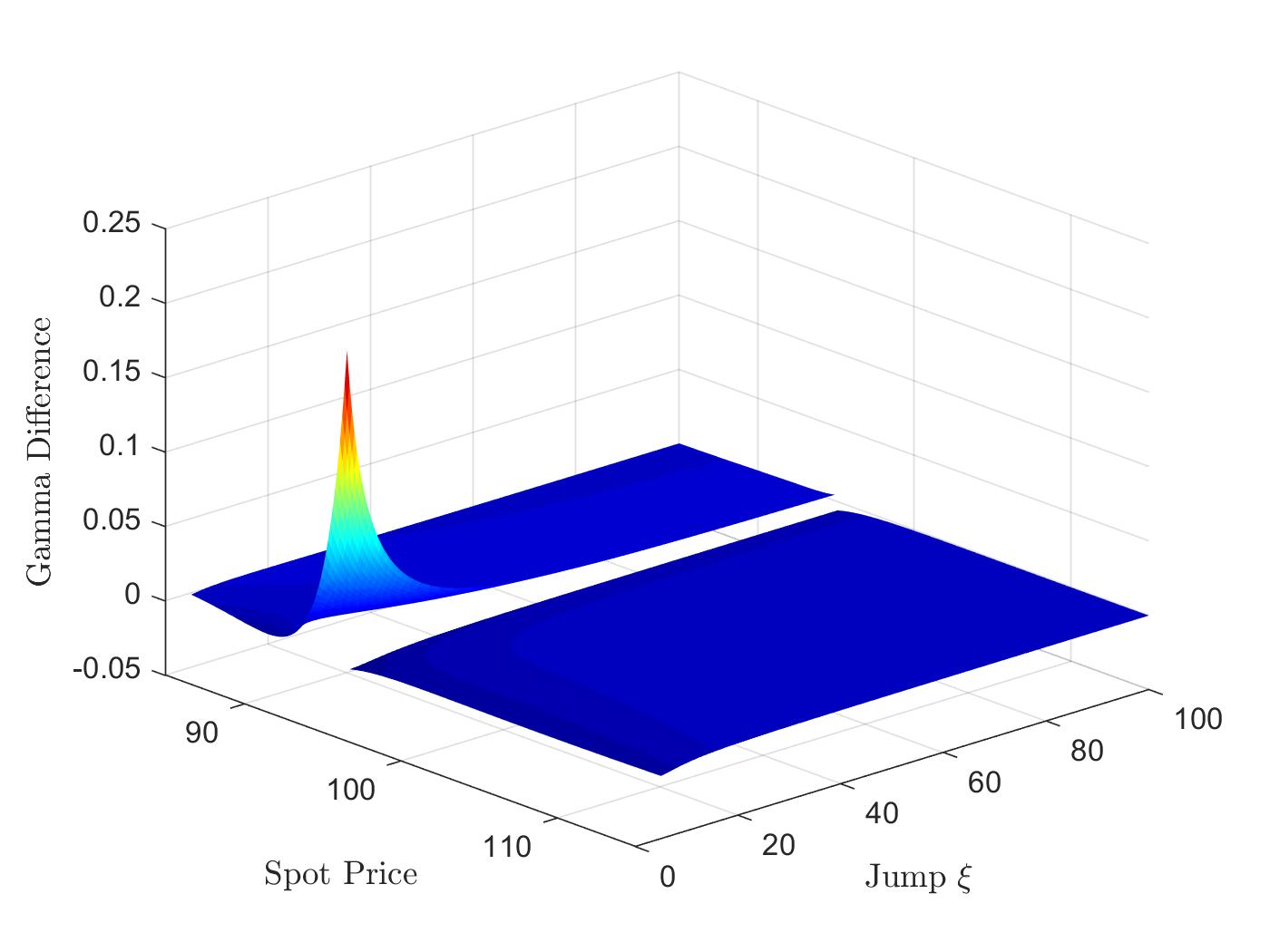}
\caption{European Gamma Difference}
\label{HEJD:FIG3:sub35}
\end{subfigure}
\begin{subfigure}{.5\linewidth}
\centering
\includegraphics[scale=.16]{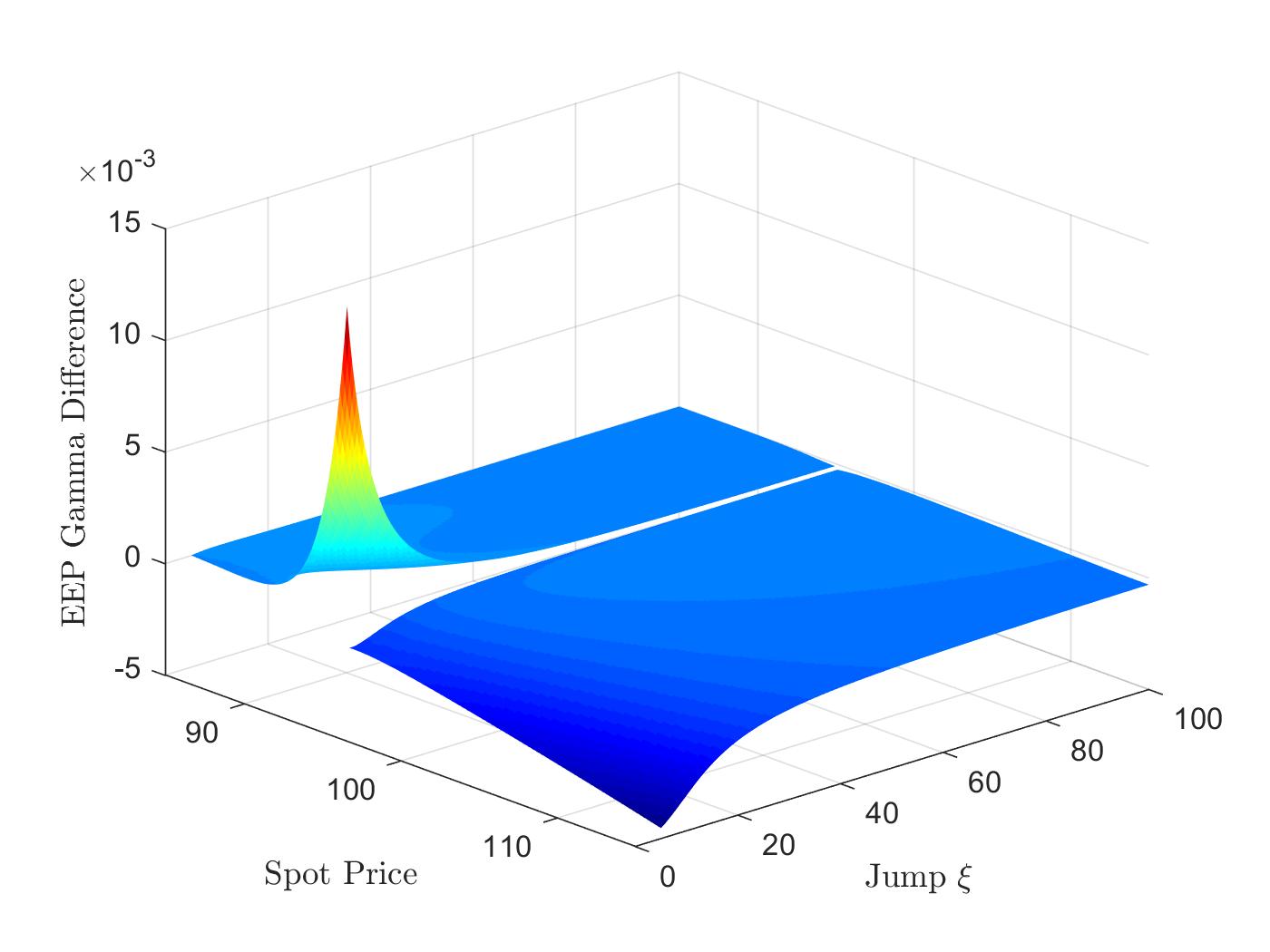}
\caption{EEP Gamma Difference}
\label{HEJD:FIG3:sub36}
\end{subfigure}
\caption{Difference in the prices, deltas, and gammas for the geometric down-and-out step calls with and without jumps as functions of the underlying price $S_{0} \in [85.115]$ and the positive jump parameter $\xi \in [5,100]$, when the remaining parameters are chosen as: $\mathcal{T} = 1.0$, $\sigma_{X} =0.2$, $r=0.05$, $\delta =0.07$, $K=100$, $L =95$, $\rho_{L} = -26.34$, $\lambda = 5$, $p=0.5$, $\eta = 50$.}
\label{HEJD:FIG3}
\end{figure}

\noindent The vast majority of the geometric step option pricing literature either studies the Black \& Scholes market (cf.~\cite{li99}, \cite{dl02}, \cite{xy13}, \cite{dlm19}) or only European-type geometric step options under more advanced models (cf.~\cite{ccw10}, \cite{cmw13}, \cite{lz16}, \cite{wzb17}). Additionally, although the inclusion of jumps naturally raises questions about their importance, no clear investigation of jump risk on the price and hedging parameters of geometric step options has been provided yet. This is the content of the next discussion. \vspace{1em} \\
\noindent We start by quantifying the impact of the jump intensity $\lambda$ on the prices and greeks of (regular) geometric down-and-out step call options and of their respective early exercise premiums. Here, we plot in Figure~\ref{HEJD:FIG2} the difference in the prices, deltas, and gammas for the geometric down-and-out step call options with and without jumps as function of the underlying price $S_{0} \in [85.115]$ and the intensity parameter $\lambda \in [0,20]$ for the following parameters: $\mathcal{T} = 1.0$, $\sigma_{X} =0.2$, $r=0.05$, $\delta =0.07$, $K=100$, $L =95$, $\rho_{L} = -26.34$, $p=0.5$, $\xi = 25$, $\eta = 50$. As expected, all differences vanish as the jump parameter approaches zero and the value of the European-type contracts increases when jumps are added (cf.~Figure~\ref{HEJD:FIG2:sub21}). However, including jumps to the asset dynamics does not necessarily increase the value of the early exercise premium. This becomes evident when looking at Figure~\ref{HEJD:FIG2:sub22} where the difference in the early exercise premiums of the geometric down-and-out step calls with and without jumps becomes negative for out of the money options. Accordingly, the difference in the deltas of the European-type geometric step options with and without jumps is always positive (cf.~Figure~\ref{HEJD:FIG2:sub23}) while the difference of the deltas for the corresponding early exercise premiums may become negative (cf.~Figure~\ref{HEJD:FIG2:sub24}). Finally, one should note that the difference in the deltas attains for both European-type options and early exercise premiums its maximum at the barrier level $L$. These findings similarly hold true for the gamma differences, where the main (positive and negative) differences are found near the barrier (cf.~Figure~\ref{HEJD:FIG2:sub25} and Figure~\ref{HEJD:FIG2:sub26}).\vspace{1em} \\
\noindent Secondly, we investigate the effect of the positive jump size $\xi$ on the prices and greeks of (regular) geometric down-and-out step call options and of their respective early exercise premiums. This is demonstrated in Figure~\ref{HEJD:FIG3} where we have plotted the difference in the prices, deltas, and gammas for the geometric down-and-out step call options with and without jumps as functions of the underlying price $S_{0} \in [85.115]$ and the positive jump parameter $\xi \in [5,100]$ for the following specification: $\mathcal{T} = 1.0$, $\sigma_{X} =0.2$, $r=0.05$, $\delta =0.07$, $K=100$, $L =95$, $\rho_{L} = -26.34$, $\lambda = 5$, $p=0.5$, $\eta = 50$. Here, for a given spot $S_{0}$ the difference in prices of the geometric down-and-out step calls with and without jumps increases with increasing average jump size~$\frac{1}{\xi}$ (cf.~Figure~\ref{HEJD:FIG3:sub31}) and the same holds true for the difference in the early exercise premiums (cf.~Figure~\ref{HEJD:FIG3:sub32}), except in parts of the payoff exercise domain, where an opposite relation is observed. While this result may seem surprising at first, it was already noticed for American-type Parisian options in \cite{cv18}, where the authors argue that the behavior is due to the structure of the early exercise premium, as difference between the intrinsic value of the option (which does not depend on the model parameters) and the corresponding European-type option price (which increases with increasing average jump size $\frac{1}{\xi}$). The same rationale also holds true in our case and the net effect then becomes negative in parts of the payoff exercise domain. Finally, an increase in the average jump size~$\frac{1}{\xi}$ also usually leads to higher sensitivities for both European-type geometric down-and-out step calls and their respective early exercise premiums, except in parts of the payoff exercise domain where the same opposite relation is observed (cf.~Figure~\ref{HEJD:FIG3:sub33}, Figure~\ref{HEJD:FIG3:sub34}, Figure~\ref{HEJD:FIG3:sub35}, and Figure~\ref{HEJD:FIG3:sub36}). 
\section{Conclusion}
\label{SecConCLUSION}
\noindent In the present article, we have extended the current literature on geometric step option pricing in several directions. Firstly, we have derived symmetry and parity relations and obtained various characterizations for both European-type and American-type geometric double barrier step options under exponential Lévy markets. In particular, we were able to translate the formalism introduced in \cite{fmv19} to the setting of geometric double barrier step options and to generalize at the same time the ideas introduced in \cite{cy13}, \cite{lv17} and \cite{cv18} to Lévy-driven markets. As a result of these extensions, we were able to derive a jump-diffusion disentanglement for the early exercise premium of American-type geometric double barrier step options and its maturity-randomized equivalent as well as to characterize the diffusion and jump contributions to these early exercise premiums separately by means of partial integro-differential equations and ordinary integro-differential equations. To illustrate the practicability and importance of our characterizations, we have subsequently derived semi-analytical pricing results for (regular) European-type and American-type geometric down-and-out step call options under hyper-exponential jump-diffusion markets. Lastly, we have used the latter results to discuss the early exercise structure of geometric step options once jumps are added and to provide an analysis of the impact of jumps on the price and hedging parameters of (European-type and American-type) geometric step contracts. \vspace{2em} \\
\noindent \acknow{The authors would like to thank Nikola Vasiljevi\'c for his helpful comments.}
\vspace{1em} \\
\newpage
\section*{Appendices}
\renewcommand{\theequation}{A.\arabic{equation}}
\subsection*{Appendix A: Proofs - Section \ref{SEC2}}
\begin{proof}[\bf Proof of Lemma \ref{lem1}]
\noindent For the sake of better exposition, we start by expanding our notation and define, for a Lévy process $(X_{t})_{t \geq 0}$, $t \geq 0$, $x \geq 0$, $\gamma \geq 0$ and given barrier level $\ell >0$,
\begin{equation}
\Gamma_{X,t,\ell}^{-}(x,\gamma) := \gamma + \int \limits_{0}^{t} \mathds{1}_{(0,\ell)}\big(xe^{X_{s}} \big) ds, \hspace{2em} \mbox{and} \hspace{2.3em}  \Gamma_{X,t,\ell}^{+}(x,\gamma) := \gamma + \int \limits_{0}^{t} \mathds{1}_{(\ell,\infty)}\big(xe^{X_{s}} \big) ds.
\label{AppNota}
\end{equation}
\noindent Then, we denote by $(\tilde{X}_{t})_{t \geq 0}$ the dual process to $(X_{t})_{t \geq 0}$, i.e.~the process defined for $t \geq 0$ by $\tilde{X}_{t} := -X_{t}$, and note that, for $t \geq 0$, $x \geq 0$, $K \geq 0$, $\gamma \geq 0$ and $\ell>0$, the following relation holds
\begin{equation}
\Gamma_{X,t,\ell}^{\pm}(x,\gamma) = \Gamma_{\tilde{X},t,\frac{xK}{\ell}}^{\mp}(K,\gamma).
\label{RelA2}
\end{equation}
\noindent Combining (\ref{RelA2}) with the change of measure defined by the ($1$-)Esscher transform\footnote{\noindent The Esscher transform was first introduced 1932 by Esscher and later established in the theory of option pricing by Gerber and Shiu (cf.~\cite{gs94}). For an economical interpretation of this pricing technique in the continuous-time framework, we refer to \cite{gs94}.}
\begin{equation}
Z_{t} : = \left. \frac{d \mathbb{Q}^{(1)}}{d \mathbb{Q}} \right|_{\mathcal{F}_{t}} := \frac{e^{1 \cdot X_{t}}}{\mathbb{E}^{\mathbb{Q}}\left[ e^{1 \cdot X_{t}}\right]} = e^{X_{t} - t \Phi_{X}(1)},
\label{EsscherT}
\end{equation}
\noindent allows us to recover (with $\delta = r- \Phi_{X}(1)$) that for any $T > 0$ and stopping time $\tau \in \mathfrak{T}_{[0,T]}$
\begin{align}
\mathcal{DSC}\big( \tau ,x, \gamma_{L}, \gamma_{H}; & \,r,\delta, K, L, H,\rho_{L}, \rho_{H}, \Psi_{X}(\cdot)\big)  \nonumber \\ 
& = \mathbb{E}^{\mathbb{Q}} \left[ B_{\tau}(r)^{-1} \exp\big\{\rho_{L} \Gamma_{X,\tau,L}^{-}(x,\gamma_{L}) + \rho_{H} \Gamma_{X,\tau,H}^{+}(x,\gamma_{H})\big \} \left(xe^{X_{\tau}} - K \right)^{+} \right] \nonumber \\
& = \mathbb{E}^{\mathbb{Q}} \Big[ Z_{\tau} B_{\tau}(\delta)^{-1} \exp\big\{\rho_{H} \Gamma_{\tilde{X},\tau,\frac{xK}{H}}^{-}(K,\gamma_{H}) + \rho_{L} \Gamma_{\tilde{X},\tau,\frac{xK}{L}}^{+}(K,\gamma_{L}) \big\} \big(x - Ke^{ \tilde{X}_{\tau}} \big)^{+} \Big] \nonumber \\
& = \mathbb{E}^{\mathbb{Q}^{(1)}} \Big[B_{\tau}(\delta)^{-1} \, \exp\big\{\rho_{H} \Gamma_{\tilde{X},\tau,\frac{xK}{H}}^{-}(K,\gamma_{H}) + \rho_{L} \Gamma_{\tilde{X},\tau,\frac{xK}{L}}^{+}(K,\gamma_{L}) \big\} \big(x - Ke^{ \tilde{X}_{\tau}} \big)^{+} \Big]
\label{SymProof}
\end{align}
holds. Therefore, if one shows that $(\tilde{X}_{t})_{t \geq 0}$ is again a Lévy process under the measure $\mathbb{Q}^{(1)}$, (\ref{SymProof}) implies that
\begin{align}
\mathcal{DSC}\big( \tau ,x, \gamma_{L}, \gamma_{H};  \,r,\delta, K, L, H,\rho_{L}, \rho_{H}, \Psi_{X}(\cdot)\big) = \mathcal{DSP}\Big(\tau,K,\gamma_{H}, \gamma_{L} ; \delta,r,x,\frac{xK}{H},\frac{xK}{L}, \rho_{H}, \rho_{L},\Psi_{\tilde{X}}^{(1)}(\cdot)\Big),
\end{align}
\noindent where $\Psi_{\tilde{X}}^{(1)}(\cdot)$ denotes the Lévy exponent of $(\tilde{X}_{t})_{t \geq 0}$ under the measure $\mathbb{Q}^{(1)}$. In fact, showing that $(\tilde{X}_{t})_{t \geq 0}$ is a Lévy process is not hard and can be done as in \cite{Ma18} (see also \cite{fm06}). To conclude, we therefore need to verify that $\Psi_{\tilde{X}}^{(1)} \equiv \Psi_{Y}$ holds, where $\Psi_{Y}(\cdot)$ satisfies (\ref{Ysatisf}) and is given as in (\ref{Yequation}). To this end, we first note that
\begin{equation}
\mathbb{E}^{\mathbb{Q}^{(1)}} \left[ e^{i \theta \tilde{X}_{1}} \right] = \mathbb{E}^{\mathbb{Q}} \left[ Z_{1} e^{-i \theta X_{1}} \right] = \mathbb{E}^{\mathbb{Q}} \left[ e^{i \left(-(\theta+i) \right) X_{1}} \right] e^{-\Phi_{X}(1)} = e^{- \left( \Psi_{X}\left(-(\theta + i) \right) + \Phi_{X}(1) \right)}.
\end{equation}
\noindent Therefore, the Lévy exponent of $(\tilde{X}_{t})_{t \geq 0}$ under $\mathbb{Q}^{(1)}$ can be recovered as
\begin{align}
\Psi_{\tilde{X}}^{(1)}(\theta) & = \Psi_{X}\left(-(\theta + i) \right) + \Phi_{X}(1) \nonumber \\
& = i \big( b_{X} + \sigma_{X}^{2} \big) \theta + \frac{1}{2} \sigma_{X}^{2} \theta^{2} + \int \limits_{\mathbb{R}} \big( e^{y} - e^{-i(\theta + i) y} - i \theta y \mathds{1}_{\{ | y | \leq 1\}} \big) \Pi_{X}(dy) \nonumber \\
& = i \Big( b_{X} + \sigma_{X}^{2} - \int \limits_{\mathbb{R}} \big( 1- e^{y} \big) y \mathds{1}_{\{ | y | \leq 1\}} \Pi_{X}(dy) \Big) \theta + \frac{1}{2} \sigma_{X}^{2} \theta^{2} + \int \limits_{\mathbb{R}} e^{y} \big( 1 - e^{i\theta(-y)} + i \theta (-y) \mathds{1}_{\{ | y | \leq 1\}} \big) \Pi_{X}(dy) \nonumber \\
& = i \Big( b_{X} + \sigma_{X}^{2} - \int \limits_{\mathbb{R}} \big( 1- e^{y} \big) y \mathds{1}_{\{ | y | \leq 1\}} \Pi_{X}(dy) \Big) \theta + \frac{1}{2} \sigma_{X}^{2} \theta^{2} + \int \limits_{\mathbb{R}} \big( 1 - e^{i\theta y} + i \theta y \mathds{1}_{\{ | y | \leq 1\}} \big) \Pi^{\star}(dy),
\end{align}
\noindent where $\Pi^{\star}(dy) := e^{-y} \,\Pi_{\tilde{X}}(dy)$ and the jump measure of the dual process $(\tilde{X}_{t})_{t \geq 0}$ satisfies $\Pi_{\tilde{X}}(dy) = \Pi_{X}(-dy)$. Lastly, we can combine these results with Equation (\ref{bXequa}) to obtain that
\begin{align}
b_{Y} & = - \Big( b_{X} + \sigma_{X}^{2} - \int \limits_{\mathbb{R}} \big( 1- e^{y} \big) y \mathds{1}_{\{ | y | \leq 1\}} \Pi_{X}(dy) \Big) = \delta - r - \frac{1}{2}\sigma_{X}^{2} + \int \limits_{\mathbb{R}} \big( 1 - e^{y} + y \mathds{1}_{\{ | y | \leq 1\}} \big) \Pi^{\star}(dy).
\end{align}
\noindent This finalizes the proof.
\end{proof}
\begin{proof}[\bf Proof of Corollary \ref{coro1}]
\noindent First, we note that Equation (\ref{Toprove1a}) is a direct consequence of Lemma \ref{lem1}, since taking $\tau \equiv \mathcal{T}$ in (\ref{EqLem1}) directly provides the result for the European-type options, while the corresponding equality for American-type options is recovered from (\ref{EqLem1}) by taking the supremum over the set $\mathfrak{T}_{[0,\mathcal{T}]}$. Therefore, we proceed with the proof of the second identity. \vspace{1em} \\
\noindent For the proof of (\ref{Toprove1b}), we note as in the proof of Lemma \ref{lem1} that, for a Lévy process $(X_{t})_{t \geq 0}$, $t \geq 0$, $x \geq 0$, $\gamma \geq 0$ and given barrier level $\ell >0$, the following identity holds
\begin{equation}
\Gamma_{X,t,\ell}^{\pm}(x,\gamma) = \Gamma_{X,t,\frac{\ell}{xK}}^{\pm}\Big(\frac{1}{K},\gamma \Big),  
\end{equation}
\noindent where we have used the notation introduced in (\ref{AppNota}). Then, combining the latter relation with Lemma \ref{lem1} allows us to recover, for $T >0$ and any stopping time $\tau \in \mathfrak{T}_{[0,T]}$, that
\begin{align}
\mathcal{DSC}\big( \tau ,x, \gamma_{L}, \gamma_{H}; & \,r,\delta, K, L, H,\rho_{L}, \rho_{H}, \Psi_{X}(\cdot)\big) \nonumber \\
& = xK \cdot \mathbb{E}^{\mathbb{Q}} \left[ B_{\tau}(r)^{-1} \exp\Big\{\rho_{L} \Gamma_{X,\tau,\frac{L}{xK}}^{-}\Big( \frac{1}{K},\gamma_{L} \Big) + \rho_{H} \Gamma_{X,\tau,\frac{H}{xK}}^{+}\Big( \frac{1}{K},\gamma_{H} \Big) \Big \} \Big(\frac{1}{K}e^{X_{\tau}} - \frac{1}{x} \Big)^{+} \right] \nonumber \\
& = xK \cdot \mathcal{DSC} \Big( \tau ,\frac{1}{K}, \gamma_{L}, \gamma_{H}; r,\delta, \frac{1}{x}, \frac{L}{xK}, \frac{H}{xK},\rho_{L}, \rho_{H}, \Psi_{X}(\cdot) \Big) \nonumber \\
& = xK \cdot \mathcal{DSP} \Big( \tau ,\frac{1}{x}, \gamma_{H}, \gamma_{L}; \delta,r, \frac{1}{K}, \frac{1}{H}, \frac{1}{L},\rho_{H}, \rho_{L}, \Psi_{Y}(\cdot) \Big).
\label{GodEQ}
\end{align}
\noindent Here, $\Psi_{Y}(\cdot)$ represents, as in Lemma \ref{lem1}, the Lévy exponent of a process $(Y_{t})_{t \geq 0}$ driving another exponential Lévy market and that satisfies the relations (\ref{Ysatisf})-(\ref{inteMeas}). \noindent Therefore, taking as earlier $\tau \equiv \mathcal{T}$ in (\ref{GodEQ}) directly provides us with the result for the European-type options, while the corresponding identity for American-type contracts is obtained from (\ref{GodEQ}) by taking the supremum over the set $\mathfrak{T}_{[0,\mathcal{T}]}$.
\end{proof}

\begin{proof}[\bf Proof of Proposition \ref{prop1}]
\noindent We start by showing the continuity of $(\mathcal{T},x) \mapsto \mathcal{DSC}^{\star}_{E}(\mathcal{T},x;K,\bm{\ell}, \bm{\rho}_{\bm{\ell}})$ on the domain $[0,T] \times [0,\infty)$ for any $K,\bm{\ell}$, and $\bm{\rho}_{\bm{\ell}} $. To do this, we first note that the continuity of the occupation times $x \mapsto \Gamma_{X,\mathcal{T},\ell}^{\pm}(x,0)$, defined for any $\mathcal{T} \in [0,T]$ and $\ell \geq 0$ as in (\ref{AppNota}), and the continuity of the function $x \mapsto (x-K)^{+}$, for $K \geq 0$, directly give by means of the dominated convergence theorem the continuity of $x \mapsto \mathcal{DSC}^{\star}_{E}(\mathcal{T},x; K, \bm{\ell}, \bm{\rho}_{\bm{\ell}} )$ for any of the parameters $\mathcal{T},K,\bm{\ell}$, and $\bm{\rho}_{\bm{\ell}}$. Therefore, to prove that $(\mathcal{T},x) \mapsto \mathcal{DSC}^{\star}_{E}(\mathcal{T},x; K, \bm{\ell}, \bm{\rho}_{\bm{\ell}} )$ is, for any parameters $K,\bm{\ell}$, and $\bm{\rho}_{\bm{\ell}}$, continuous on $[0,T] \times [0,\infty)$, it is enough to show that $\mathcal{T} \mapsto \mathcal{DSC}^{\star}_{E}(\mathcal{T},x; K, \bm{\ell}, \bm{\rho}_{\bm{\ell}} )$ is, for any parameters $x,K,\bm{\ell}$, and $\bm{\rho}_{\bm{\ell}}$, uniformly continuous on~$[0,T]$. To obtain this property, we fix times to maturity $0 \leq u < t \leq T$, recall that $\rho_{L}, \rho_{H} \leq 0$ and derive that
\begin{align}
\big| & \mathcal{DSC}^{\star}_{E}(t,x; K, \bm{\ell}, \bm{\rho}_{\bm{\ell}} ) - \mathcal{DSC}^{\star}_{E}(u,x; K, \bm{\ell}, \bm{\rho}_{\bm{\ell}} ) \big| \nonumber \\
& \leq \mathbb{E}_{x}^{\mathbb{Q}} \bigg[ \,e^{-ru + \rho_{L} \Gamma_{u,L}^{-} + \rho_{H} \Gamma_{u,H}^{+}} \, \Big| e^{-\int_{u}^{t} \left(r-\rho_{L} \mathds{1}_{(0,L)}(S_{s}) - \rho_{H} \mathds{1}_{(H,\infty)}(S_{s}) \right) ds }  (S_{t} - K )^{+} - (S_{u} - K)^{+} \Big| \, \bigg]  \nonumber \\
& \leq \mathbb{E}_{x}^{\mathbb{Q}} \bigg[ \,\Big| e^{-\int_{u}^{t} \left(r-\rho_{L} \mathds{1}_{(0,L)}(S_{s}) - \rho_{H} \mathds{1}_{(H,\infty)}(S_{s}) \right) ds }  (S_{t} - K ) - (S_{u} - K) \Big| \, \bigg]  \nonumber \\
& \leq \mathbb{E}_{x}^{\mathbb{Q}} \bigg[  S_{u} \, \Big| S_{t} S_{u}^{-1}e^{-\int_{u}^{t} \left(r-\rho_{L} \mathds{1}_{(0,L)}(S_{s}) - \rho_{H} \mathds{1}_{(H,\infty)}(S_{s}) \right) ds } -1 \Big| \,\bigg] + K \mathbb{E}_{x}^{\mathbb{Q}} \bigg[ \, \Big| e^{-\int_{u}^{t} \left(r-\rho_{L} \mathds{1}_{(0,L)}(S_{s}) - \rho_{H} \mathds{1}_{(H,\infty)}(S_{s}) \right) ds } -1 \Big| \, \bigg]  \nonumber \\
& \leq \mathbb{E}^{\mathbb{Q}} \left[  xe^{X_{u}}  \right] \bigg( \mathbb{E}^{\mathbb{Q}}\Big[ \big| e^{X_{t-u} + \lambda^{\star} (t-u)} -1 \big| \,\Big] + \mathbb{E}^{\mathbb{Q}}\Big[ \big| e^{X_{t-u} - \lambda^{\star} (t-u)} -1 \big| \,\Big] \bigg) + K \big( 1 - e^{-\lambda^{\star} (t-u)} \big)\nonumber \\
& \leq x \max \big\{ 1,\, e^{\Phi_{X}(1) T} \big\} \bigg( \mathbb{E}^{\mathbb{Q}}\Big[ \big| e^{X_{t-u} + \lambda^{\star} (t-u)} -1 \big| \,\Big] + \mathbb{E}^{\mathbb{Q}}\Big[ \big| e^{X_{t-u} - \lambda^{\star} (t-u)} -1 \big| \,\Big] \bigg) + K \big( 1 - e^{-\lambda^{\star} (t-u)} \big), 
\end{align}
\noindent where $\lambda^{\star} := r-\rho_{L} - \rho_{H}$. Consequently, the right-continuity of the process $(X_{t})_{t \in [0,T]}$ implies the convergence 
\begin{equation}
\mathcal{DSC}^{\star}_{E}(t,x; K, \bm{\ell}, \bm{\rho}_{\bm{\ell}} ) - \mathcal{DSC}^{\star}_{E}(u,x; K, \bm{\ell}, \bm{\rho}_{\bm{\ell}} ) \rightarrow 0, \hspace{1.5em} \mbox{whenever} \; \; t-u \rightarrow 0.
\end{equation}
\noindent This shows that the function $\mathcal{T} \mapsto \mathcal{DSC}^{\star}_{E}(\mathcal{T},x; K, \bm{\ell}, \bm{\rho}_{\bm{\ell}} )$ is, for any parameters $x,K,\bm{\ell},$ and $\bm{\rho}_{\bm{\ell}}$, uniformly continuous over $[0,T]$ and the proof of the initial claim is complete. \vspace{1em} \\
\noindent We now prove that $\mathcal{DSC}^{\star}_{E}(\cdot)$ solves Equation (\ref{GSCEuPIDE1}) on $(0,T] \times [0,\infty)$ with initial condition (\ref{GSCEuPIDE2}). Here, we start by noting that, for any parameters $\mathcal{T}, x, K, \bm{\ell},$ and  $\bm{\rho}_{\bm{\ell}}$, geometric double barrier step options can be rewritten in the simpler form
\begin{equation}
\mathcal{DSC}^{\star}_{E}(\mathcal{T},x; K, \bm{\ell}, \bm{\rho}_{\bm{\ell}} ) = \mathbb{E}_{x}^{\mathbb{Q}} \left[ B_{\mathcal{T}}(r)^{-1} \, e^{\rho_{L} \Gamma_{\mathcal{T},L}^{-} \, + \, \rho_{H} \Gamma_{\mathcal{T},H}^{+}} \,\left(S_{\mathcal{T}} - K \right)^{+} \right] = \mathbb{E}_{x}^{\mathbb{Q}} \left[ \left(\bar{S}_{\mathcal{T}} - K \right)^{+} \right],
\end{equation}
\noindent where $(\bar{S}_{t})_{t \in [0,T]}$ refers to the (strong) Markov process\footnote{It is well-known (cf.~\cite{pe06}) that the process $(\bar{S}_{t})_{t \in [0,T]}$ defined this way preserves the (strong) Markov property of the underlying process $(S_{t})_{t \in [0,T]}$.} obtained by ``killing'' the sample path of $(S_{t})_{t \in [0,T]}$ at the proportional rate $\lambda(x) := r - \bm{\rho}_{\bm{\ell}} \cdot \bigg( \begin{array}{c}
 \mathds{1}_{(0,L)}(x) \\
 \mathds{1}_{(H,\infty)}(x) 
\end{array} \bigg)$. The process' transition probabilities are then given by
\begin{equation}
\mathbb{Q}_{x} \left( \bar{S}_{t} \in A \right) = \mathbb{E}_{x}^{\mathbb{Q}} \left[ e^{- \int_{0}^{t} \lambda(S_{s}) ds} \, \mathds{1}_{A}(S_{t}) \right]
\label{TPKilling}
\end{equation}
\noindent and we identify its cemetery state, without loss of generality, with $\partial \equiv 0$. Consequently, for any initial value $z= (\mathbf{t},x) \in [0,T] \times [0,\infty) $, the process $(Z_{t})_{t \in [0,\mathbf{t}]}$ defined via $Z_{t}:= (\mathbf{t}-t, \bar{S}_{t})$, $\bar{S}_{0}=x$, is a strong Markov process with state domain given by $\mathcal{D}_{\mathbf{t}}:= [0,\mathbf{t}] \times [0,\infty)$. Additionally, $\mathcal{DSC}_{E}^{\star}(\cdot)$ can be re-expressed, for any $K,\bm{\ell}$, and $\bm{\rho}_{\bm{\ell}}$, as
\begin{equation}
\mathcal{DSC}^{\star}_{E}(\mathcal{T},x; K, \bm{\ell}, \bm{\rho}_{\bm{\ell}} ) = V_{E}\big((\mathcal{T},x)\big),
\label{IMeq1}
\end{equation}
\noindent where the value function $V_{E}(\cdot)$ has the following representation under the measure $\mathbb{Q}_{z}^{Z}$ having initial distribution $Z_{0} = z$:
\begin{align}
V_{E}(z) : = \mathbb{E}^{\mathbb{Q}^{Z}}_{z} \big[ G(Z_{\tau_{\mathcal{S}}}) \big], \hspace{1.5em}  G(z)  := (x -K)^{+},
\end{align}
\noindent and $\tau_{\mathcal{S}} := \inf \{ t \geq 0: Z_{t} \in \mathcal{S} \}$, $\mathcal{S} := \big(\{0 \} \times [0,\infty) \big) \cup \big( [0,\mathbf{t}] \times \{0 \} \big)$, is a stopping time that satisfies $\tau_{\mathcal{S}} \leq \mathbf{t}$, under $\mathbb{Q}_{z}^{Z}$ with $z = (\mathbf{t},x)$. Furthermore, the stopping region $\mathcal{S}$ is for any $\mathbf{t} \in [0,T]$ a closed set in $\mathcal{D}_{\mathbf{t}}$. Therefore, standard arguments based on the strong Markov property of $(Z_{t})_{t \in [0,\mathbf{t}]}$ (cf.~\cite{pe06}) imply that $V_{E}(\cdot)$ satisfies the following problem
\begin{align}
\mathcal{A}_{Z} V_{E}(z) & = 0, \hspace{2em} \mbox{on} \,\,  \mathcal{D}_{T} \setminus \mathcal{S}, \\
V_{E}(z) & = G(z), \hspace{1.5em} \mbox{on} \,\,  \mathcal{S},
\end{align}
\noindent where $\mathcal{A}_{Z}$ denotes the infinitesimal generator of the process $(Z_{t})_{t \in [0,\mathbf{t}]}$. To complete the proof, we note that (for any suitable function $V: \mathcal{D}_{\mathbf{t}} \rightarrow \mathbb{R}$) the infinitesimal generator $\mathcal{A}_{Z}$ can be re-expressed as
\begin{align}
\mathcal{A}_{Z} V\big((\mathbf{t},x) \big) & = -\partial_{\mathbf{t}} V\big((\mathbf{t},x) \big) + \mathcal{A}_{\bar{S}} V\big((\mathbf{t},x) \big) \nonumber \\
& = -\partial_{\mathbf{t}} V\big((\mathbf{t},x) \big) + \mathcal{A}_{S} V\big((\mathbf{t},x) \big) - \lambda(x) V\big((\mathbf{t},x) \big).
\label{IgEnE}
\end{align}
\noindent Therefore, recovering $\mathcal{DSC}^{\star}_{E}(\cdot)$ via (\ref{IMeq1}) finally gives the required equation and initial condition.
\end{proof}

\begin{proof}[\bf Proof of Proposition \ref{prop2}]
First, we note that the continuity of $x \mapsto \mathcal{DSC}^{\star}_{A}(\mathcal{T},x; K, \bm{\ell}, \bm{\rho}_{\bm{\ell}} )$ for any $\mathcal{T}, K, \bm{\ell}$, and $\bm{\rho_{\bm{\ell}}}$, follows, just like the continuity of $x \mapsto \mathcal{DSC}^{\star}_{E}(\mathcal{T},x; K, \bm{\ell}, \bm{\rho}_{\bm{\ell}} )$ for $\mathcal{T}, K, \bm{\ell}$, and $\bm{\rho_{\bm{\ell}}}$, by means of the dominated convergence theorem while noticing the continuity of the occupation times $x \mapsto \Gamma_{X,\mathcal{T},\ell}^{\pm}(x,0)$, defined for any $\mathcal{T} \in [0,T]$ and $\ell \geq 0$ as in (\ref{AppNota}), and the continuity of the function $x \mapsto (x-K)^{+}$, for $K \geq 0$. Therefore, to prove that $(\mathcal{T},x) \mapsto \mathcal{DSC}^{\star}_{A}(\mathcal{T},x; K, \bm{\ell}, \bm{\rho}_{\bm{\ell}} )$ is, for any parameters $K,\bm{\ell}$, and $\bm{\rho}_{\bm{\ell}}$, continuous on $[0,T] \times [0,\infty)$, it is enough to show that $\mathcal{T} \mapsto \mathcal{DSC}^{\star}_{A}(\mathcal{T},x; K, \bm{\ell}, \bm{\rho}_{\bm{\ell}} )$ is, for any parameters $x,K,\bm{\ell}$ and $\bm{\rho}_{\bm{\ell}}$, uniformly continuous on $[0,T]$. To derive this property, we fix times to maturity $0 \leq u < t \leq T$, denote by $\tau_{2}$ the optimal stopping time for $\mathcal{DSC}^{\star}_{A}(t,x; K, \bm{\ell}, \bm{\rho}_{\bm{\ell}} )$ and set $\tau_{1} := \tau_{2} \wedge u$. Then, noting that $\mathcal{T} \mapsto \mathcal{DSC}_{A}^{\star}(\mathcal{T},x;K, \bm{\ell}, \bm{\rho}_{\bm{\ell}} )$ is a non-decreasing function\footnote{This directly follows since, for $0\leq \mathcal{T}_{1} \leq \mathcal{T}_{2} \leq T$, any stopping time $\tau \in \mathfrak{T}_{[0,\mathcal{T}_{1}]}$ also satisfies $\tau \in \mathfrak{T}_{[0,\mathcal{T}_{2}]}$.}~while recalling that $\rho_{L}, \rho_{H} \leq 0$ holds and that $\tau_{1}$ is not necessarily optimal for the time to maturity~$u$, we obtain that
\begin{align}
0 & \leq \mathcal{DSC}^{\star}_{A}(t,x; K, \bm{\ell}, \bm{\rho}_{\bm{\ell}} ) - \mathcal{DSC}^{\star}_{A}(u,x; K, \bm{\ell}, \bm{\rho}_{\bm{\ell}} ) \nonumber \\
& \leq \mathbb{E}_{x}^{\mathbb{Q}} \bigg[ \,e^{-r \tau_{1} + \rho_{L} \Gamma_{\tau_{1},L}^{-} + \rho_{H} \Gamma_{\tau_{1},H}^{+}} \Big(  e^{-\int_{\tau_{1}}^{\tau_{2}} \left(r-\rho_{L} \mathds{1}_{(0,L)}(S_{s}) - \rho_{H} \mathds{1}_{(H,\infty)}(S_{s}) \right) ds }  (S_{\tau_{2}} - K )^{+} - (S_{\tau_{1}} - K)^{+} \Big) \bigg]  \nonumber \\
& \leq \mathbb{E}_{x}^{\mathbb{Q}} \bigg[ \, \Big| e^{-\int_{\tau_{1}}^{\tau_{2}} \left(r-\rho_{L} \mathds{1}_{(0,L)}(S_{s}) - \rho_{H} \mathds{1}_{(H,\infty)}(S_{s}) \right) ds }  (S_{\tau_{2}} - K ) - (S_{\tau_{1}} - K) \Big| \, \bigg]  \nonumber \\
& \leq \mathbb{E}_{x}^{\mathbb{Q}} \bigg[  S_{\tau_{1}} \, \Big| S_{\tau_{2}} S_{\tau_{1}}^{-1}e^{-\int_{\tau_{1}}^{\tau_{2}} \left(r-\rho_{L} \mathds{1}_{(0,L)}(S_{s}) - \rho_{H} \mathds{1}_{(H,\infty)}(S_{s}) \right) ds } -1 \Big| \,\bigg]   + K \big( 1 - e^{-\lambda^{\star} (t-u)} \big)  \nonumber \\
& \leq x \max \big\{ 1,\, e^{\Phi_{X}(1) T} \big\} \bigg( \mathbb{E}^{\mathbb{Q}}\Big[ \big| e^{X_{\tau_{2}-\tau_{1}} + \lambda^{\star} (t-u)} -1 \big| \,\Big] + \mathbb{E}^{\mathbb{Q}}\Big[ \big| e^{X_{\tau_{2}-\tau_{1}} - \lambda^{\star} (t-u)} -1 \big| \,\Big] \bigg) + K \big( 1 - e^{-\lambda^{\star} (t-u)} \big), 
\end{align}
\noindent where $\lambda^{\star} := r-\rho_{L} - \rho_{H}$. Therefore, since we have that $\tau_{2} - \tau_{1} \rightarrow 0$, for $t-u \rightarrow 0$, we obtain, by means of the dominated convergence theorem, the convergence
\begin{equation}
\mathcal{DSC}^{\star}_{A}(t,x; K, \bm{\ell}, \bm{\rho}_{\bm{\ell}} ) - \mathcal{DSC}^{\star}_{A}(u,x; K, \bm{\ell}, \bm{\rho}_{\bm{\ell}} ) \rightarrow 0, \hspace{1.5em} \mbox{whenever} \; \; t-u \rightarrow 0.
\end{equation}
\noindent This finally shows that the function $\mathcal{T} \mapsto \mathcal{DSC}^{\star}_{A}(\mathcal{T},x; K, \bm{\ell}, \bm{\rho}_{\bm{\ell}} )$ is, for any parameters $x,K,\bm{\ell}$, and $\bm{\rho}_{\bm{\ell}}$, uniformly continuous over $[0,T]$ and the proof of the initial claim is complete. \vspace{1em} \\
\noindent To prove that $\mathcal{DSC}_{A}^{\star}(\cdot)$ satisfies the Cauchy-type problem (\ref{GSCAmePIDE1}), (\ref{GSCAmePIDE2}), we consider again, for any initial value $z = (\mathbf{t},x) \in [0,T] \times [0,\infty)$, the (strong) Markov process $(Z_{t})_{t \in [0,\mathbf{t}]}$ defined via $Z_{t} := (\mathbf{t} -t , \bar{S}_{t})$, $\bar{S}_{0} = x$, and make use of the fact that
\begin{equation}
\mathcal{DSC}^{\star}_{A}(\mathcal{T},x; K, \bm{\ell}, \bm{\rho}_{\bm{\ell}} ) = V_{A}\big((\mathcal{T},x)\big),
\label{RARAppputa}
\end{equation}
\noindent where $V_{A}(\cdot)$ is defined, under the measure $\mathbb{Q}_{z}^{Z}$ having initial distribution $Z_{0} = z$, by
\begin{align}
V_{A}(z) : = \mathbb{E}^{\mathbb{Q}^{Z}}_{z} \big[ G(Z_{\tau_{\mathcal{D}_{s}}}) \big], \hspace{1.5em}  G(z)  := (x -K)^{+},
\end{align}
\noindent and $\tau_{\mathcal{D}_{s}}$ refers to the optimal stopping time defined according to (\ref{OTime}). Since $\tau_{\mathcal{D}_{s}} \leq T$ and the stopping region $\mathcal{D}_{s}$ is a closed set in the domain $[0,T] \times [0,\infty)$,\footnote{This directly follows from Representation (\ref{Sregion}) and the continuity of $(\mathcal{T},x) \mapsto \mathcal{DSC}^{\star}_{A}(\mathcal{T},x;K,\bm{\ell}, \bm{\rho}_{\bm{\ell}})$ on $[0,T] \times [0,\infty)$ for any $K,\bm{\ell}$, and $\bm{\rho}_{\bm{\ell}} $.}~this leads via standard arguments based on the strong Markov property of $(Z_{t})_{t \in [0,\mathbf{t}]}$ (cf.~\cite{pe06}) to the following problem
\begin{align}
\mathcal{A}_{Z} V_{A}(z) & = 0, \hspace{3em} \mbox{on} \,\,  \mathcal{D}_{c} , \\
V_{A}(z) & = G(z), \hspace{1.5em} \mbox{on} \,\,  \mathcal{D}_{s},
\end{align}
\noindent and finally allows to recover the required equations (\ref{GSCAmePIDE1}) and (\ref{GSCAmePIDE2}) by means of Relations (\ref{RARAppputa}) and (\ref{IgEnE}).
\end{proof}

\begin{proof}[\bf Proof of Proposition \ref{prop3}]
\noindent To start, we note that the strong Markov property of the process $(\bar{S}_{t})_{t \in [0,T]}$ together with the optimality of the stopping time $\tau_{\mathcal{D}_{s}}$ defined, for any (fixed) $\mathcal{T} \in [0,T]$, according to (\ref{OTime}) imply that the diffusion and jump contributions to the early exercise premium of the geometric double barrier step call, $\mathcal{E}_{\mathcal{DSC}}^{0,\star}(\cdot)$ and $\mathcal{E}_{\mathcal{DSC}}^{\mathcal{J},\star}(\cdot)$ respectively, can be written in the form 
\begin{align}
\mathcal{E}_{\mathcal{DSC}}^{0,\star}(\mathcal{T},x; K, \bm{\ell}, \bm{\rho}_{\bm{\ell}} ) & = \mathbb{E}_{x}^{\mathbb{Q}} \left[ \Big( \left(\bar{S}_{\tau_{\mathcal{D}_{s}}} - K \right)^{+} - \mathbb{E}_{\bar{S}_{\tau_{\mathcal{D}_{s}}}}^{\mathbb{Q}} \left[ (\bar{S}_{\mathcal{T}-\tau_{\mathcal{D}_{s}}} - K )^{+} \right] \Big) \, \mathds{1}_{ \partial \mathcal{D}_{s}} \big( (\mathcal{T}-\tau_{\mathcal{D}_{s}}, \bar{S}_{\tau_{\mathcal{D}_{s}}}) \big) \right] \nonumber \\
& = \mathbb{E}_{x}^{\mathbb{Q}} \bigg[ \Big( \left(\bar{S}_{\tau_{\mathcal{D}_{s}}} - K \right)^{+} - \mathcal{DSC}^{\star}_{E}(\mathcal{T}-\tau_{\mathcal{D}_{s}}, \bar{S}_{\tau_{\mathcal{D}_{s}}} ; K, \bm{\ell}, \bm{\rho}_{\bm{\ell}} ) \Big) \, \mathds{1}_{\partial \mathcal{D}_{s}} \big( (\mathcal{T}-\tau_{\mathcal{D}_{s}}, \bar{S}_{\tau_{\mathcal{D}_{s}}}) \big) \bigg], \\
\mathcal{E}_{\mathcal{DSC}}^{\mathcal{J},\star}(\mathcal{T},x; K, \bm{\ell}, \bm{\rho}_{\bm{\ell}} ) & = \mathbb{E}_{x}^{\mathbb{Q}} \left[ \Big( \left(\bar{S}_{\tau_{\mathcal{D}_{s}}} - K \right)^{+} - \mathbb{E}_{\bar{S}_{\tau_{\mathcal{D}_{s}}}}^{\mathbb{Q}} \left[ (\bar{S}_{\mathcal{T}-\tau_{\mathcal{D}_{s}}} - K )^{+} \right] \Big) \, \mathds{1}_{\mathcal{D}_{s}^{\circ} } \big( (\mathcal{T}-\tau_{\mathcal{D}_{s}}, \bar{S}_{\tau_{\mathcal{D}_{s}}}) \big) \right] \nonumber \\
& = \mathbb{E}_{x}^{\mathbb{Q}} \bigg[ \Big( \left(\bar{S}_{\tau_{\mathcal{D}_{s}}} - K \right)^{+} - \mathcal{DSC}^{\star}_{E}(\mathcal{T}-\tau_{\mathcal{D}_{s}}, \bar{S}_{\tau_{\mathcal{D}_{s}}} ; K, \bm{\ell}, \bm{\rho}_{\bm{\ell}} ) \Big) \, \mathds{1}_{\mathcal{D}_{s}^{\circ}} \big( (\mathcal{T}-\tau_{\mathcal{D}_{s}}, \bar{S}_{\tau_{\mathcal{D}_{s}}}) \big) \bigg].
\end{align}
\noindent Therefore, to prove that $\mathcal{E}_{\mathcal{DSC}}^{0,\star}(\cdot)$ and $\mathcal{E}_{\mathcal{DSC}}^{\mathcal{J},\star}(\cdot)$ satisfy Problem (\ref{GSCAmeEEPPIDE1})-(\ref{GSCAmeEEPPIDE1-2}) and (\ref{GSCAmeEEPPIDE2})-(\ref{GSCAmeEEPPIDE2-2}) respectively, we consider again, for any initial value $z = (\mathbf{t},x) \in [0,T] \times [0,\infty)$, the (strong) Markov process $(Z_{t})_{t \in [0,\mathbf{t}]}$ defined via $Z_{t} := (\mathbf{t} -t , \bar{S}_{t})$, $\bar{S}_{0} = x$, and make use of the fact that
\begin{align}
\mathcal{E}_{\mathcal{DSC}}^{0,\star}(\mathcal{T},x; K, \bm{\ell}, \bm{\rho}_{\bm{\ell}} ) = V_{\mathcal{E}}^{0} \big((\mathcal{T},x)\big), \hspace{1.2em} \mbox{and} \hspace{1.5em}  \mathcal{E}_{\mathcal{DSC}}^{\mathcal{J},\star}(\mathcal{T},x; K, \bm{\ell}, \bm{\rho}_{\bm{\ell}} ) = V_{\mathcal{E}}^{\mathcal{J}} \big((\mathcal{T},x)\big),
\label{RARAputa}
\end{align}
\noindent where $V_{\mathcal{E}}^{0}(\cdot)$ and $V_{\mathcal{E}}^{\mathcal{J}}(\cdot)$ are defined, under the measure $\mathbb{Q}_{z}^{Z}$ having initial distribution $Z_{0} = z = (\mathbf{t},x)$, by
\begin{align}
V_{\mathcal{E}}^{0}(z) : = \mathbb{E}^{\mathbb{Q}^{Z}}_{z} \big[ G_{0}(Z_{\tau_{\mathcal{D}_{s}}}) \big], & \hspace{1.5em}  G_{0}\big((\mathbf{t},x)\big)  := \big( (x -K)^{+} - \mathcal{DSC}^{\star}_{E}(\mathbf{t}, x ; K, \bm{\ell}, \bm{\rho}_{\bm{\ell}} ) \big) \, \mathds{1}_{\partial \mathcal{D}_{s}} ( (\mathbf{t},x) ), \\
 V_{\mathcal{E}}^{\mathcal{J}}(z) : = \mathbb{E}^{\mathbb{Q}^{Z}}_{z} \big[ G_{\mathcal{J}}(Z_{\tau_{\mathcal{D}_{s}}}) \big], & \hspace{1.5em}  G_{\mathcal{J}}\big((\mathbf{t},x)\big)  := \big( (x -K)^{+} - \mathcal{DSC}^{\star}_{E}(\mathbf{t}, x ; K, \bm{\ell}, \bm{\rho}_{\bm{\ell}} ) \big) \, \mathds{1}_{ \mathcal{D}_{s}^{\circ}} ((\mathbf{t},x)).
\end{align}
As earlier, since $\tau_{\mathcal{D}_{s}} \leq T$ and the stopping region $\mathcal{D}_{s}$ is a closed set in the domain $[0,T] \times [0,\infty)$, this leads via standard arguments based on the strong Markov property of $(Z_{t})_{t \in [0,\mathbf{t}]}$ (cf.~\cite{pe06}) to the following problems
\begin{align}
\mathcal{A}_{Z} V_{\mathcal{E}}^{0}(z) & = 0, \hspace{3.5em} \mbox{on} \,\,  \mathcal{D}_{c} , \\
V_{\mathcal{E}}^{0}(z) & = G_{0}(z), \hspace{1.5em} \mbox{on} \,\,  \mathcal{D}_{s},
\end{align}
\noindent and
\begin{align}
\mathcal{A}_{Z} V_{\mathcal{E}}^{\mathcal{J}}(z) & = 0, \hspace{3.8em} \mbox{on} \,\,  \mathcal{D}_{c} , \\
V_{\mathcal{E}}^{\mathcal{J}}(z) & = G_{\mathcal{J}}(z), \hspace{1.5em} \mbox{on} \,\,  \mathcal{D}_{s},
\end{align}
\noindent and finally allows to recover the required equations (\ref{GSCAmeEEPPIDE1})-(\ref{GSCAmeEEPPIDE1-2}) and (\ref{GSCAmeEEPPIDE2})-(\ref{GSCAmeEEPPIDE2-2}) by means of Relations (\ref{RARAputa}) and (\ref{IgEnE}).
\end{proof}

\begin{proof}[\bf Proof of Proposition \ref{prop4}]
\noindent We start the proof of Proposition \ref{prop4} by noting that the continuity of the function \mbox{$x \mapsto \widehat{\mathcal{DSC}^{\star}_{E}}(\vartheta,x;K,\bm{\ell}, \bm{\rho}_{\bm{\ell}})$} on $[0,\infty)$ directly follows from (\ref{LCTMREuro1}) and the continuity of $x \mapsto \mathcal{DSC}^{\star}_{E}(\mathcal{T},x; K, \bm{\ell}, \bm{\rho}_{\bm{\ell}} )$ for $\mathcal{T}, K, \bm{\ell}$ and $\bm{\rho}_{\bm{\ell}}$, by means of the dominated convergence theorem.\footnote{Recall that we have assumed the integrability of the underlying price process $(S_{t})_{t \geq 0}$.}~Therefore, we only need to establish that $\widehat{\mathcal{DSC}^{\star}_{E}}(\cdot)$ solves Equation (\ref{MRGSCEuOIDE1}) on $ (0,\infty)$ with initial condition (\ref{MRGSCEuOIDE2}). To this end, we first recall that the (independent) exponentially distributed random time $\mathcal{T}_{\vartheta}$ can be viewed as the (first) jump time of a corresponding Poisson process $(N_{t})_{t \geq 0}$ with intensity $\vartheta > 0$. Hence, for a fixed $\vartheta > 0 $, we consider the process $(Z_{t})_{t \geq 0}$ defined, for any initial value $z = (n,x) \in \mathbb{N}_{0} \times [0,\infty)$, via $Z_{t} := (n + N_{t} , \bar{S}_{t})$, $\bar{S}_{0} = x$, and note that it is a strong Markov process with state domain $\mathcal{D} := \mathbb{N}_{0} \times [0,\infty)$. Additionally, $\widehat{\mathcal{DSC}_{E}^{\star}}(\cdot)$ can be re-expressed, for $\vartheta, K,\bm{\ell}$ and $\bm{\rho}_{\bm{\ell}}$, as
\begin{equation}
\widehat{\mathcal{DSC}^{\star}_{E}}(\vartheta,x; K, \bm{\ell}, \bm{\rho}_{\bm{\ell}} ) = \widehat{V_{E}}\big((0,x)\big),
\label{MRIMeq}
\end{equation}
\noindent where the value function $\widehat{V_{E}}(\cdot)$ has the following representation under the measure $\mathbb{Q}_{z}^{Z}$ having initial distribution $Z_{0} = z$:
\begin{align}
\widehat{V_{E}}(z) : = \mathbb{E}^{\mathbb{Q}^{Z}}_{z} \big[ G(Z_{\tau_{\mathcal{S}}}) \big], \hspace{1.5em}  G(z)  := (x -K)^{+},
\end{align}
\noindent and $\tau_{\mathcal{S}} := \inf \{ t \geq 0: Z_{t} \in \mathcal{S} \}$, $\mathcal{S} := \big(\mathbb{N} \times [0,\infty) \big) \cup \big( \mathbb{N}_{0} \times \{0 \} \big)$, is a $\mathbb{Q}_{z}^{Z}$-almost surely finite stopping time for any $z = (n,x) \in \mathcal{D}$.\footnote{The finiteness of this stopping time directly follows from the finiteness of the first moment of any exponential distribution.}~Furthermore, the stopping region $\mathcal{S}$ forms (under an appropriate product-metric) a closed set in $\mathcal{D}$.\footnote{We note that several choices of a product-metric on $\mathcal{D}$ give the closedness of the set $\mathcal{S}$. In particular, one may choose on $\mathbb{N}_{0}$ the following metric
$$ d_{\mathbb{N}_{0}}(m,n) := \left \{ \begin{array}{lc}
1 + |2^{-m} - 2^{-n} |, & m\neq n, \\
0, & m =n,
\end{array} \right. $$
\noindent and consider the product-metric on $\mathcal{D}$ obtained by combining $d_{\mathbb{N}_{0}}(\cdot,\cdot)$ on $\mathbb{N}_{0}$ with the Euclidean metric on $[0,\infty)$. \label{footnoteMETRIC}}~Therefore, standard arguments based on the strong Markov property of the process $(Z_{t})_{t \geq 0}$ (cf.~\cite{pe06}) imply that $\widehat{V_{E}}(\cdot)$ satisfies the following problem
\begin{align}
\mathcal{A}_{Z} \widehat{V_{E}}(z) & = 0, \hspace{2em} \mbox{on} \,\,  \mathcal{D} \setminus \mathcal{S}, \\
\widehat{V_{E}}(z) & = G(z), \hspace{1.5em} \mbox{on} \,\,  \mathcal{S},
\end{align}
\noindent where $\mathcal{A}_{Z}$ denotes the infinitesimal generator of the process $(Z_{t})_{t \geq 0}$. To complete the proof, we note that the infinitesimal generator $\mathcal{A}_{Z}$ can be re-expressed (for any suitable function $V:\mathcal{D} \rightarrow \mathbb{R}$) as
\begin{align}
\mathcal{A}_{Z} V\big((n,x)\big) & = \mathcal{A}_{N}^{n} V\big((n,x) \big) + \mathcal{A}_{\bar{S}}^{x} V \big((n,x) \big) \nonumber \\
& = \vartheta \left( V\big( (n+1,x) \big) - V\big( (n,x) \big) \right) + \mathcal{A}_{S}^{x} V\big( (n,x) \big) - \lambda(x) V\big( (n,x) \big) , \label{MRIgEn}
\end{align}
\noindent where $\mathcal{A}_{N}$ denotes the infinitesimal generator of the Poisson process $(N_{t})_{ t \geq 0}$ and the notation $\mathcal{A}_{N}^{n}$, $\mathcal{A}_{\bar{S}}^{x}$, and $\mathcal{A}_{S}^{x}$ is used to indicate that the generators are applied to $n$ and $x$, respectively. Therefore, recovering $\widehat{\mathcal{DSC}^{\star}_{E}}(\cdot)$ via (\ref{MRIMeq}) while noting Relation (\ref{MRIgEn}) and the fact that for any $x \in [0,\infty)$ we have
\begin{equation}
\widehat{V_{E}}\big((1,x)\big) = G\big((1,x)\big) = (x-1)^{+} 
\end{equation}
\noindent finally completes the proof.
\end{proof}

\begin{proof}[\bf Proof of Proposition \ref{prop5}]
\noindent First, we note that the discussion preceding Proposition \ref{prop5} implies that the optimal stopping problem (\ref{NEWproBL}) can be re-expressed, under the measure $\mathbb{Q}_{z}^{Z}$ having initial distribution $Z_{0} = z \in \mathcal{D}$, as
\begin{align}
\widehat{V_{A}}(z)  = \mathbb{E}^{\mathbb{Q}^{Z}}_{z} \left[ G\Big(Z_{\tau_{\widehat{\mathcal{D}_{s}^{Gen.}}}}^{\mathcal{S}_{J}}\Big) \right],
\label{PROOFNEWproBL}
\end{align}
\noindent where $\tau_{\widehat{\mathcal{D}_{s}^{Gen.}}}$ is defined as in (\ref{NeWWOTime}) and $G(z)  := (x - K)^{+}$, for $z \in \mathcal{D}$. Additionally, the finiteness of the first moment of the exponential distribution for any $\vartheta >0$ implies that this stopping time is $\mathbb{Q}_{z}^{Z}$-almost surely finite for any $z \in \mathcal{D}$, and combining this property with the closedness\footnote{As earlier, this property can be obtained under the product-metric considered in Footnote~\ref{footnoteMETRIC}.}~of the stopping domain $\widehat{\mathcal{D}_{s}^{Gen.}}$ gives~(cf.~\cite{pe06}) that $\widehat{V_{A}}(\cdot)$ satisfies the following problem
\begin{align}
\mathcal{A}_{Z} \widehat{V_{A}}(z) & = 0, \hspace{3em} \mbox{on} \,\,  \widehat{\mathcal{D}_{c}^{Gen.}} , \\
\widehat{V_{A}}(z) & = G(z), \hspace{1.5em} \mbox{on} \,\,  \widehat{\mathcal{D}_{s}^{Gen.}}.
\end{align}
\noindent Consequently, recovering $\widehat{\mathcal{DSC}^{\star}_{A}}(\cdot)$ by means of Relation (\ref{MRAmerIMeq}) while noting Identity (\ref{MRIgEn}) and the fact that
\begin{equation}
\widehat{\mathcal{D}_{s}^{Gen.}} = \mathcal{S}_{J} \cup \big( \{ 0 \} \times \widehat{\mathcal{D}}_{\vartheta,s} \big)
\label{ImpStopRegio}                    
\end{equation}
\noindent and
\begin{equation}
\widehat{V_{A}}\big((1,x)\big) = G\big((1,x)\big) = (x-1)^{+} 
\end{equation}
\noindent finally gives the required Equations (\ref{MRGSCAmerOIDE1}) and (\ref{MRGSCAmerOIDE2}). \vspace{1em} \\
\noindent The continuity of $x \mapsto \widehat{\mathcal{DSC}^{\star}_{A}}(\vartheta,x;K,\bm{\ell}, \bm{\rho}_{\bm{\ell}})$ on $[0,\infty)$ for $\vartheta, K, \bm{\ell},$ and $\bm{\rho}_{\bm{\ell}}$ is an easy consequence of the continuity of $x \mapsto (x-K)^{+}$ and the dominated convergence theorem. This concludes the proof of the proposition.
\end{proof}

\begin{proof}[\bf Proof of Proposition \ref{prop6}]
\noindent Following the ideas outlined in the previous proofs, we re-consider, for any $\vartheta > 0 $ and initial value $z = (n,x) \in \mathbb{N}_{0} \times [0,\infty)$, the process $(Z_{t})_{t \geq 0}$ defined on the state domain $\mathcal{D} := \mathbb{N}_{0} \times [0,\infty)$ via $Z_{t} := (n + N_{t} , \bar{S}_{t})$, $\bar{S}_{0} = x$, as well as its stopped version, $(Z_{t}^{\mathcal{S}_{J}})_{t \geq 0}$, defined according to (\ref{STOPPEDprocDEF}) and note that the diffusion and jump contributions to the maturity-randomized early exercise premium of the geometric double barrier step call, $\widehat{\mathcal{E}_{\mathcal{DSC}}^{0,\star}}(\cdot)$ and $\widehat{\mathcal{E}_{\mathcal{DSC}}^{\mathcal{J},\star}}(\cdot)$ respectively, can be re-expressed, using these processes, in the form 
\begin{align}
\widehat{\mathcal{E}_{\mathcal{DSC}}^{0,\star}}(\vartheta,x; K, \bm{\ell}, \bm{\rho}_{\bm{\ell}} ) = \widehat{V_{\mathcal{E}}^{0}} \big((0,x)\big), \hspace{1.2em} \mbox{and} \hspace{1.5em}  \widehat{\mathcal{E}_{\mathcal{DSC}}^{\mathcal{J},\star}}(\vartheta,x; K, \bm{\ell}, \bm{\rho}_{\bm{\ell}} ) = \widehat{V_{\mathcal{E}}^{\mathcal{J}}} \big((0,x)\big),
\label{MRRARAputaBup}
\end{align}
\noindent where $\widehat{V_{\mathcal{E}}^{0}}(\cdot)$ and $\widehat{V_{\mathcal{E}}^{\mathcal{J}}}(\cdot)$ are defined, under the measure $\mathbb{Q}_{z}^{Z}$ having initial distribution $Z_{0} = z$, by
\begin{align}
\widehat{V_{\mathcal{E}}^{0}}(z) : = \mathbb{E}^{\mathbb{Q}^{Z}}_{z} \left[ \widehat{G_{0}}\Big(Z_{\tau_{\widehat{\mathcal{D}_{s}^{Gen.}}}}^{\mathcal{S}_{J}}\Big) \right], & \hspace{1.5em}  \widehat{G_{0}}\big((n,x)\big)  := \big( (x -K)^{+} - \widehat{V_{E}}\big((n,x)\big) \big) \, \mathds{1}_{\partial \widehat{\mathcal{D}_{s}^{Gen.}}}\big((n,x)\big), \\
\widehat{V_{\mathcal{E}}^{\mathcal{J}}}(z) : = \mathbb{E}^{\mathbb{Q}^{Z}}_{z} \left[ \widehat{G_{\mathcal{J}}}\Big(Z_{\tau_{\widehat{\mathcal{D}_{s}^{Gen.}}}}^{\mathcal{S}_{J}}\Big) \right], & \hspace{1.5em}  \widehat{G_{\mathcal{J}}}\big((n,x)\big)  := \big( (x -K)^{+} - \widehat{V_{E}}\big((n,x)\big) \big) \, \mathds{1}_{ \big(\widehat{\mathcal{D}_{s}^{Gen.}}\big)^{\circ}}\big((n,x)\big).
\end{align}
As earlier, the $\mathbb{Q}_{z}^{Z}$-almost sure finiteness of the stopping time $\tau_{\widehat{\mathcal{D}_{s}^{Gen.}}}$ for any $z \in \mathcal{D}$ and the closedness\footnote{e.g.~under the product-metric considered in Footnote~\ref{footnoteMETRIC}.}~of the stopping domain $\widehat{\mathcal{D}_{s}^{Gen.}}$ lead via standard arguments (cf.~\cite{pe06}) to the following problems
\begin{align}
\mathcal{A}_{Z} \widehat{V_{\mathcal{E}}^{0}}(z) & = 0, \hspace{3.5em} \mbox{on} \,\,  \widehat{\mathcal{D}_{c}^{Gen.}} , \\
\widehat{V_{\mathcal{E}}^{0}}(z) & = \widehat{G_{0}}(z), \hspace{1.5em} \mbox{on} \,\,  \widehat{\mathcal{D}_{s}^{Gen.}},
\end{align}
\noindent and
\begin{align}
\mathcal{A}_{Z} \widehat{V_{\mathcal{E}}^{\mathcal{J}}}(z) & = 0, \hspace{3.8em} \mbox{on} \,\,  \widehat{\mathcal{D}_{c}^{Gen.}}, \\
\widehat{V_{\mathcal{E}}^{\mathcal{J}}}(z) & = \widehat{G_{\mathcal{J}}}(z), \hspace{1.5em} \mbox{on} \,\,  \widehat{\mathcal{D}_{s}^{Gen.}}.
\end{align}
\noindent Finally, in view of (\ref{ImpStopRegio}), it is clear that\footnote{e.g.~under the product-metric considered in Footnote~\ref{footnoteMETRIC}.}
\begin{equation}
\partial \widehat{\mathcal{D}_{s}^{Gen.}} = \mathcal{S}_{J} \cup \big( \{0 \} \times \partial \widehat{\mathcal{D}}_{\vartheta,s} \big), \hspace{1,5em} \mbox{and} \hspace{1.7em} \big(\widehat{\mathcal{D}_{s}^{Gen.}}\big)^{\circ} = \mathcal{S}_{J} \cup \big( \{0 \} \times \widehat{\mathcal{D}}_{\vartheta,s}^{\circ} \big),
\end{equation}
\noindent so that
\begin{equation}
\widehat{V_{\mathcal{E}}^{0}}\big((1,x)\big) =  \widehat{G_{0}}\big((1,x)\big) = 0 = \widehat{G_{\mathcal{J}}}\big((1,x)\big) = \widehat{V_{\mathcal{E}}^{\mathcal{J}}}\big((1,x)\big).
\end{equation}
\noindent Therefore, combining these properties with Relations (\ref{MRRARAputaBup}) and (\ref{MRIgEn}) finally allows to recover the required equations (\ref{GSCAmeEEPOIDE1})-(\ref{GSCAmeEEPOIDE1-2}) and (\ref{GSCAmeEEPOIDE2})-(\ref{GSCAmeEEPOIDE2-2}). This completes the proof.
\end{proof}

\subsection*{Appendix B: Proofs - Section \ref{SEC3}}
\begin{proof}[\bf Proof of Proposition \ref{PropEurHEJD}]
\noindent For simplicity, we rewrite the price of the maturity-randomized European-type down-and-out step contract $\widehat{\mathcal{DOSC}^{\star}_{E}}(\cdot)$ as function of the log-price $\bm{x} := \log(x)$ and the log-strike $\bm{k} := \log(K)$ via $\overline{\mathcal{DOSC}}_{E}^{\star}(\cdot)$, i.e.~we rely on the following relation
\begin{equation}
\overline{\mathcal{DOSC}}_{E}^{\star}(\vartheta,\bm{x}; \bm{k}, L, \rho_{L}) := \widehat{\mathcal{DOSC}^{\star}_{E}}(\vartheta,e^{\bm{x}}; e^{\bm{k}}, L,\rho_{L}).
\end{equation}
\noindent This transforms (\ref{MRGSCEuOIDE1}) into the following equation 
\begin{align}
\vartheta (e^{\bm{x}} - e^{\bm{k}})^{+} + \mathcal{A}_{X} \overline{\mathcal{DOSC}}_{E}^{\star}(\vartheta,\bm{x}; \bm{k}, L, \rho_{L})  - \big( r + \vartheta - \rho_{L} \mathds{1}_{(0,L)}(e^{\bm{x}}) \big) \overline{\mathcal{DOSC}}_{E}^{\star}(\vartheta,\bm{x}; \bm{k}, L, \rho_{L}) = 0,
\end{align}
\noindent with $\mathcal{A}_{X}$ denoting the infinitesimal generator of $(X_{t})_{t \geq 0}$, i.e.
\begin{equation}
\mathcal{A}_{X} V(\bm{x}) := \frac{1}{2} \sigma^{2}_{X} \partial_{\bm{x}}^{2} V(\bm{x}) + \Big( r -\delta -\lambda \zeta - \frac{1}{2}\sigma_{X}^{2} \Big) \partial_{\bm{x}} V(\bm{x}) +  \lambda \int \limits_{\mathbb{R}}  \big( V(\bm{x}+\bm{y}) - V(\bm{x}) \big) f_{J_{1}}(\bm{y})d\bm{y}.
\label{NewGen}
\end{equation}
\noindent Equivalently, this can be written in the following system of three equations
\begin{align}
\mathcal{A}_{X} \overline{\mathcal{DOSC}}_{E}^{\star}(\vartheta,\bm{x}; \bm{k}, L, \rho_{L})  - ( r + \vartheta - \rho_{L}) & \overline{\mathcal{DOSC}}_{E}^{\star}(\vartheta,\bm{x}; \bm{k}, L, \rho_{L}) = 0, \hspace{1.5em} \mbox{for} \; -\infty < \bm{x} < \ell^{\ast}, \label{EQfirst1}\\
\mathcal{A}_{X} \overline{\mathcal{DOSC}}_{E}^{\star}(\vartheta,\bm{x}; \bm{k}, L, \rho_{L})  - ( r + \vartheta) & \overline{\mathcal{DOSC}}_{E}^{\star}(\vartheta,\bm{x}; \bm{k}, L, \rho_{L})  = 0, \hspace{2em} \mbox{for} \; \ell^{\ast} \leq \bm{x} \leq \bm{k},  \label{EQfirst2}\\
\mathcal{A}_{X} \overline{\mathcal{DOSC}}_{E}^{\star}(\vartheta,\bm{x}; \bm{k}, L, \rho_{L})  - ( r + \vartheta) \overline{\mathcal{DOSC}}_{E}^{\star}&(\vartheta,\bm{x}; \bm{k}, L, \rho_{L})  = \vartheta(e^{\bm{k}} - e^{\bm{x}}), \hspace{1.5em} \mbox{for} \; \bm{k} < \bm{x} < \infty,  \label{EQlast1}
\end{align}
\noindent where we have set $\ell^{\ast} := \log(L)$. Combining the arguments provided in \cite{ck11} (cf.~also \cite{lv17}, \cite{cv18}, \cite{fmv19}) with the fact that
$$ P_{1}(x) := \vartheta\left( \frac{e^{\bm{x}}}{\delta + \vartheta} -\frac{e^{\bm{k}}}{r + \vartheta} \right) $$
\noindent is a particular solution to (\ref{EQlast1}) implies that the general solution to (\ref{EQfirst1})-(\ref{EQlast1}) takes the following form
\begin{equation}
\overline{\mathcal{DOSC}}_{E}^{\star}(\vartheta,\bm{x}; \bm{k}, L, \rho_{L}) = \left \{ \begin{array}{lc}
\sum \limits_{s=1}^{m+1} A_{s}^{+} e^{\beta_{s,(r+\vartheta-\rho_{L})} \cdot (\bm{x}-\ell^{\ast})}, & -\infty < \bm{x} < \ell^{\ast} ,\\
\sum \limits_{s=1}^{m+1} B_{s}^{+} e^{\beta_{s,(r+\vartheta)} \cdot (\bm{x}-\ell^{\ast})} + \sum \limits_{u=1}^{n+1} B_{u}^{-} e^{\gamma_{u,(r+\vartheta)} \cdot (\bm{x}-\bm{k})}, & \ell^{\ast} \leq \bm{x} \leq \bm{k}, \\
 \sum \limits_{u=1}^{n+1} C_{u}^{-} e^{\gamma_{u,(r+\vartheta)} \cdot (\bm{x}-\bm{k})} + \vartheta \left( \frac{e^{\bm{x}}}{\delta + \vartheta} -\frac{e^{\bm{k}}}{r + \vartheta} \right) , & \bm{k} < \bm{x} < \infty ,
\end{array} \right.
\end{equation}
\noindent where the coefficients $(A_{s}^{+})_{s=1,\ldots,m+1}$, $(B_{s}^{+})_{s=1,\ldots,m+1}$, $(B_{u}^{-})_{u=1,\ldots,n+1}$ and $(C_{u}^{-})_{u=1,\ldots,n+1}$ are subsequently determined by analyzing the solution under the respective equations and in the different regions. This is done next.\vspace{1em} \\
\noindent \underline{\bf STEP 1: $-\infty < \bm{x} < \ell^{\ast}$.} \\
\noindent To start we derive that 
\begin{align}
& \int \limits_{\mathbb{R}} \overline{\mathcal{DOSC}}_{E}^{\star}(\vartheta,\bm{x}+y; \bm{k}, L, \rho_{L}) f_{J_{1}}(y) dy \nonumber \\
& = \sum \limits_{s=1}^{m+1} \sum \limits_{j=1}^{n} q_{j} \eta_{j} e^{-\eta_{j} \bm{x}} A_{s}^{+} \int \limits_{-\infty}^{\bm{x}} e^{\eta_{j} z} e^{\beta_{s,(r+\vartheta-\rho_{L})} \cdot (z-\ell^{\ast})} \, dz  + \sum \limits_{s=1}^{m+1} \sum \limits_{i=1}^{m} p_{i} \xi_{i} e^{\xi_{i} \bm{x}} A_{s}^{+} \int \limits_{\bm{x}}^{\ell^{\ast}} e^{-\xi_{i} z} e^{\beta_{s,(r+\vartheta-\rho_{L})} \cdot (z-\ell^{\ast})} \, dz \nonumber \\
& \hspace{0.3em} + \sum \limits_{s=1}^{m+1} \sum \limits_{i=1}^{m} p_{i} \xi_{i} e^{\xi_{i} \bm{x}} B_{s}^{+} \int \limits_{\ell^{\ast}}^{\bm{k}} e^{-\xi_{i} z} e^{\beta_{s,(r+\vartheta)} \cdot (z-\ell^{\ast})} \, dz + \sum \limits_{u=1}^{n+1} \sum \limits_{i=1}^{m} p_{i} \xi_{i} e^{\xi_{i} \bm{x}} B_{u}^{-} \int \limits_{\ell^{\ast}}^{\bm{k}} e^{-\xi_{i} z} e^{\gamma_{u,(r+\vartheta)} \cdot (z-\bm{k})} \, dz \nonumber \\
& \hspace{0.3em} + \sum \limits_{u=1}^{n+1} \sum \limits_{i=1}^{m} p_{i} \xi_{i} e^{\xi_{i} \bm{x}} C_{u}^{-} \int \limits_{\bm{k}}^{\infty} e^{-\xi_{i}z} e^{\gamma_{u,(r+\vartheta)} \cdot (z-\bm{k})} \, dz + \sum \limits_{i=1}^{m} p_{i} \xi_{i} e^{\xi_{i} \bm{x}} \Bigg( \frac{\vartheta}{\delta + \vartheta} \int \limits_{\bm{k}}^{\infty} e^{-(\xi_{i}-1) \cdot z} \, dz  -  \frac{\vartheta e^{\bm{k}}}{r + \vartheta} \int \limits_{\bm{k}}^{\infty} e^{-\xi_{i} z} \, dz  \Bigg).
\label{INTEalgebra}
\end{align}
\noindent After some algebra, Equation (\ref{EQfirst1}) can be transformed to obtain
\begin{equation}
\sum \limits_{s=1}^{m+1} A_{s}^{+} e^{\beta_{s,(r+\vartheta-\rho_{L})} \cdot (\bm{x}-\ell^{\ast})} \underbrace{\left( \Phi_{X}\big(\beta_{s,(r+\vartheta-\rho_{L})} \big) - (r+\vartheta-\rho_{L}) \right)}_{=0} + \lambda \sum \limits_{i=1}^{m} p_{i} \xi_{i} e^{\xi_{i} (\bm{x}-\ell^{\ast})} \, \mathcal{R}_{i}^{1}(\vartheta;\bm{k},L, \rho_{L}) = 0,
\end{equation}
\noindent where, for $i=1,\ldots,m$,
\begin{align}
\mathcal{R}_{i}^{1}(\vartheta;\bm{k},L, \rho_{L}) & := - \sum \limits_{s=1}^{m+1} \Bigg( A_{s}^{+} \frac{1}{\xi_{i} - \beta_{s,(r+\vartheta-\rho_{L})}} - B_{s}^{+} \frac{1 - e^{(\beta_{s,(r+\vartheta)}-\xi_{i}) (\bm{k}-\ell^{\ast})}}{ {\xi_{i} - \beta_{s,(r+\vartheta)}}} \Bigg) \nonumber \\
& \hspace{3.5em} + \sum \limits_{u=1}^{n+1} \Bigg( B_{u}^{-} \frac{e^{-\gamma_{u,(r+\vartheta)} (\bm{k} -\ell^{\ast})} - e^{-\xi_{i} (\bm{k}-\ell^{\ast})}}{ {\xi_{i} - \gamma_{u,(r+\vartheta)}}} + C_{u}^{-} \frac{e^{-\xi_{i} (\bm{k}-\ell^{\ast})}}{ {\xi_{i} - \gamma_{u,(r+\vartheta)}}} \Bigg) \nonumber \\
& \hspace{3.5em} + \frac{\vartheta e^{\bm{k}} e^{-\xi_{i}(\bm{k}-\ell^{\ast})}}{(\xi_{i}-1)(\delta+\vartheta)} -\frac{\vartheta e^{\bm{k}} e^{-\xi_{i}(\bm{k}-\ell^{\ast})}}{\xi_{i}(r+\vartheta)} .
\end{align}
\noindent Therefore, since the parameters $\xi_{1}, \ldots, \xi_{m}$ are all different from each other, we conclude that
\begin{equation}
\mathcal{R}_{i}^{1}(\vartheta; \bm{k}, L, \rho_{L}) = 0, \hspace{1.5em} \mbox{for}\; \; i =1, \ldots, m .
\end{equation}
\noindent \underline{\bf STEP 2: $\ell^{\ast} \leq \bm{x} \leq \bm{k}$.} \\
\noindent Combining similar arguments to the ones used in (\ref{INTEalgebra}) with Equation (\ref{EQfirst2}), we derive that
\begin{align}
& \sum \limits_{s=1}^{m+1} B_{s}^{+} e^{\beta_{s,(r+\vartheta)} \cdot (\bm{x}-\ell^{\ast})} \underbrace{\left( \Phi_{X}\big(\beta_{s,(r+\vartheta)} \big) - (r+\vartheta) \right)}_{=0} + \sum \limits_{u=1}^{n+1} B_{u}^{-} e^{\gamma_{u,(r+\vartheta)} \cdot (\bm{x}-\bm{k})} \underbrace{\left( \Phi_{X}\big(\gamma_{u,(r+\vartheta)} \big) - (r+\vartheta) \right)}_{=0} \nonumber \hspace{10em} \\
& \hspace{10em} +\lambda \Bigg( \sum \limits_{i=1}^{m} p_{i} \xi_{i} e^{\xi_{i} (\bm{x}-k)} \, \mathcal{R}_{i}^{2,+}(\vartheta;\bm{k},L, \rho_{L}) +  \sum \limits_{j=1}^{n} q_{j} \eta_{j} e^{-\eta_{j} (\bm{x}-\ell^{\ast})} \, \mathcal{R}_{j}^{2,-}(\vartheta;\bm{k},L, \rho_{L}) \Bigg) = 0,
\end{align}
\noindent where, for $i=1,\ldots,m$ and $j=1, \ldots,n$, 
\begin{align}
\mathcal{R}_{i}^{2,+}(\vartheta;\bm{k},L, \rho_{L}) & := - \sum \limits_{s=1}^{m+1} B_{s}^{+} \frac{e^{\beta_{s,(r+\vartheta)} (\bm{k}-\ell^{\ast}) }}{\xi_{i} - \beta_{s,(r+\vartheta)}} - \sum \limits_{u=1}^{n+1} \big( B_{u}^{-} - C_{u}^{-} \big) \frac{1}{\xi_{i} - \gamma_{u,(r+\vartheta)}} + \frac{\vartheta e^{\bm{k}}}{(\xi_{i}-1)(\delta+\vartheta)} -\frac{\vartheta e^{\bm{k}}}{\xi_{i}(r+\vartheta)}, \\
\mathcal{R}_{j}^{2,-}(\vartheta;\bm{k},L, \rho_{L}) & := \sum \limits_{s=1}^{m+1} \Bigg( A_{s}^{+} \frac{1}{\eta_{j} + \beta_{s,(r+\vartheta-\rho_{L})}} - B_{s}^{+} \frac{1}{\eta_{j} + \beta_{s,(r+\vartheta)}} \Bigg) - \sum \limits_{u=1}^{n+1} B_{u}^{-} \frac{e^{- \gamma_{u,(r+\vartheta)} (\bm{k} - \ell^{\ast})}}{\eta_{j} + \gamma_{u,(r+\vartheta)}}.
\end{align}
\noindent Hence, since the parameters $\xi_{1}, \ldots, \xi_{m}, \eta_{1}, \ldots, \eta_{n}$ are all different from each other, we conclude that
\begin{align}
\mathcal{R}_{i}^{2,+}(\vartheta; \bm{k}, L, \rho_{L}) & = 0, \hspace{1.5em} \mbox{for}\; \; i =1, \ldots, m,  \\
\mathcal{R}_{j}^{2,-}(\vartheta; \bm{k}, L, \rho_{L}) & = 0, \hspace{1.5em} \mbox{for}\; \; j =1, \ldots, n .
\end{align} 
\noindent \underline{\bf STEP 3: $\bm{k} < \bm{x} < \infty$.} \\
\noindent Following the line of the arguments used in STEP 1 and STEP 2, we rewrite Equation (\ref{EQlast1}) as
\begin{align}
& \sum \limits_{u=1}^{n+1} C_{u}^{-} e^{\gamma_{u,(r+\vartheta)} \cdot (\bm{x}-\bm{k})} \underbrace{\left( \Phi_{X}\big(\gamma_{u,(r+\vartheta)} \big) - (r+\vartheta) \right)}_{=0}  \nonumber \\
&  + \underbrace{\Bigg( \frac{\vartheta e^{\bm{x}}}{\delta + \vartheta} \Big( \frac{\sigma^{2}}{2} + b_{X} + \lambda \zeta -(r + \vartheta) \Big) + (r+\vartheta) \frac{\vartheta e^{\bm{k}}}{r+\vartheta} - \vartheta e^{\bm{k}} + \vartheta e^{\bm{x}} \Bigg)}_{=0} + \lambda \sum \limits_{j=1}^{n} q_{j} \eta_{j} e^{-\eta_{j} (\bm{x}-\bm{k})} \, \mathcal{R}_{j}^{3}(\vartheta;\bm{k},L, \rho_{L}) = 0,
\end{align}
\noindent where, for $j=1,\ldots,n$,
\begin{align}
\mathcal{R}_{j}^{3}(\vartheta;\bm{k},L, \rho_{L}) & := \sum \limits_{s=1}^{m+1} \Bigg( A_{s}^{+} \frac{e^{-\eta_{j} (\bm{k}-\ell^{\ast}) }}{\eta_{j} + \beta_{s,(r+\vartheta-\rho_{L})}} + B_{s}^{+} \frac{e^{\beta_{s,(r+\vartheta)}) (\bm{k}-\ell^{\ast})} - e^{-\eta_{j}(\bm{k}-\ell^{\ast})}}{ \eta_{j} + \beta_{s,(r+\vartheta)}} \Bigg) \nonumber \\
& \hspace{3.5em} + \sum \limits_{u=1}^{n+1} \Bigg( B_{u}^{-} \frac{1 - e^{-(\eta_{j}+\gamma_{u,(r+\vartheta)}) (\bm{k}-\ell^{\ast})}}{ {\eta_{j} + \gamma_{u,(r+\vartheta)}}} - C_{u}^{-} \frac{1}{\eta_{j} + \gamma_{u,(r+\vartheta)}} \Bigg) \nonumber \\
& \hspace{3.5em} - \frac{\vartheta e^{\bm{k}}}{(\eta_{j}+1)(\delta+\vartheta)} +\frac{\vartheta e^{\bm{k}}}{\eta_{j}(r+\vartheta)} .
\end{align}
\noindent Therefore, since the parameters $\eta_{1}, \ldots, \eta_{n}$ are all different from each other, we conclude that
\begin{equation}
\mathcal{R}_{j}^{3}(\vartheta; \bm{k}, L, \rho_{L}) = 0, \hspace{1.5em} \mbox{for}\; \; j =1, \ldots, n .
\end{equation}
\noindent \underline{\bf STEP 4:} \\
To close the system of equations, we impose smooth-fit conditions and obtain the following four identities:
\begin{align}
& \hspace{5em} \sum \limits_{s=1}^{m+1} \big( A_{s}^{+} - B_{s}^{+} \big) - \sum \limits_{u=1}^{n+1} B_{u}^{-} e^{-\gamma_{u,(r+\vartheta)} \cdot (\bm{k} -\ell^{\ast})}  = 0, \\
& \hspace{4em}\sum \limits_{s=1}^{m+1} B_{s}^{+} e^{\beta_{s,(r+\vartheta)} \cdot (\bm{k}-\ell^{\ast})} + \sum \limits_{u=1}^{n+1} \big( B_{u}^{-} - C_{u}^{-} \big)  = \frac{\vartheta e^{\bm{k}}}{\delta + \vartheta} - \frac{\vartheta e^{\bm{k}}}{r + \vartheta} , \\
& \sum \limits_{s=1}^{m+1} \Big( A_{s}^{+} \beta_{s,(r+\vartheta-\rho_{L})} - B_{s}^{+} \beta_{s,(r+\vartheta)} \Big) - \sum \limits_{u=1}^{n+1} B_{u}^{-} \gamma_{u,(r+\vartheta)} e^{-\gamma_{u,(r+\vartheta)} \cdot (\bm{k}-\ell^{\ast})}  = 0, \label{smoothpasting1} \\
& \hspace{0.9em} \sum \limits_{s=1}^{m+1} B_{s}^{+} \beta_{s,(r+\vartheta)} e^{\beta_{s,(r+\vartheta)} \cdot (\bm{k}-\ell^{\ast})} + \sum \limits_{u=1}^{n+1} \Big( B_{u}^{-} \gamma_{u,(r+\vartheta)} - C_{u}^{-} \gamma_{u,(r+\vartheta)} \Big)  = \frac{\vartheta e^{\bm{k}}}{\delta + \vartheta} . \label{smoothpasting2}
\end{align}
\noindent Although we do not further comment on the appropriateness of the smooth-fit conditions (\ref{smoothpasting1}) and (\ref{smoothpasting2}), we emphasize that smooth-pasting is very natural under hyper-exponential jump-diffusion markets and refer for similar results, e.g.~to~\cite{ccw10}, \cite{xy13}, \cite{lz16}. \vspace{1em} \\
\noindent \underline{\bf STEP 5:} \\
\noindent To finalize our derivations, we combine the results obtained in STEP 1 - STEP 4. This leads to the following system of equations
\begin{equation}
\mathbf{Q_{E}} \mathbf{v} = \mathbf{q_{E}},
\label{AppendixBSysEq}
\end{equation}
\noindent where $\mathbf{v} := (A_{1}^{+}, \ldots, A_{m+1}^{+},B_{1}^{+}, \ldots, B_{m+1}^{+}, B_{1}^{-}, \ldots, B_{n+1}^{-}, C_{1}^{-}, \ldots, C_{n+1}^{-})^{\intercal}$. Here, $\mathbf{q_{E}} = (\mathbf{q_{E}^{1}}, \ldots, \mathbf{q_{E}^{8}})^{\intercal}$ is a $(2m+2n+4)$-dimensional column vector, whose elements are defined in the following way:
\begin{itemize} \setlength \itemsep{-0.2em}
\item[$i)$] $\mathbf{q_{E}^{1}}$ and $\mathbf{q_{E}^{2}}$ are $1 \times m$ vectors given by 
\begin{align}
\big(\mathbf{q_{E}^{1}}\big)_{i} := \frac{\vartheta e^{\bm{k}} e^{-\xi_{i}(\bm{k}-\ell^{\ast})}}{\xi_{i}(r+\vartheta)} & - \frac{\vartheta e^{\bm{k}} e^{-\xi_{i}(\bm{k}-\ell^{\ast})}}{(\xi_{i}-1)(\delta+\vartheta)}  , \hspace{1.5em} i = 1, \ldots, m , \\
\big(\mathbf{q_{E}^{2}}\big)_{i}   := \frac{\vartheta e^{k}}{\xi_{i}(r+\vartheta)} - & \frac{\vartheta e^{k}}{(\xi_i -1)(\delta + \vartheta)}, \hspace{1.5em} i = 1, \ldots, m, 
\end{align}
\item[$ii)$] $\mathbf{q_{E}^{3}}$ and $\mathbf{q_{E}^{4}}$ are $1 \times n$ vectors given by
\begin{align}
\big(\mathbf{q_{E}^{3}}\big)_{j} & := 0, \hspace{1.5em} j = 1, \ldots, n, \\
\big(\mathbf{q_{E}^{4}}\big)_{j} := - \frac{\vartheta e^{\bm{k}}}{\eta_{j}(r+\vartheta)} & + \frac{\vartheta e^{\bm{k}}}{(\eta_{j}+1)(\delta+\vartheta)}, \hspace{1.5em} j = 1, \ldots, n , 
\end{align}
\item[$iii)$] $\mathbf{q_{E}^{5}}$, $\mathbf{q_{E}^{6}}$, $\mathbf{q_{E}^{7}}$ and $\mathbf{q_{E}^{8}}$ are real values given by
\begin{align}
\mathbf{q_{E}^{5}} := 0, \hspace{1.5em} \mathbf{q_{E}^{6}}:= \frac{\vartheta e^{\bm{k}}}{\delta + \vartheta} - \frac{\vartheta e^{\bm{k}}}{r + \vartheta}, \hspace{1.5em} \mathbf{q_{E}^{7}} := 0, \hspace{1.5em} \mathbf{q_{E}^{8}} := \frac{\vartheta e^{\bm{k}}}{\delta + \vartheta}.
\end{align}                     
\end{itemize}
\noindent Finally, $\mathbf{Q_{E}}$ is a $(2m+2n+4)$-dimensional square matrix
\begin{equation}
\mathbf{Q_{E}} = \left( \begin{array}{cccc}
\mathbf{Q_{E}^{11}} & \mathbf{Q_{E}^{12}} & \mathbf{Q_{E}^{13}} & \mathbf{Q_{E}^{14}} \\
\mathbf{Q_{E}^{21}} & \mathbf{Q_{E}^{22}} & \mathbf{Q_{E}^{23}} & \mathbf{Q_{E}^{24}}  \\
\vdots & \vdots & \vdots & \vdots \\
\mathbf{Q_{E}^{81}} & \mathbf{Q_{E}^{82}} & \mathbf{Q_{E}^{83}} & \mathbf{Q_{E}^{84}}
\end{array} \right)
\end{equation}
\noindent that is defined in the following way: 
\begin{itemize} \setlength \itemsep{-0.2em}
\item[$i)$] $\mathbf{Q_{E}^{11}}$, $\mathbf{Q_{E}^{12}}$ and $\mathbf{Q_{E}^{13}}$, $\mathbf{Q_{E}^{14}}$ are respectively $m\times (m+1)$ and $m \times (n+1)$ matrices given, for $i=1,\ldots,m$, $s=1,\ldots, m+1$, and $u=1,\ldots, n+1$, by
\begin{align}
&  \big( \mathbf{Q_{E}^{11}} \big)_{is}  := - \frac{1}{\xi_{i} - \beta_{s,(r+\vartheta-\rho_{L})}}, \hspace{1.5em} \big( \mathbf{Q_{E}^{12}} \big)_{is}  := \frac{1 - e^{(\beta_{s,(r+\vartheta)}-\xi_{i}) (\bm{k}-\ell^{\ast})}}{ {\xi_{i} - \beta_{s,(r+\vartheta)}}},  \\
& \big( \mathbf{Q_{E}^{13}} \big)_{iu}  := \frac{e^{-\gamma_{u,(r+\vartheta)} (\bm{k} -\ell^{\ast})} - e^{-\xi_{i} (\bm{k}-\ell^{\ast})}}{ {\xi_{i} - \gamma_{u,(r+\vartheta)}}} , \hspace{1.5em} \big( \mathbf{Q_{E}^{14}} \big)_{iu}  := \frac{e^{-\xi_{i} (\bm{k}-\ell^{\ast})}}{ {\xi_{i} - \gamma_{u,(r+\vartheta)}}} ,
\end{align}
\item[$ii)$] $\mathbf{Q_{E}^{21}}$, $\mathbf{Q_{E}^{22}}$ and $\mathbf{Q_{E}^{23}}$, $\mathbf{Q_{E}^{24}}$ are respectively $m\times (m+1)$ and $m \times (n+1)$ matrices given, for $i=1,\ldots,m$, $s=1,\ldots, m+1$, and $u=1,\ldots, n+1$, by
\begin{align}
&  \hspace{2em} \big( \mathbf{Q_{E}^{21}} \big)_{is}  := 0, \hspace{1.5em} \big( \mathbf{Q_{E}^{22}} \big)_{is}  := -\frac{e^{\beta_{s,(r+\vartheta)} (\bm{k}-\ell^{\ast}) }}{\xi_{i} - \beta_{s,(r+\vartheta)}},  \\
&  \big( \mathbf{Q_{E}^{23}} \big)_{iu}  := -\frac{1}{\xi_{i} - \gamma_{u,(r+\vartheta)}} , \hspace{1.5em} \big( \mathbf{Q_{E}^{24}} \big)_{iu}  := -\big( \mathbf{Q_{E}^{23}} \big)_{iu} ,
\end{align}
\item[$iii)$] $\mathbf{Q_{E}^{31}}$, $\mathbf{Q_{E}^{32}}$ and $\mathbf{Q_{E}^{33}}$, $\mathbf{Q_{E}^{34}}$ are respectively $n\times (m+1)$ and $n \times (n+1)$ matrices given, for $j=1,\ldots,n$, $s=1,\ldots, m+1$, and $u=1,\ldots, n+1$, by
\begin{align}
&  \big( \mathbf{Q_{E}^{31}} \big)_{js}  := \frac{1}{\eta_{j} + \beta_{s,(r+\vartheta-\rho_{L})}}, \hspace{1.5em} \big( \mathbf{Q_{E}^{32}} \big)_{js}  := -\frac{1}{\eta_{j} + \beta_{s,(r+\vartheta)}},  \\
&  \hspace{3em} \big( \mathbf{Q_{E}^{33}} \big)_{ju}  := -\frac{e^{- \gamma_{u,(r+\vartheta)} (\bm{k} - \ell^{\ast})}}{\eta_{j} + \gamma_{u,(r+\vartheta)}} , \hspace{1.5em} \big( \mathbf{Q_{E}^{34}} \big)_{ju}  := 0 ,
\end{align}
\item[$iv)$] $\mathbf{Q_{E}^{41}}$, $\mathbf{Q_{E}^{42}}$ and $\mathbf{Q_{E}^{43}}$, $\mathbf{Q_{E}^{44}}$ are respectively $n\times (m+1)$ and $n \times (n+1)$ matrices given, for $j=1,\ldots,n$, $s=1,\ldots, m+1$, and $u=1,\ldots, n+1$, by
\begin{align}
&  \big( \mathbf{Q_{E}^{41}} \big)_{js}  := \frac{e^{-\eta_{j} (\bm{k}-\ell^{\ast}) }}{\eta_{j} + \beta_{s,(r+\vartheta-\rho_{L})}}, \hspace{1.5em} \big( \mathbf{Q_{E}^{42}} \big)_{js}  := \frac{e^{\beta_{s,(r+\vartheta)}) (\bm{k}-\ell^{\ast})} - e^{-\eta_{j}(\bm{k}-\ell^{\ast})}}{ \eta_{j} + \beta_{s,(r+\vartheta)}},  \\
&  \hspace{1em} \big( \mathbf{Q_{E}^{43}} \big)_{ju}  := \frac{1 - e^{-(\eta_{j}+\gamma_{u,(r+\vartheta)}) (\bm{k}-\ell^{\ast})}}{ {\eta_{j} + \gamma_{u,(r+\vartheta)}}} , \hspace{1.5em} \big( \mathbf{Q_{E}^{44}} \big)_{ju}  := -\frac{1}{\eta_{j} + \gamma_{u,(r+\vartheta)}} ,
\end{align}
\item[$v)$] $\mathbf{Q_{E}^{51}}$, $\mathbf{Q_{E}^{52}}$ and $\mathbf{Q_{E}^{53}}$, $\mathbf{Q_{E}^{54}}$ are respectively $1\times (m+1)$ and $1 \times (n+1)$ vectors given, for $s=1,\ldots, m+1$, and $u=1,\ldots, n+1$, by
\begin{align}
&  \big( \mathbf{Q_{E}^{51}} \big)_{s}  := 1, \hspace{1.5em} \big( \mathbf{Q_{E}^{52}} \big)_{s}  := -1, \hspace{1.5em} \big( \mathbf{Q_{E}^{53}} \big)_{u} := -e^{-\gamma_{u,(r+\vartheta)} \cdot (\bm{k}-\ell^{\ast})} , \hspace{1.5em}  \big( \mathbf{Q_{E}^{54}} \big)_{u} := 0 ,
\end{align}
\item[$vi)$] $\mathbf{Q_{E}^{61}}$, $\mathbf{Q_{E}^{62}}$ and $\mathbf{Q_{E}^{63}}$, $\mathbf{Q_{E}^{64}}$ are respectively $1\times (m+1)$ and $1 \times (n+1)$ vectors given, for $s=1,\ldots, m+1$, and $u=1,\ldots, n+1$, by
\begin{align}
&  \big( \mathbf{Q_{E}^{61}} \big)_{s}  := 0, \hspace{1.5em} \big( \mathbf{Q_{E}^{62}} \big)_{s}  := e^{\beta_{s,(r+\vartheta)} \cdot (\bm{k}-\ell^{\ast})}, \hspace{1.5em} \big( \mathbf{Q_{E}^{63}} \big)_{u} := 1 , \hspace{1.5em}  \big( \mathbf{Q_{E}^{64}} \big)_{u} := -1 ,
\end{align}
\item[$vii)$] $\mathbf{Q_{E}^{71}}$, $\mathbf{Q_{E}^{72}}$ and $\mathbf{Q_{E}^{73}}$, $\mathbf{Q_{E}^{74}}$ are respectively $1\times (m+1)$ and $1 \times (n+1)$ vectors given, for $s=1,\ldots, m+1$, and $u=1,\ldots, n+1$, by
\begin{align}
&  \hspace{1.2em} \big( \mathbf{Q_{E}^{71}} \big)_{s}  := \beta_{s,(r+\vartheta-\rho_{L})} , \hspace{1.5em} \big( \mathbf{Q_{E}^{72}} \big)_{s}  := -\beta_{s,(r+\vartheta)} , \\
&  \big( \mathbf{Q_{E}^{73}} \big)_{u} := - \gamma_{u,(r+\vartheta)} e^{-\gamma_{u,(r+\vartheta)} \cdot (\bm{k}-\ell^{\ast})} , \hspace{1.5em}  \big( \mathbf{Q_{E}^{74}} \big)_{u} := 0 ,
\end{align}
\item[$viii)$] $\mathbf{Q_{E}^{81}}$, $\mathbf{Q_{E}^{82}}$ and $\mathbf{Q_{E}^{83}}$, $\mathbf{Q_{E}^{84}}$ are respectively $1\times (m+1)$ and $1 \times (n+1)$ vectors given, for $s=1,\ldots, m+1$, and $u=1,\ldots, n+1$, by
\begin{align}
&  \big( \mathbf{Q_{E}^{81}} \big)_{s}  := 0, \hspace{1.5em} \big( \mathbf{Q_{E}^{82}} \big)_{s}  := \beta_{s,(r+\vartheta)} e^{\beta_{s,(r+\vartheta)} \cdot (\bm{k}-\ell^{\ast})}, \\
&  \hspace{1.2em} \big( \mathbf{Q_{E}^{83}} \big)_{u} := \gamma_{u,(r+\vartheta)} , \hspace{1.5em}  \big( \mathbf{Q_{E}^{84}} \big)_{u} := -\big( \mathbf{Q_{E}^{83}} \big)_{u} .
\end{align}
\end{itemize}
\end{proof}    

\begin{proof}[\bf Proof of Proposition \ref{PropAmerHEJD}]    
\noindent We proceed as in the proof of Proposition \ref{PropEurHEJD}, i.e.~we first rewrite the value of the maturity-randomized early exercise premium $\widehat{\mathcal{E}_{\mathcal{DOSC}}^{\star}}(\cdot)$ as function of the log-price $\bm{x} := \log(x)$ and of the log-strike $\bm{k} := \log(K)$ via $\overline{\mathcal{E}_{\mathcal{DOSC}}^{\star}}(\cdot)$ by relying on the following relation
\begin{equation}
\overline{\mathcal{E}_{\mathcal{DOSC}}^{\star}}(\vartheta,\bm{x}; \bm{k}, L, \rho_{L}) := \widehat{\mathcal{E}_{\mathcal{DOSC}}^{\star}}(\vartheta,e^{\bm{x}}; e^{\bm{k}}, L,\rho_{L}).
\end{equation}
\noindent This transforms (\ref{MRGSCAmerOIDE1}), (\ref{MRGSCAmerOIDE2}) into the following problem
\begin{align}
& \mathcal{A}_{X} \overline{\mathcal{E}_{\mathcal{DOSC}}^{\star}}(\vartheta,\bm{x}; \bm{k}, L, \rho_{L}) - \big( r + \vartheta - \rho_{L} \mathds{1}_{(0,L)}(e^{\bm{k}}) \big) \overline{\mathcal{E}_{\mathcal{DOSC}}^{\star}}(\vartheta,\bm{x}; \bm{k}, L, \rho_{L}) = 0, \hspace{1.5em} \mbox{for} \; -\infty < \bm{x} < b^{\ast} ,\\
& \hspace{7.4em} \overline{\mathcal{E}_{\mathcal{DOSC}}^{\star}}(\vartheta,\bm{x}; \bm{k}, L, \rho_{L}) = e^{\bm{x}} - e^{\bm{k}} - \overline{\mathcal{DOSC}}_{E}^{\star}(\vartheta,\bm{x}; \bm{k}, L, \rho_{L}), \hspace{2em} \mbox{for} \; b^{\ast} \leq \bm{x} < \infty,
\end{align}
\noindent with $\mathcal{A}_{X}$ given as in (\ref{NewGen}) and $b^{\ast}$ denoting the log early exercise boundary, i.e.~$b^{\ast} := \log(\mathfrak{b}_{s})$.
\noindent Equivalently, this can be written in the following system of three equations
\begin{align}
\mathcal{A}_{X} \overline{\mathcal{E}_{\mathcal{DOSC}}^{\star}}(\vartheta,\bm{x}; \bm{k}, L, \rho_{L})  - ( r + \vartheta - \rho_{L}) & \overline{\mathcal{E}_{\mathcal{DOSC}}^{\star}}(\vartheta,\bm{x}; \bm{k}, L, \rho_{L}) = 0, \hspace{2.1em} \mbox{for} \; -\infty < \bm{x} < \ell^{\ast}, \label{EQfirst21}\\
\mathcal{A}_{X} \overline{\mathcal{E}_{\mathcal{DOSC}}^{\star}}(\vartheta,\bm{x}; \bm{k}, L, \rho_{L})  - ( r + \vartheta) & \overline{\mathcal{E}_{\mathcal{DOSC}}^{\star}}(\vartheta,\bm{x}; \bm{k}, L, \rho_{L})  = 0, \hspace{2.7em} \mbox{for} \; \ell^{\ast} \leq \bm{x} < b^{\ast},  \label{EQfirst22}\\
\overline{\mathcal{E}_{\mathcal{DOSC}}^{\star}}(\vartheta,\bm{x}; \bm{k}, L, \rho_{L}) = e^{\bm{x}} - e^{\bm{k}} & -\overline{\mathcal{DOSC}}_{E}^{\star}(\vartheta,\bm{x}; \bm{k}, L, \rho_{L}), \hspace{2.6em} \mbox{for} \; b^{\ast} \leq \bm{x} < \infty,  \label{EQlast21}
\end{align}
\noindent where we have set $\ell^{\ast} := \log(L)$. Consequently, following the arguments in the proof of Proposition \ref{PropEurHEJD}, we obtain that the general solution to (\ref{EQfirst21})-(\ref{EQlast21}) takes the following form
\begin{equation}
\overline{\mathcal{E}_{\mathcal{DOSC}}^{\star}}(\vartheta,\bm{x}; \bm{k}, L, \rho_{L}) = \left \{ \begin{array}{lc}
\sum \limits_{s=1}^{m+1} D_{s}^{+} e^{\beta_{s,(r+\vartheta-\rho_{L})} \cdot (\bm{x}-\ell^{\ast})}, & -\infty < \bm{x} < \ell^{\ast} ,\\
\sum \limits_{s=1}^{m+1} F_{s}^{+} e^{\beta_{s,(r+\vartheta)} \cdot (\bm{x}-\ell^{\ast})} + \sum \limits_{u=1}^{n+1} F_{u}^{-} e^{\gamma_{u,(r+\vartheta)} \cdot (\bm{x}-b^{\ast})}, & \ell^{\ast} \leq \bm{x} < b^{\ast}, \\
e^{\bm{x}} - e^{\bm{k}} -\overline{\mathcal{DOSC}}_{E}^{\star}(\vartheta,\bm{x}; \bm{k}, L, \rho_{L}), & b^{\ast} \leq \bm{x} < \infty,
\end{array} \right.
\end{equation}
\noindent where the coefficients $(D_{s}^{+})_{s=1,\ldots,m+1}$, $(F_{s}^{+})_{s=1,\ldots,m+1}$, $(F_{u}^{-})_{u=1,\ldots,n+1}$ and the free-boundary $b^{\ast}$ are subsequently determined by analyzing the solution under the respective equations and in the different regions. Here, following the steps outlined in the proof of Proposition \ref{PropEurHEJD}, we arrive at the following system of equation
\begin{equation}
\mathbf{Q_{A}}\mathbf{w} = \mathbf{q_{A}},
\label{AppendixBAmerSysEq}
\end{equation}
\noindent where $\mathbf{w} := (D_{1}^{+}, \ldots, D_{m+1}^{+}, F_{1}^{+}, \ldots, F_{m+1}^{+}, F_{1}^{-}, \ldots, F_{n+1}^{-})^{\intercal}$. The vector $\mathbf{q_{A}} = (\mathbf{q_{A}^{1}}, \ldots, \mathbf{q_{A}^{6}})^{\intercal}$ is a $(2m+n+3)$-dimensional column vector, whose elements are defined in the following way:
\begin{itemize} \setlength \itemsep{-0.2em}
\item[$i)$] $\mathbf{q_{A}^{1}}$ and $\mathbf{q_{A}^{2}}$ are $1 \times m$ vectors given by 
\begin{align}
\big(\mathbf{q_{A}^{1}}\big)_{i} := \sum \limits_{u=1}^{n+1} C_{u}^{-} \frac{e^{- \xi_{i} (b^{\ast}-\ell^{\ast})} e^{\gamma_{u,(r+\vartheta)} \cdot (b^{\ast} - \bm{k})}}{\xi_{i} - \gamma_{u,(r+\vartheta)}} & + \frac{r e^{\bm{k}} e^{-\xi_{i} (b^{\ast}-\ell^{\ast})}}{\xi_{i}(r+\vartheta)}  - \frac{\delta e^{b^{\ast}} e^{-\xi_{i} (b^{\ast}-\ell^{\ast})}}{(\xi_{i}-1)(\delta+\vartheta)}, \hspace{1.5em} i = 1, \ldots, m , \\
\big(\mathbf{q_{A}^{2}}\big)_{i} := \sum \limits_{u=1}^{n+1} C_{u}^{-} \frac{ e^{\gamma_{u,(r+\vartheta)} \cdot (b^{\ast} - \bm{k})}}{\xi_{i} - \gamma_{u,(r+\vartheta)}} & + \frac{r e^{\bm{k}}}{\xi_{i}(r+\vartheta)}  - \frac{\delta e^{b^{\ast}} }{(\xi_{i}-1)(\delta+\vartheta)}, \hspace{1.5em} i = 1, \ldots, m ,
\end{align}
\item[$ii)$] $\mathbf{q_{A}^{3}}$ is a $1 \times n$ vector given by $\big(\mathbf{q_{A}^{3}}\big)_{j} := 0, \;  j = 1, \ldots, n,$
\item[$iii)$] $\mathbf{q_{A}^{4}}$, $\mathbf{q_{A}^{5}}$, $\mathbf{q_{A}^{6}}$ are real values given by
\begin{align}
\mathbf{q_{A}^{4}} := 0, \hspace{1.5em} \mathbf{q_{A}^{5}}:= \frac{\delta e^{b^{\ast}}}{\delta + \vartheta} - \frac{r e^{\bm{k}}}{r + \vartheta} - \sum \limits_{u=1}^{n+1} C_{u}^{-} e^{\gamma_{u,(r+\vartheta)} \cdot (b^{\ast}-\bm{k})}, \hspace{1.5em} \mathbf{q_{A}^{6}} := 0.
\end{align}                     
\end{itemize}
\noindent Additionally, $\mathbf{Q_{A}}$ is a $(2m+n+3)$-dimensional square matrix
\begin{equation}
\mathbf{Q_{A}} = \left( \begin{array}{ccc}
\mathbf{Q_{A}^{11}} & \mathbf{Q_{A}^{12}} & \mathbf{Q_{A}^{13}}  \\
\mathbf{Q_{A}^{21}} & \mathbf{Q_{A}^{22}} & \mathbf{Q_{A}^{23}}   \\
\vdots & \vdots & \vdots \\
\mathbf{Q_{A}^{61}} & \mathbf{Q_{A}^{62}} & \mathbf{Q_{A}^{63}}
\end{array} \right)
\end{equation}
\noindent that is defined in the following way: 
\begin{itemize} \setlength \itemsep{-0.2em}
\item[$i)$] $\mathbf{Q_{A}^{11}}$, $\mathbf{Q_{A}^{12}}$ and $\mathbf{Q_{A}^{13}}$ are respectively $m\times (m+1)$ and $m \times (n+1)$ matrices given, for $i=1,\ldots,m$, $s=1,\ldots, m+1$, and $u=1,\ldots, n+1$, by
\begin{align}
&  \big( \mathbf{Q_{A}^{11}} \big)_{is}  := - \frac{1}{\xi_{i} - \beta_{s,(r+\vartheta-\rho_{L})}}, \hspace{1.5em} \big( \mathbf{Q_{A}^{12}} \big)_{is}  := \frac{1-e^{(\beta_{s,(r+\vartheta)}-\xi_{i}) (b^{\ast} -\ell^{\ast})}}{ \xi_{i} - \beta_{s,(r+\vartheta)}},  \\
& \hspace{5.5em} \big( \mathbf{Q_{A}^{13}} \big)_{iu}  := \frac{e^{-\gamma_{u,(r+\vartheta)} \cdot (b^{\ast}-\ell^{\ast})} - e^{-\xi_{i} (b^{\ast}-\ell^{\ast})}}{ \xi_{i} - \gamma_{u,(r+\vartheta)}} ,
\end{align}
\item[$ii)$] $\mathbf{Q_{A}^{21}}$, $\mathbf{Q_{A}^{22}}$ and $\mathbf{Q_{A}^{23}}$ are respectively $m\times (m+1)$ and $m \times (n+1)$ matrices given, for $i=1,\ldots,m$, $s=1,\ldots, m+1$, and $u=1,\ldots, n+1$, by
\begin{align}
&  \big( \mathbf{Q_{A}^{21}} \big)_{is}  := 0, \hspace{1.5em} \big( \mathbf{Q_{A}^{22}} \big)_{is}  := -\frac{e^{\beta_{s,(r+\vartheta)} \cdot (b^{\ast}-\ell^{\ast}) }}{\xi_{i} - \beta_{s,(r+\vartheta)}}, \hspace{1.5em} \big( \mathbf{Q_{A}^{23}} \big)_{iu}  := -\frac{1}{ \xi_{i} - \gamma_{u,(r+\vartheta)}} ,
\end{align}
\item[$iii)$] $\mathbf{Q_{A}^{31}}$, $\mathbf{Q_{A}^{32}}$ and $\mathbf{Q_{A}^{33}}$ are respectively $n\times (m+1)$ and $n \times (n+1)$ matrices given, for $j=1,\ldots,n$, $s=1,\ldots, m+1$, and $u=1,\ldots, n+1$, by
\begin{align}
&  \big( \mathbf{Q_{A}^{31}} \big)_{js}  := \frac{1}{\eta_{j} + \beta_{s,(r+\vartheta-\rho_{L})}}, \hspace{1.5em} \big( \mathbf{Q_{A}^{32}} \big)_{js}  := -\frac{1}{\eta_{j} + \beta_{s,(r+\vartheta)}},  \\
&  \hspace{6em} \big( \mathbf{Q_{A}^{33}} \big)_{ju}  := -\frac{e^{- \gamma_{u,(r+\vartheta)} \cdot (b^{\ast} -\ell^{\ast})}}{\eta_{j} + \gamma_{u,(r+\vartheta)}} , 
\end{align}
\item[$iv)$] $\mathbf{Q_{A}^{41}}$, $\mathbf{Q_{A}^{42}}$ and $\mathbf{Q_{A}^{43}}$  are respectively $1\times (m+1)$ and $1 \times (n+1)$ vectors given, for $s=1,\ldots, m+1$, and $u=1,\ldots, n+1$, by
\begin{align}
&  \big( \mathbf{Q_{A}^{41}} \big)_{s}  := 1, \hspace{1.5em} \big( \mathbf{Q_{A}^{42}} \big)_{s}  := -1, \hspace{1.5em} \big( \mathbf{Q_{A}^{43}} \big)_{u} := -e^{-\gamma_{u,(r+\vartheta)} \cdot (b^{\ast}-\ell^{\ast})} ,
\end{align}
\item[$v)$] $\mathbf{Q_{A}^{51}}$, $\mathbf{Q_{A}^{52}}$ and $\mathbf{Q_{A}^{53}}$ are respectively $1\times (m+1)$ and $1 \times (n+1)$ vectors given, for $s=1,\ldots, m+1$, and $u=1,\ldots, n+1$, by
\begin{align}
&  \big( \mathbf{Q_{A}^{51}} \big)_{s}  := 0, \hspace{1.5em} \big( \mathbf{Q_{A}^{52}} \big)_{s}  := e^{\beta_{s,(r+\vartheta)} \cdot (b^{\ast}-\ell^{\ast})}, \hspace{1.5em} \big( \mathbf{Q_{A}^{53}} \big)_{u} := 1 ,
\end{align}
\item[$vi)$] $\mathbf{Q_{A}^{61}}$, $\mathbf{Q_{A}^{62}}$ and $\mathbf{Q_{A}^{63}}$ are respectively $1\times (m+1)$ and $1 \times (n+1)$ vectors given, for $s=1,\ldots, m+1$, and $u=1,\ldots, n+1$, by
\begin{align}
&  \big( \mathbf{Q_{A}^{61}} \big)_{s}  := \beta_{s,(r+\vartheta-\rho_{L})} , \hspace{1.5em} \big( \mathbf{Q_{A}^{62}} \big)_{s}  := -\beta_{s,(r+\vartheta)} , \hspace{1.5em} \big( \mathbf{Q_{A}^{63}} \big)_{u} := - \gamma_{u,(r+\vartheta)} e^{-\gamma_{u,(r+\vartheta)} \cdot (b^{\ast}-\ell^{\ast})}.
\end{align}
\end{itemize}
\noindent Finally, the free-boundary $b^{\ast}$ can be recovered by combining (\ref{AppendixBAmerSysEq}) with the usual smooth-fit condition at the boundary level:\footnote{We emphasize that smooth-fit at the boundary $b^{\ast}$ can be proved using the same approach as the one outlined in \cite{Ma18}; See also \cite{pe06}, \cite{lm11}.}
\begin{equation}
\sum \limits_{s=1}^{m+1} F_{s}^{+} \beta_{s,(r+\vartheta)} e^{\beta_{s,(r+\vartheta)} \cdot (b^{\ast}-\ell^{\ast})} + \sum \limits_{u=1}^{n+1} F_{u}^{-} \gamma_{u,(r+\vartheta)} = \frac{\delta e^{b^{\ast}}}{\delta + \vartheta} - \sum \limits_{u=1}^{n+1} C_{u}^{-} \gamma_{u,(r+\vartheta)} e^{\gamma_{u,(r+\vartheta)} \cdot (b^{\ast}-\bm{k})}.
\label{SPuseful}
\end{equation}
\end{proof}                   

\begin{proof}[\bf Proof of Proposition \ref{PropEepHEJD}]    
\noindent To derive Representations (\ref{DiffEEP_MR1}) and (\ref{JumpEEP_MR1}), we mainly rely on the proof of Proposition~\ref{PropAmerHEJD}. As earlier, we write 
\begin{align}
\overline{\mathcal{E}_{\mathcal{DOSC}}^{0,\star}}(\vartheta,\bm{x}; \bm{k}, L, \rho_{L}) &:= \widehat{\mathcal{E}_{\mathcal{DOSC}}^{0,\star}}(\vartheta,e^{\bm{x}}; e^{\bm{k}}, L,\rho_{L}), \\
\overline{\mathcal{E}_{\mathcal{DOSC}}^{\mathcal{J},\star}}(\vartheta,\bm{x}; \bm{k}, L, \rho_{L}) & := \widehat{\mathcal{E}_{\mathcal{DOSC}}^{\mathcal{J},\star}}(\vartheta,e^{\bm{x}}; e^{\bm{k}}, L,\rho_{L}),
\end{align}
\noindent and obtain, by the same arguments as in the proof of Proposition \ref{PropAmerHEJD}, that $\overline{\mathcal{E}_{\mathcal{DOSC}}^{0,\star}}(\cdot)$ and $\overline{\mathcal{E}_{\mathcal{DOSC}}^{\mathcal{J},\star}}(\cdot)$ take the form
\begin{equation}
\overline{\mathcal{E}_{\mathcal{DOSC}}^{0,\star}}(\vartheta,\bm{x}; \bm{k}, L, \rho_{L}) = \left \{ \begin{array}{lc}
\sum \limits_{s=1}^{m+1} D_{s}^{0,+} e^{\beta_{s,(r+\vartheta-\rho_{L})} \cdot (\bm{x}-\ell^{\ast})}, & -\infty < \bm{x} < \ell^{\ast} ,\\
\sum \limits_{s=1}^{m+1} F_{s}^{0,+} e^{\beta_{s,(r+\vartheta)} \cdot (\bm{x}-\ell^{\ast})} + \sum \limits_{u=1}^{n+1} F_{u}^{0,-} e^{\gamma_{u,(r+\vartheta)} \cdot (\bm{x}-b^{\ast})}, & \ell^{\ast} \leq \bm{x} < b^{\ast}, \\
e^{\bm{x}} - e^{\bm{k}} -\overline{\mathcal{DOSC}}_{E}^{\star}(\vartheta,\bm{x}; \bm{k}, L, \rho_{L}), &  \bm{x} = b^{\ast}, \\
0, & b^{\ast} < \bm{x} < \infty,
\end{array} \right.
\end{equation}
\begin{equation}
\overline{\mathcal{E}_{\mathcal{DOSC}}^{\mathcal{J},\star}}(\vartheta,\bm{x}; \bm{k}, L, \rho_{L}) = \left \{ \begin{array}{lc}
\sum \limits_{s=1}^{m+1} D_{s}^{\mathcal{J},+} e^{\beta_{s,(r+\vartheta-\rho_{L})} \cdot (\bm{x}-\ell^{\ast})}, & -\infty < \bm{x} < \ell^{\ast} ,\\
\sum \limits_{s=1}^{m+1} F_{s}^{\mathcal{J},+} e^{\beta_{s,(r+\vartheta)} \cdot (\bm{x}-\ell^{\ast})} + \sum \limits_{u=1}^{n+1} F_{u}^{\mathcal{J},-} e^{\gamma_{u,(r+\vartheta)} \cdot (\bm{x}-b^{\ast})}, & \ell^{\ast} \leq \bm{x} < b^{\ast}, \\
0, &  \bm{x} = b^{\ast}, \\
e^{\bm{x}} - e^{\bm{k}} -\overline{\mathcal{DOSC}}_{E}^{\star}(\vartheta,\bm{x}; \bm{k}, L, \rho_{L}), & b^{\ast} < \bm{x} < \infty.
\end{array} \right.
\end{equation}
\noindent Here, analogous derivations to the ones in the proof of Proposition \ref{PropAmerHEJD} show that the vectors of coefficients 
$$ \mathbf{w_{0}} := (D_{1}^{0,+}, \ldots, D_{m+1}^{0,+}, F_{1}^{0,+}, \ldots, F_{m+1}^{0,+}, F_{1}^{0,-}, \ldots, F_{n+1}^{0,-})^{\intercal} $$
\noindent and
$$\mathbf{w_{J}} := (D_{1}^{\mathcal{J},+}, \ldots, D_{m+1}^{\mathcal{J},+}, F_{1}^{\mathcal{J},+}, \ldots, F_{m+1}^{\mathcal{J},+}, F_{1}^{\mathcal{J},-}, \ldots, F_{n+1}^{\mathcal{J},-})^{\intercal}$$ solve the following system of equations, respectively, 
\begin{equation}
\mathbf{Q_{A}}\mathbf{w_{0}} = \mathbf{q_{A,0}}, \hspace{1.5em} \mbox{and} \hspace{1.5em} \mathbf{Q_{A}}\mathbf{w_{J}} = \mathbf{q_{A,J}},
\label{AppendixBEepSysEq}
\end{equation}
\noindent where $\mathbf{q_{A,0}} = (\mathbf{q_{A,0}^{1}}, \ldots, \mathbf{q_{A,0}^{6}})^{\intercal}$ and $\mathbf{q_{A,J}} = (\mathbf{q_{A,J}^{1}}, \ldots, \mathbf{q_{A,J}^{6}})^{\intercal}$ are $(2m+n+3)$-dimensional column vectors, whose elements are defined by:
\begin{itemize} \setlength \itemsep{-0.2em}
\item[$i)$] $\mathbf{q_{A,0}^{1}}$ and $\mathbf{q_{A,0}^{2}}$ are $1 \times m$ vectors given by $\big(\mathbf{q_{A,0}^{1}}\big)_{i} := 0, \; \big(\mathbf{q_{A,0}^{2}}\big)_{i} := 0,  \; i= 1, \ldots, m,$

\item[$ii)$] $\mathbf{q_{A,0}^{3}}$ is a $1 \times n$ vector given by $\big(\mathbf{q_{A,0}^{3}}\big)_{j} := 0, \;  j = 1, \ldots, n,$
\item[$iii)$] $\mathbf{q_{A,0}^{4}}$, $\mathbf{q_{A,0}^{5}}$, $\mathbf{q_{A,0}^{6}}$ are real values given by
\begin{align}
\mathbf{q_{A,0}^{4}} := 0, \hspace{1.5em} \mathbf{q_{A,0}^{5}}:= \frac{\delta e^{b^{\ast}}}{\delta + \vartheta} - \frac{r e^{\bm{k}}}{r + \vartheta} - \sum \limits_{u=1}^{n+1} C_{u}^{-} e^{\gamma_{u,(r+\vartheta)} \cdot (b^{\ast}-\bm{k})}, \hspace{1.5em} \mathbf{q_{A,0}^{6}} := 0.
\end{align}                     
\end{itemize}
\noindent and
\begin{itemize} \setlength \itemsep{-0.2em}
\item[$i)$] $\mathbf{q_{A,J}^{1}}$ and $\mathbf{q_{A,J}^{2}}$ are $1 \times m$ vectors given by 
\begin{align}
\big(\mathbf{q_{A,J}^{1}}\big)_{i} := \sum \limits_{u=1}^{n+1} C_{u}^{-} \frac{e^{- \xi_{i} (b^{\ast}-\ell^{\ast})} e^{\gamma_{u,(r+\vartheta)} \cdot (b^{\ast} - \bm{k})}}{\xi_{i} - \gamma_{u,(r+\vartheta)}} & + \frac{r e^{\bm{k}} e^{-\xi_{i} (b^{\ast}-\ell^{\ast})}}{\xi_{i}(r+\vartheta)}  - \frac{\delta e^{b^{\ast}} e^{-\xi_{i} (b^{\ast}-\ell^{\ast})}}{(\xi_{i}-1)(\delta+\vartheta)}, \hspace{1.5em} i = 1, \ldots, m , \\
\big(\mathbf{q_{A,J}^{2}}\big)_{i} := \sum \limits_{u=1}^{n+1} C_{u}^{-} \frac{ e^{\gamma_{u,(r+\vartheta)} \cdot (b^{\ast} - \bm{k})}}{\xi_{i} - \gamma_{u,(r+\vartheta)}} & + \frac{r e^{\bm{k}}}{\xi_{i}(r+\vartheta)}  - \frac{\delta e^{b^{\ast}} }{(\xi_{i}-1)(\delta+\vartheta)}, \hspace{1.5em} i = 1, \ldots, m ,
\end{align}
\item[$ii)$] $\mathbf{q_{A,J}^{3}}$ is a $1 \times n$ vector given by $\big(\mathbf{q_{A,J}^{3}}\big)_{j} := 0, \;  j = 1, \ldots, n,$
\item[$iii)$] $\mathbf{q_{A,J}^{4}}$, $\mathbf{q_{A,J}^{5}}$, $\mathbf{q_{A,J}^{6}}$ are real values given by
\begin{align}
\mathbf{q_{A,J}^{4}} := 0, \hspace{1.5em} \mathbf{q_{A,J}^{5}}:= 0, \hspace{1.5em} \mathbf{q_{A,J}^{6}} := 0.
\end{align}                     
\end{itemize}
\noindent As a final remark, it is worth mentioning that the above values for $\mathbf{q_{A,0}^{5}}$ and $\mathbf{q_{A,J}^{5}}$ only hold under the assumption that $\sigma_{X} >0$. In fact, whenever $\sigma_{X} = 0$, hyper-exponential jump-diffusion processes reduce to finite activity pure jump processes and the corresponding continuous-fit conditions do not anymore hold at the boundary level $b^{\ast}$ (cf.~\cite{fmv19}). 
\end{proof}

\end{document}